\documentclass[11pt,letterpaper]{article}
\usepackage{jheppub}
\usepackage{style}

\begin{document}

\title{Suppressed Quantum Effects of Weakly Coupled Waves}

\author[1,2,3,4]{Yunjia Bao,}
\author[1,2,3,4]{Dhong Yeon Cheong,}
\author[5,6]{Nicholas L. Rodd,}
\author[5,6]{Joey Takach,}
\author[1,2,3,4]{Lian-Tao Wang,}
\author[5,6]{and Kevin Zhou}

\affiliation[1]{\footnotesize Enrico Fermi Institute, The University of Chicago, Chicago, IL 60637, USA}
\affiliation[2]{\footnotesize Department of Physics, The University of Chicago, Chicago, IL 60637, USA}
\affiliation[3]{\footnotesize Leinweber Institute for Theoretical Physics, The University of Chicago, Chicago,
IL 60637, USA}
\affiliation[4]{\footnotesize Kavli Institute for Cosmological Physics, The University of Chicago, Chicago,
IL 60637, USA}
\affiliation[5]{\footnotesize Theory Group, Lawrence Berkeley National Laboratory, Berkeley, CA 94720, USA}
\affiliation[6]{\footnotesize Leinweber Institute for Theoretical Physics, University of California, Berkeley, CA 94720, USA}

\emailAdd{yunjia.bao@uchicago.edu}
\emailAdd{dycheong@uchicago.edu}
\emailAdd{nrodd@lbl.gov}
\emailAdd{joseph\_takach@berkeley.edu}
\emailAdd{liantaow@uchicago.edu}
\emailAdd{kzhou7@berkeley.edu}

\abstract{Precision experiments increasingly target weakly coupled waves, including axion dark matter and gravitational radiation.
Such waves are commonly described as classical fields, yet they could exist in quantum states with no classical counterpart.
We exhibit two severe obstructions to detecting nonclassical effects, both independent of the mode occupancy.
First, realistic detectors cannot resolve the fundamental modes of a field; instead they couple to coarse-grained ``effective'' modes, which often washes out nonclassical effects.
Second, all nonclassical effects are suppressed by extra powers of the weak coupling, making them much harder to detect than the waves themselves.
We prove this in general, and explicitly show how the suppression arises for quadrature and number statistics, entanglement, and decoherence.
The suppression can in principle be overcome given suitable quantum resources, such as highly squeezed detector states, but the required parameters are far beyond current experimental capabilities.
We use the axion cavity haloscope as an explicit example, although our conclusions apply to many ultralight dark matter searches, and rule out proposals to establish the quantization of gravity from observations of gravitational waves.}

\maketitle
\setcounter{page}{2}

\vspace{-6mm}
\paragraph{Conventions.} We use natural units, $\hbar = c = k_B = 1$, and a mostly-negative spacetime metric.
Integrals and delta functions over coherent states are written as $d\alpha = d \, \text{Re}(\alpha) \, d \, \text{Im}(\alpha)$ and $\delta(\alpha) = \delta(\text{Re}(\alpha)) \, \delta(\text{Im}(\alpha))$.
Unless otherwise specified, a ``Gaussian state'' is the thermal Gaussian in Eq.~\eqref{eq:gaussian_state_def}.

\section{Introduction}
\label{sec:intro}
Modern experiments can now target weakly coupled waves of extraordinarily high mode occupancy.
Examples include gravitational wave (GW) detectors such as LIGO~\cite{LIGOScientific:2014pky,LIGOScientific:2016aoc}, and searches for ultralight dark matter (DM), exemplified by ADMX's search for the axion~\cite{ADMX:2001dbg,ADMX:2003rdr,ADMX:2009iij,ADMX:2018gho,ADMX:2019uok,ADMX:2021nhd}. 
Such experiments are deploying quantum measurement techniques~\cite{Fang:2024ple} such as squeezing of the detector's state~\cite{LIGOScientific:2013pcc,HAYSTAC:2020kwv,Jia:2024iqe} to reach beyond the standard quantum limit.
It is natural to ask if they could show that the field they seek to detect is quantized, or whether their results can be explained semiclassically, with a quantum detector coupled to a classical field.

Remarkably, discussions in the axion and GW communities suggest opposing conclusions, even though the two cases are closely analogous.
In the 1980s, the foundational works on axion DM~\cite{Abbott:1982af,Preskill:1982cy} treated the axion as a classical field. 
In the most-cited modern reviews~\cite{Jaeckel:2010ni,Marsh:2015xka,Irastorza:2018dyq}, this is justified by the fact that for $m_a \ll 10 \, \mathrm{eV}$, many axions occupy each field mode.
By contrast, in the GW literature, there have been recent high-profile claims~\cite{Parikh:2020kfh,Tobar:2023ksi,Schutzhold:2025vti} that GW detectors provide a path to establishing the quantization of gravity by, for instance, performing number measurements in the detector or detecting deviations from coherent state statistics.

In this work, we argue that both of these perspectives are incomplete.
First, high occupancy does not guarantee classical behavior; this is well-known in quantum optics~\cite{mandel1995optical,loudon2000quantum,barnett2002methods}, and has been emphasized by a subset of us for axion DM~\cite{Cheong:2024ose}.
Second, to determine if a measured result is inherently nonclassical, one must show that the same measurement statistics could not have arisen from any ensemble of classical fields.

For concreteness, we focus on resonant microwave cavity haloscopes for axion DM, in particular, the cylindrical cavities employed by the ADMX~\cite{ADMX:2001dbg,ADMX:2003rdr,ADMX:2009iij,ADMX:2018gho,ADMX:2019uok,ADMX:2021nhd}, HAYSTAC~\cite{HAYSTAC:2020kwv,HAYSTAC:2023cam,HAYSTAC:2024jch}, and CAPP/IBS-DMAG~\cite{CAPP:2020utb,CAPP:2024dtx,Ahn:2026ssw} collaborations. 
Consider a toy model of a haloscope in which the axion is modeled as a harmonic oscillator with lowering operator $a$, and the detector cavity is modeled as a harmonic oscillator with lowering operator $c$, interacting with Hamiltonian
\be \label{eq:H-toy}
H = \omega(c^\dag c + a^\dag a) + ig(c^\dag a - c a^\dag).
\ee
If the axion is classical, we can replace $a$ with a $c$-number $\alpha$, and a cavity in the vacuum state is excited to a coherent state with an amplitude proportional to $gt \alpha$.
Here $t$ is the maximum time over which the signal can accumulate, and for weakly coupled waves one always has $gt \ll 1$.
Nevertheless, they can still be detectable by virtue of having large amplitude $\alpha$.

If the axion is quantum and in a coherent state $\ket{\alpha}$, the cavity state evolves in exactly the same way, at first order in $gt$.
Thus, for weakly coupled waves, coherent states act like classical field values. 
More generally, if the axion is in a mixture of coherent states, 
\be \label{eq:GS_P}
\rho = \int d\alpha \, P(\alpha) \ket{\alpha}\bra{\alpha}
\ee
for $P(\alpha) \geq 0$, then it acts like a probabilistic ensemble of classical field values.
As for nonclassical effects, Glauber and Sudarshan showed~\cite{Glauber:1963tx,Sudarshan:1963ts} that \textit{any} state can be written in the form of Eq.~\eqref{eq:GS_P} for a real $P$.
However, one must generalize $P$ to include cases where it is not a valid probability distribution, either because it takes negative values or contains distributions more singular than a delta function.\footnote{The distributions in the latter case are effectively negative, in the sense that they can yield negative values when integrated against a nonnegative test function.
Thus, for brevity we will simply refer to both cases as ``negative $P$'' in this work.}
States with negative $P(\alpha)$ are intrinsically quantum, in the sense that they can yield measurement statistics that cannot be realized by any classical ensemble.
This broad definition of nonclassicality encompasses squeezing, entanglement, and Wigner negativity, which is related to quantum magic~\cite{Bravyi:2004isx}.
Examples of nonclassical states include squeezed states, Fock states, and ``cat'' states such as $\ket{\psi} \propto \ket{\alpha} + \ket{-\alpha}$, all of which can have arbitrarily high occupancy.

\begin{figure}[t!]
\centering
\includegraphics[width=0.6\linewidth]{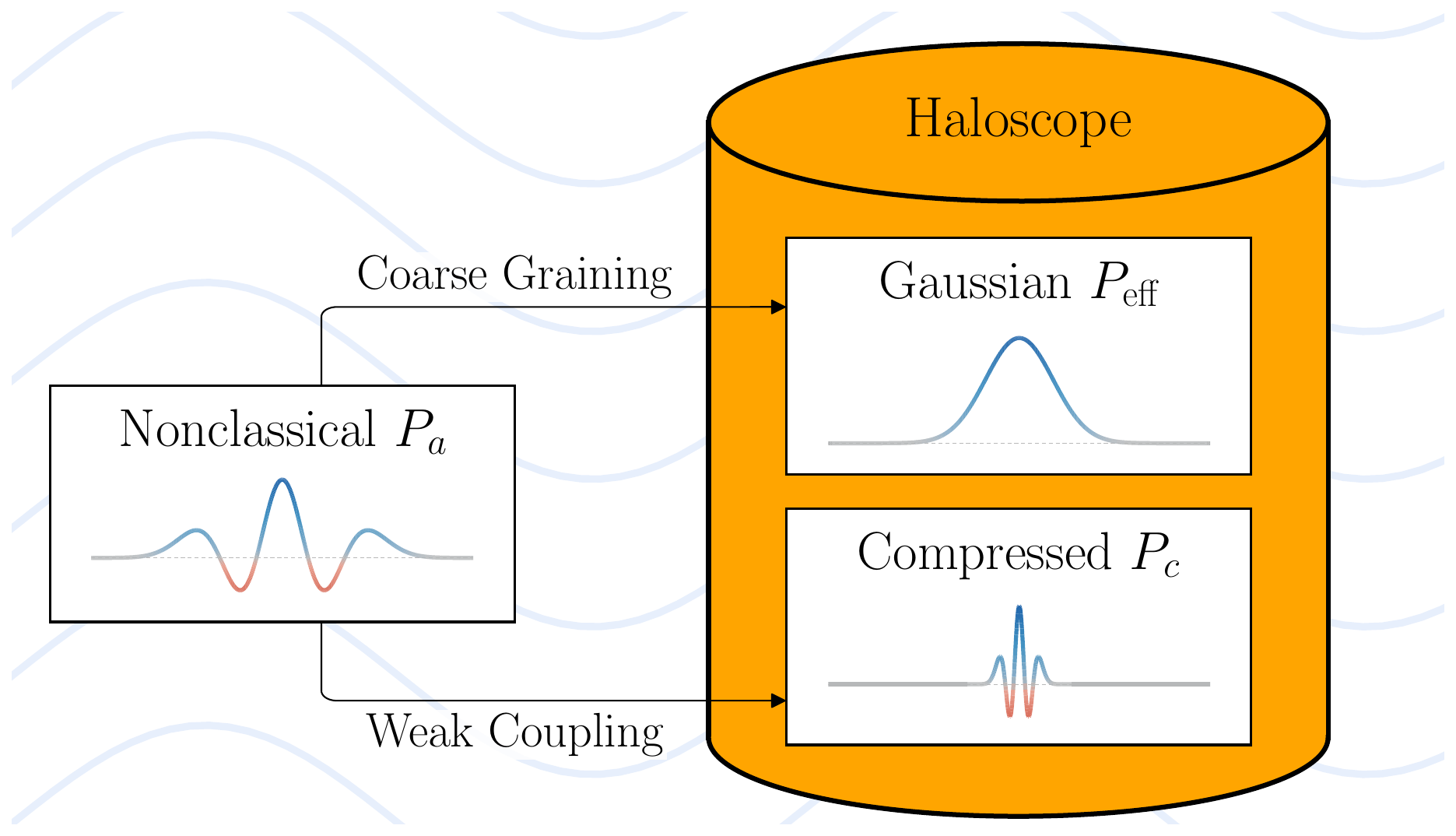}
\vspace{-0.3cm}
\caption{Two obstructions to observing intrinsically nonclassical effects from weakly coupled waves, such as axion dark matter.
Even if the axion state $P_a$ is nonclassical, the haloscope only couples to a coarse-grained effective mode $P_\eff$, which is often in a thermal Gaussian state, erasing nonclassical effects.
Weak coupling maps the axion state onto the cavity $P_c$ in a compressed form with finely spaced sign oscillations, suppressing the magnitude of observable nonclassical effects.
These issues are neither caused nor avoided by high mode occupancy.}
\label{fig:intro_key_arguments}
\end{figure}

Given this definition of nonclassicality, there are two independent obstructions to observing nonclassical behavior even if the field exists in such a state, illustrated in Fig.~\ref{fig:intro_key_arguments}.
First, the axion (or gravitational) field actually has infinitely many modes.
However, detectors cannot resolve these modes individually; instead one can regard them as coupling to ``effective'' modes, each constructed from many approximately plane-wave fundamental modes.
When the (quantum) central limit theorem (CLT) applies, this coarse-graining generically erases negativity in the effective mode's $P$-function $P_{\text{eff}}(\alpha)$.

Second, intrinsically quantum modifications to measurement statistics are always penalized by \textit{additional} powers of $gt$, as we showed in Ref.~\cite{Bao:2025nsd} for the simplest nonclassicality measures, entering without a compensating power of $\alpha$.
In this work we greatly extend these results, showing that for any nonclassical state, there always exists a classical ensemble whose statistics are identical up to effects that are higher-order in $gt$.
This implies that detecting nonclassical features of a wave is always parametrically more difficult than detecting the wave itself.
However, this fact was not accounted for in dozens of previous studies, which either compared a nonclassical state to a single fixed classical state, or considered signatures that were not intrinsically quantum.

We organize our discussion around these two obstructions as follows.
In Sec.~\ref{sec:toy} we study the toy Hamiltonian in Eq.~\eqref{eq:H-toy} in detail; we solve the model explicitly, compute the evolution of the cavity $P$-function, and introduce examples of nonclassical states.
In Sec.~\ref{sec:modes} we generalize to a realistic cavity haloscope, which couples to the full axion field.
The dynamics of each cavity mode are equivalent to the toy model for a suitably defined effective mode with occupancy $N_\eff$.
Assuming DM populates a large number of independent modes, as might be expected under standard virialization, the quantum CLT implies the effective mode's $P$-function is a thermal Gaussian, which is nonnegative.
This would already erase all nonclassical effects, though there are also alternative scenarios where the CLT does not apply.

Assuming this first barrier is overcome, we turn to the difficulty of observing nonclassical effects at weak coupling.
We use an idealized measurement model in which the cavity is repeatedly prepared in a fixed state, interacts with the effective mode, and is projectively measured after a time $t_m$. 
This is a reasonable model for some ultralight DM experiments; for example, transmon qubits can be used to measure the photon number inside a cavity~\cite{Chakram:2021bxb,Dixit:2020ymh,Gu:2025pms}, and to prepare the cavity in a Fock state~\cite{Agrawal:2023umy} or a cat state~\cite{Zheng:2025qgv}.
(Other axion experiments are better described by continuous measurement.
We treat this case in detail in a companion work~\cite{continuous_paper} and find similar conclusions.)

In Sec.~\ref{sec:projective}, we show that many nonclassicality signatures involving number, quadrature, and entanglement are suppressed by the small axion-photon conversion efficiency $\eta \sim g^2 t_m^2$.
While nonclassical effects are readily observable in quantum optics, where $\eta \sim 1$, for axions and GWs the extremely small value of $\eta$ (e.g.~$\eta \sim 10^{-21}$ for a cavity haloscope) strongly suppresses nonclassical effects.
We also present a general argument for any single-cavity measurement, and show that detecting nonclassical effects requires either integration times or quantum resources scaling as powers of $1/\eta$, both of which are far outside the reach of existing experiments. 
In Sec.~\ref{sec:decoherence}, we consider the decoherence induced by axion DM, but again find that nonclassical states only extend the range by an amount suppressed by $\eta$.
These results greatly generalize previous results for axion DM~\cite{Bao:2025nsd} and GW detection~\cite{Carney:2023nzz,Carney:2024dsj}, which focused on the simplest nonclassicality measures.

Finally, in Sec.~\ref{sec:discussion} we outline how our conclusions extend to other ultralight DM experiments and GWs.
We critically examine claims concerning detectable quantum axion DM or GW effects, and conclude that even when highly sensitive experiments successfully detect these external, weakly coupled waves, it is currently impossible for them to establish the quantization of the corresponding field.

\section{A Toy Model for Wave Dark Matter Detection}
\label{sec:toy}
In the simplest possible model of a cavity haloscope, the axion field and the cavity mode are each modeled as a single harmonic oscillator, coupled by a weak interaction between their dimensionless quadratures, $H_\text{int} = 2 g Y_c X_a$, where $g$ has units of frequency, and
\be \label{eq:quad_defs}
X_a = \frac{1}{\sqrt 2}(a+a^\dag), \qquad Y_c = \frac{-i}{\sqrt 2}(c-c^\dag).
\ee
We let both modes have angular frequency $\omega$, so that their interaction is resonantly enhanced.\footnote{The utility of this simple model was also emphasized in Ref.~\cite{Brubaker:2017ohw}.}
The coupling $g$ governs the rate of energy exchange between the modes.
This model neglects cavity dissipation, which is appropriate on timescales $t \lesssim Q_c/\omega$, where $Q_c$ is the cavity quality factor.
We assume this condition holds throughout this work, and in Ref.~\cite{continuous_paper} we discuss how dissipation can be incorporated for continuous measurements.

\paragraph{Mode Evolution.}
%
The Heisenberg equations of motion for the lowering operators are 
\begin{subequations} \label{eq:Heisenberg_eom}
\begin{align}
\dot c &= i [H, c] = -i\omega c + g a + g a^\dagger, \\
\dot a &= i [H, a] = -i\omega a - g c + g c^\dagger.
\end{align}
\end{subequations}
In the absence of the interaction, both $a$ and $c$ rotate with angular frequency $\omega$.
Then in each line of Eq.~\eqref{eq:Heisenberg_eom}, the second term rotates in the same way, and hence can have a resonantly enhanced effect, while the final term rotates with angular frequency $-\omega$ and is thus far off resonance.
We thus perform the \textit{rotating wave approximation}, dropping the latter terms. 
This is equivalent to dropping terms proportional to $ac$ and $a^\dagger c^\dagger$ in $H_{\text{int}}$, which do not conserve the total excitation number, thus recovering the Hamiltonian in Eq.~\eqref{eq:H-toy}.

\begin{figure}[t!]
\centering
\includegraphics[width=0.8\linewidth]{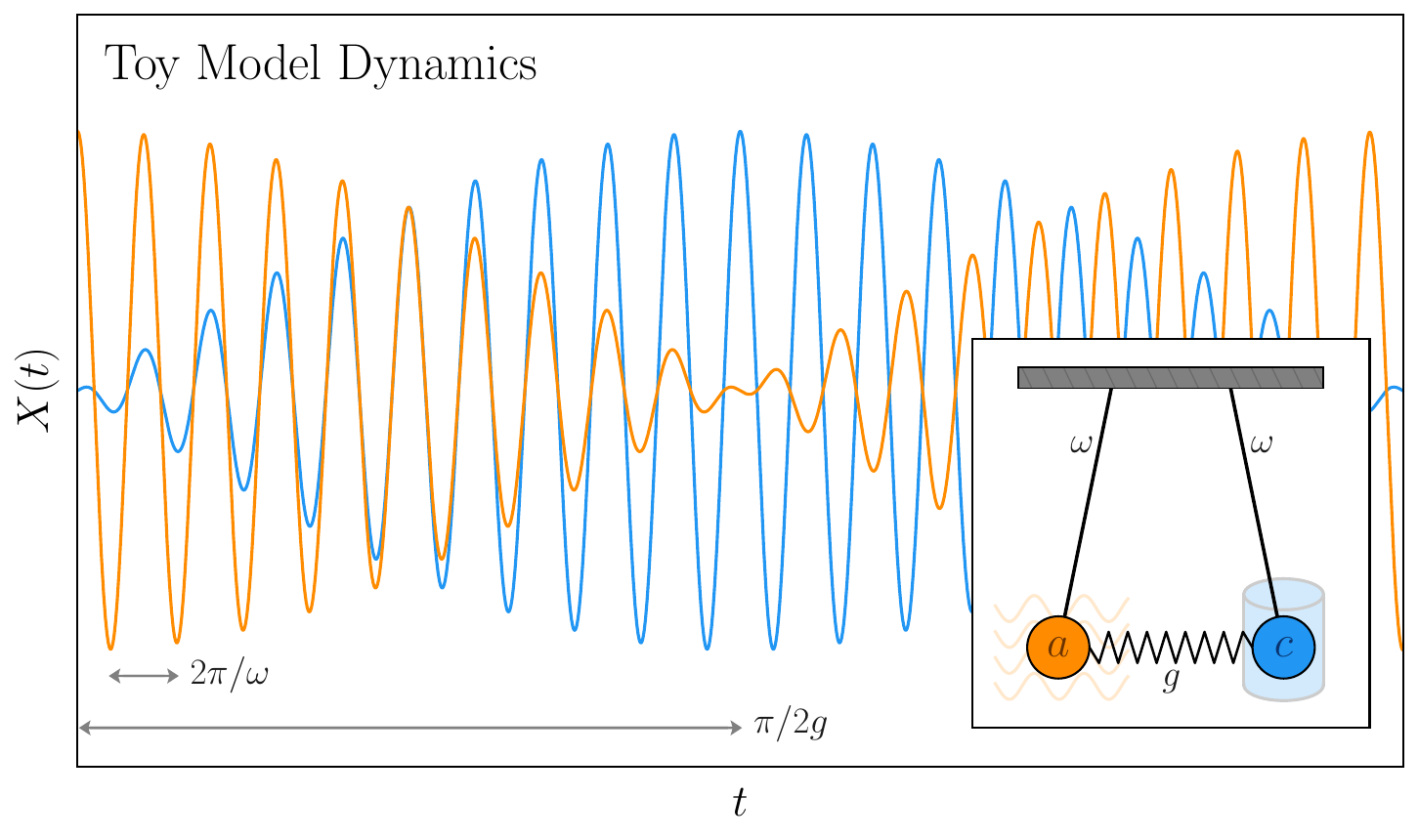}
\vspace{-0.3cm}
\caption{The dynamics of the toy model in Eq.~\eqref{eq:H-toy}, which describes the evolution of two weakly coupled oscillators of angular frequency $\omega \simeq m_a$.
We show throughout this work that this model can be an accurate effective description of certain axion DM detection strategies.
The plot takes $g/\omega = 1/40$, whereas for DM the coupling is \textit{much} weaker; ADMX targets axions with $g/\omega \sim g_{a\gamma\gamma} B_0 / \omega \sim 10^{-16}$.}
\label{fig:ToyModel}
\end{figure}

It is then straightforward to solve the equations of motion, yielding\footnote{If we had not performed the rotating wave approximation, there would be additional $\mathcal{O}(g/\omega)$ terms on the right-hand sides.
At $t \sim Q_c / \omega$, they would be $Q_c$ times smaller than the resonantly enhanced terms.}
\begin{subequations} \label{eq:acHP}
\begin{align}
c(t) &= e^{-i\omega t}\bigl[c(0) \cos gt + a(0) \sin gt \bigr], \\
a(t) &= e^{-i\omega t}\bigl[a(0) \cos gt - c(0) \sin gt \bigr].
\end{align}
\end{subequations}
As for two coupled classical oscillators, the axion and cavity exchange energy on a timescale of order $1/g$, though in practice this would be cut off by their finite quality factors.
The simple dynamics of the model are shown for a quadrature of the cavity and axion in Fig.~\ref{fig:ToyModel}.
(It is even possible to exactly solve the toy model with $N$ modes, as we show in App.~\ref{app:exact_solution}.)

\paragraph{Quantum State Evolution.} 
%
The joint density matrix of the axion mode and cavity mode can be written in terms of the joint $P$-function
\be
\rho = \int d\alpha \, d\gamma \, P(\alpha,\gamma) \ket{\alpha,\gamma}\bra{\alpha,\gamma}
\ee
with $\alpha$ and $\gamma$ labeling the axion and cavity, respectively.
To compute the evolution of the $P$-function, we use the fact that the ladder operator evolution in Eq.~\eqref{eq:acHP} only mixes the annihilation operators, so that time evolution maps coherent states to other coherent states.
Explicitly, in the Schr\"odinger picture, the joint coherent state $\ket{\alpha, \gamma}$ evolves to 
\be \label{eq:joint_coherent}
U(t) \ket{\alpha, \gamma} = \big\vert e^{-i \omega t} (\alpha \cos gt - \gamma \sin gt),\, e^{- i \omega t} (\gamma \cos gt + \alpha \sin gt) \big\rangle.
\ee
This can be confirmed by acting on both sides with either $a$ or $c$ and noting that, e.g.~$c\, U(t) = U(t) c(t)$, where $c(t)$ is given by Eq.~\eqref{eq:acHP}.
For the rest of this section we move to the interaction picture, where the trivial $e^{- i \omega t}$ phase factors are removed from the state evolution.

We take the initial axion and cavity states to be independent, so that the initial $P$-function factorizes as $P(\alpha,\gamma) = P_a(\alpha) P_c(\gamma)$. 
Then upon applying Eq.~\eqref{eq:joint_coherent} and performing a change of variables, we find that the joint density matrix evolves to
\be \label{eq:joint_state}
\rho(t) = \int d\alpha \, d\gamma \, P_a(\alpha \cos gt + \gamma \sin gt) P_c(\gamma \cos gt - \alpha \sin gt) \,\ket{\alpha,\gamma}\bra{\alpha,\gamma}.
\ee
Experiments probe only the cavity state $\rho_c(t)$, constructed by tracing out the axion mode, 
\be
\rho_c(t) = \int d\gamma \, P_c(\gamma,t) \ket{\gamma}\bra{\gamma}.
\ee
Here the final cavity state $P$-function is
\be \label{eq:final_cavity_P}
P_c(\gamma,t) = \int d\alpha \, P_a(\alpha \cos gt + \gamma \sin gt) P_c(\gamma \cos gt - \alpha \sin gt).
\ee
Given the weakness of DM couplings, we will always be in the limit $gt \ll 1$ even for the highest achievable quality factors, so it is useful to define the small conversion efficiency
\be \label{eq:efficiency}
\eta = \sin^2 (gt) \ll 1.
\ee
We may then equivalently write Eq.~\eqref{eq:final_cavity_P} as 
\be \label{eq:approx_cavity_P}
P_c(\gamma, t) = \int d\alpha \, \frac{P_a(\alpha / \sqrt{\eta})}{\eta} \frac{P_c((\gamma - \alpha) / \sqrt{1-\eta})}{1 - \eta} \simeq \int d\alpha \, \frac{P_a(\alpha / \sqrt{\eta})}{\eta} P_c(\gamma - \alpha).
\ee
That is, to leading order in $\eta$, the final cavity $P$-function is a convolution of the original cavity $P$-function with the axion $P$-function, with the latter scaled down by $\sqrt{\eta}$.

\paragraph{Classical State Evolution.}
%
The preceding derivation treated the axion quantum mechanically.
To treat the axion classically, we replace $a$ and $a^\dagger$ with the c-numbers $\alpha$ and $\alpha^*$, where $\alpha$ has a probability distribution $P_a(\alpha)$.
The interaction Hamiltonian becomes
\be \label{eq:H_classical}
H_{\text{int}} = ig \big(c^\dag \alpha - c \alpha^* \big).
\ee
If $\alpha$ has a definite value, then an initial cavity coherent state $\ket{\gamma}$ evolves to $U(t) \ket{\gamma} \propto \ket{\gamma + \alpha g t}$ in the interaction picture, which is simply Eq.~\eqref{eq:joint_coherent} without the trigonometric factors.
Then by analogous logic, the evolution of the cavity $P$-function is 
\be \label{eq:classical_cavity_P}
P_c(\gamma, t) = \int d\alpha \, \frac{P_a(\alpha / gt)}{g^2 t^2} P_c(\gamma - \alpha)
\ee
which matches the result in Eq.~\eqref{eq:approx_cavity_P} for a quantum axion up to corrections suppressed by $\eta$.

These corrections represent the fact that when one treats the axion as a classical background, there is no backreaction onto the axion field (i.e.~no depletion from conversion to photons, or excitation from converting photons to axions).
Indeed, in the classical approximation it is not clear how to account for backreaction at all, since there is no canonical, generally consistent prescription to drive a classical mode with a quantum one.
We return to this issue in Sec.~\ref{sec:discussion}.
However, we can already state that for cavity haloscopes the effective value of $\eta$ is extremely small, so this difference is negligible.
Thus, for all practical purposes, quantum axion states with nonnegative $P_a(\alpha)$ act like ensembles of classical fields.

\paragraph{Computing Observables.}
%
We can use $P_c(\gamma, t)$ to calculate any observable of interest at time $t$.
For instance, the expectation value of an operator $\mathcal{O}$ is given by
\be \label{eq:O_expt}
\la \mathcal O(t)\ra = \tr\!\big(\rho_c(t) \mathcal O \big) = \int d\gamma \, P_c(\gamma,t) \bra{\gamma}\mathcal O \ket{\gamma}.
\ee
In particular, if $\mathcal{O}(c^\dag, c)$ is normally ordered, then $\bra{\gamma}\mathcal{O}(c^\dag, c) \ket{\gamma} = \mathcal{O}(\gamma^*,\gamma)$, a result known as the optical equivalence theorem~\cite{mandel1995optical}.

As another example, if the cavity mode starts in the vacuum state, $P_c(\gamma) = \delta(\gamma)$, then the expected number of photons in the cavity at time $t$ is
\bea \label{eq:nc-na-calc}
\la n_c(t) \ra &= \int d\gamma\,|\gamma|^2 \,P_c(\gamma,t)
= \int d\alpha \, d\gamma \,|\gamma|^2\, \frac{P_a(\alpha / \sqrt{\eta})}{\eta} \delta(\gamma - \alpha) \\
&= \eta \int d\alpha \,|\alpha|^2 P_a(\alpha)
= \eta \la n_a \ra.
\eea
That is, for an initial vacuum cavity state, $\eta$ is the expected fraction of axions that convert to cavity photons, which is why it was introduced as the conversion efficiency above.

\paragraph{Coherent Axion State.}
%
If the axion begins in a coherent state $\ket{\alpha_0}$, corresponding to $P_a(\alpha) = \delta(\alpha - \alpha_0)$, then the cavity state evolves to 
\be \label{eq:coherent_evolution}
P_c(\gamma, t) = \frac{1}{1-\eta} \, P_c \!\left( \frac{\gamma-\sqrt{\eta} \, \alpha_0}{\sqrt{1-\eta}} \right) \simeq P_c(\gamma-\sqrt{\eta} \, \alpha_0).
\ee
Up to a negligible scaling, the $P$-function of the cavity is simply translated by $\sqrt{\eta} \, \alpha_0$, and we would get the same result if the axion had an initial classical value $\alpha_0$.

Let us briefly review basic properties of coherent states for later use. 
Coherent states are pure states, with overlap $|\la \alpha | \beta \ra|^2 = e^{-|\alpha - \beta|^2}$.
The coherent state $\ket{\alpha}$ has a Poisson number distribution with mean $|\alpha|^2$.
Its dimensionless quadratures have expectation values $(\langle X \rangle, \langle Y \rangle) = \sqrt{2} \, (\text{Re}(\alpha), \text{Im}(\alpha))$, and the quadrature variances are $\var(X) = \var(Y) = 1/2$.

\paragraph{Mixtures of Coherent States.}
%
We can construct other axion states as probabilistic mixtures of coherent states, in which case the axion acts as a classical field with unknown value.
For example, if we know the magnitude of the axion field but not its phase, 
\be \label{eq:phaseMixtureP}
P_a(\alpha) = \frac{1}{\pi} \, \delta \Big(|\alpha|^2 - |\alpha_0|^2 \Big).
\ee
This state has the same mean occupancy and Poisson number distribution as the coherent state $\ket{\alpha_0}$, yet it has a vanishing expectation value for both quadratures.
Unlike the coherent state, the phase-randomized coherent state obeys $P(\alpha)=P(|\alpha|)$ and is therefore stationary (see App.~\ref{app:stationarity}).
Using tools to be introduced in Sec.~\ref{sec:gaussian_clt}, one can show the purity of the state is 
\be \label{eq:phase_averaged_purity}
\tr (\rho_a^2) = e^{-2|\alpha_0|^2} I_0(2|\alpha_0|^2),
\ee
where $I_0$ is the modified Bessel function of the first kind.
For $|\alpha_0| > 0$, this is less than one, indicating the state is mixed, and it monotonically decreases with increasing $|\alpha_0|$.

Another important example is the Gaussian state, which also has uncertain magnitude,
\be \label{eq:gaussian_state_def}
P_a(\alpha) = \frac{1}{\pi |\alpha_0|^2} \, \exp(- \frac{|\alpha|^2}{|\alpha_0|^2}).
\ee
Like the previous examples, this state has mean occupancy $|\alpha_0|^2$. It is mixed, with purity $1/\big(1+2|\alpha_0|^2\big)$, and it has a Bose--Einstein (geometric) number distribution, $p_n = |\alpha_0|^{2n}/\big(1+|\alpha_0|^2\big)^{n+1}$. 
It is also called a ``thermal'' state, since it corresponds to a thermal density matrix for an effective temperature $T_\eff$ satisfying $|\alpha_0|^2 = 1 / \big(e^{\omega / T_\eff} - 1\big)$.

\begin{figure}[t!]
\centering
\includegraphics[width=0.85\linewidth]{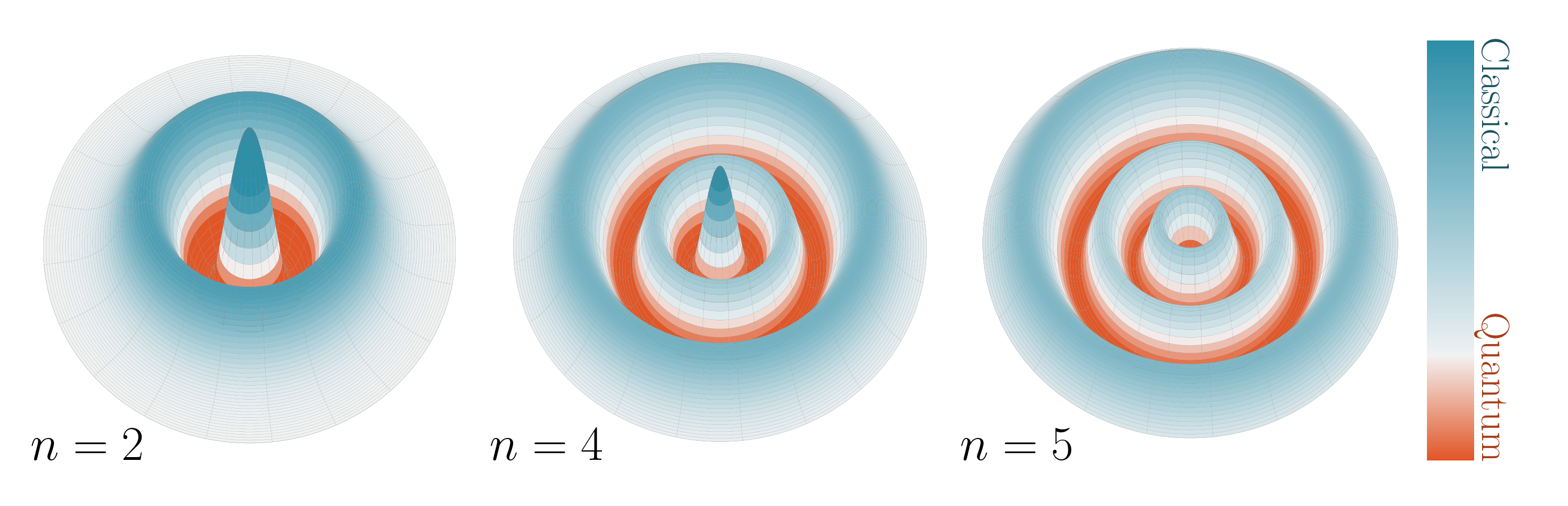}
\vspace{-0.3cm}
\caption{$P$-functions for nonclassical DM states, constructed by adding $n$ axions to the Gaussian state of Eq.~\eqref{eq:gaussian_state_def} with mean occupancy $|\alpha_0|^2=1.2$.
Note the rapid sign oscillations for higher $n$.
Negative regions shown in red yield effects that cannot be reproduced with a classical axion model.}
\label{fig:Pexamples}
\end{figure}

\paragraph{Intrinsically Nonclassical States.}
%
The preceding examples are all effectively classical, because they involve nonnegative $P$-functions; in all these cases the axion can be treated as a random c-number with a classical probability distribution $P_a(\alpha)$.
However, many simple states, such as Fock states, squeezed states, and the cat state $\ket{\alpha} + \ket{-\alpha}$, cannot be described this way, and thus can yield measurement statistics unattainable by any classical state.
For example, as we discuss in detail in Sec.~\ref{sec:number_meas}, for classical states the variance of the number is always at least as great as the mean, but Fock states have zero number variance.

In Fig.~\ref{fig:Pexamples}, we show the $P$-functions of nonclassical states constructed by adding $n$ axions to a Gaussian state $\rho_G$ with a mean occupancy of $|\alpha_0|^2 = 1.2$, corresponding to the density matrix $\rho_a \propto (a^\dagger)^n \rho_G \, a^n$.
This illustrates that negativity can survive a finite amount of Gaussian noise.
As $n$ increases, the $P$-function displays many sign oscillations; this is a common phenomenon for high-occupancy nonclassical states. 
In general, the negativity of any $P$-function is washed out by smoothing over an $\order{1}$ scale in the dimensionless quadratures, and we show in Sec.~\ref{sec:general_suppression} that this generically suppresses the visibility of quantum effects.

In the limit $|\alpha_0|^2 \to 0$, the oscillations occur more and more tightly about the origin, and for $\alpha_0 = 0$ we have a Fock state, whose $P$-function is a highly singular distribution located at the origin.
This is a common phenomenon for simple idealized nonclassical states; the $P$-function of a squeezed state is a singular distribution as well.
In these cases the $P$-function cannot be readily visualized, but as shown in Sec.~\ref{sec:projective}, quantum effects remain suppressed.

\section{Effective Modes and Gaussian Fields}
\label{sec:modes}
Here we move from the toy model to the axion-photon coupling in a cavity haloscope.
The main difference is that, although we can generally isolate a single cavity mode of interest, that mode couples to an approximate continuum of plane-wave axion modes.
Nevertheless, in Sec.~\ref{sec:effective_intro} we show that each cavity mode can be regarded as coupling to a single \textit{effective} axion mode, so that the dynamics of a cavity haloscope can be mapped back onto the toy model of Sec.~\ref{sec:toy}.

In Sec.~\ref{sec:gaussian_clt}, we show how to compute the quantum state of the axion effective mode.
We argue that in many circumstances the detector sees a mixed Gaussian state described by Eq.~\eqref{eq:gaussian_state_def}, even if the axion plane-wave modes are in pure or nonclassical states.
This is a fundamental obstruction to observing nonclassical behavior of axion DM.

\subsection{Effective Modes in a Cavity Haloscope}
\label{sec:effective_intro}

In a cavity of volume $V_c$, the axion-photon coupling $\Lag \supset - \ga \phi F_{\mu\nu} \tilde{F}^{\mu\nu}/4$ corresponds to the interaction Hamiltonian
\be \label{eq:Hint_initial}
H_{\text{int}} = - \ga \int_{V_c} \hspace{-0.1cm} d^3 \bx \, \phi(\bx)\, \bE(\bx) \cdot \bB(\bx),
\ee
where $\bB = B_0 \, \hat{\vb{z}}$ is a strong classical background magnetic field.
The axion field in the Schr\"odinger picture can be quantized with the usual plane wave mode expansion,\footnote{As usual, one is free to pass to a discrete set of modes by placing the DM in a finite volume $\mathcal{V}$.
In either case, the use of plane waves for the basis functions is appropriate when the relevant spatial structure of the DM is on scales far larger than probed by the experiment, e.g.~if it were set by the gravitational potential of the galaxy.
There are situations where this need not be the case, such as when gravitational focusing by the Sun is relevant~\cite{Kim:2021yyo}, or if a component of DM were in a Bose--Einstein condensate (BEC) around the Sun~\cite{Budker:2023sex}.}
\be \label{eq:phi_x}
\phi(\bx) = \int\! \frac{d^3 \bk}{(2\pi)^3}\, \frac{1}{\sqrt{2\omega_\bk}} \Big(a_\bk e^{i \bk \cdot \bx} + \hc\Big),
\ee
with $\omega_\bk^2 = m_a^2 + |\bk|^2$ and the axion ladder operators obeying $[a_\bk, a_{\bk'}^\dagger] = (2 \pi)^3 \delta^{(3)}(\bk - \bk')$.
The electric field is restricted to the cavity volume and can accordingly be decomposed into a discrete set of modes indexed by $\ell$.
We review the explicit construction of cavity modes in App.~\ref{app:cavity_modes} and here simply state the result for the electric field,
\be \label{eqn:cavEModeExpansion}
\bE(\bx) = \sum_\ell i \omega_\ell \bu_\ell^*(\bx) c_\ell + \hc,
\ee
where for each cavity mode $\omega_\ell$ is the angular frequency, $c_\ell$ is the annihilation operator, and $\bu_\ell(\bx)$ is the spatial profile, obeying the orthogonality relation\footnote{In Ref.~\cite{Bao:2025nsd} we instead used unit-normalized mode profiles, $\tilde{\bE}_\ell = \sqrt{2\omega_\ell} \, \bu_\ell$.}
\be \label{eq:orthog_relation}
\int_{V_c} \hspace{-0.1cm} d^3 \bx\, \bu_\ell^*(\bx) \cdot \bu_{\ell'}(\bx) = \frac{\delta_{\ell\ell'}}{2\omega_\ell}.
\ee

\paragraph{Defining Effective Modes.}
%
After direct substitution, the Hamiltonian becomes
\be \label{eq:Hint_expanded}
H_{\text{int}} = i \ga B_0
\sum_\ell \int \!\frac{d^3 \bk}{(2\pi)^3}\, c^\dag_\ell\, a_\bk \left[ \frac{1}{2} \omega_\ell\, \sqrt{\frac{2}{\omega_\bk}}
\,\int_{V_c} \hspace{-0.1cm} d^3 \bx\, u_\ell(\bx) e^{i \bk \cdot \bx} \right] + \hc,
\ee
where $u_\ell(\bx) = \bu_\ell(\bx)\cdot \hat{\vb{z}}$ and we have applied the rotating wave approximation by dropping terms with two creation or annihilation operators, as their effects are not resonantly enhanced.

Up to a sum over modes $\ell$ and $\bk$, the Hamiltonian displays a qualitative similarity to the toy model of Eq.~\eqref{eq:H-toy}, with interaction rate $g \sim \ga B_0$.
The bracketed expression contains a dimensionless form factor
\be \label{eq:formfactor}
C_\ell(\bk) = \frac{\sqrt{2}\, \omega_\ell}{\sqrt{\omega_\bk V_c}}
\int_{V_c} \hspace{-0.1cm} d^3 \bx\, u_\ell(\bx) e^{i \bk \cdot \bx}
\ee
which quantifies the overlap between the cavity mode and an axion plane-wave mode.
For non-relativistic DM we can focus on $\bk = \vb{0}$, and taking the cavity mode on resonance, $\omega_\ell=m_a$, we have the usual cavity haloscope form factor
\be
|C_\ell(\vb{0})|^2 = \frac{2\, \omega_\ell}{V_c}
\left|\int_{V_c} \hspace{-0.1cm} d^3 \bx\, \bu_\ell(\bx) \cdot \hat{\bf{z}} \right|^2\!.
\ee
More generally, $C_\ell(\bk)$ is the cavity form factor for an arbitrary axion momentum $\bk$.\footnote{A similar form factor appears in the context of searches for the relativistic cosmic axion background~\cite{Dror:2021nyr,ADMX:2023rsk}.}

Rewriting Eq.~\eqref{eq:Hint_expanded} using the form factor, we have
\be
H_{\text{int}} = i \ga B_0
\sum_\ell c^\dag_\ell\, \left[ \frac{1}{2} \sqrt{V_c} \int \!\frac{d^3 \bk}{(2\pi)^3}\, C_\ell(\bk) a_\bk \right]\! + \hc
\ee
Each cavity mode couples to a linear combination of axion modes, weighted by their overlap $C_\ell(\bk)$.
We can formalize this by introducing \textit{effective modes} for the axion, $a_\ell$, defined by
\be \label{eq:effective_mode_def}
a_\ell = \frac{1}{2} \sqrt{\frac{V_c}{\Omega_\ell}} \int\! \frac{d^3\bk}{(2\pi)^3} \, C_\ell(\bk)\,a_\bk, \qquad 
\Omega_\ell = \frac{V_c}{4}\int \!\frac{d^3\bk}{(2\pi)^3} \, |C_\ell(\bk)|^2.
\ee
Here $\Omega_\ell$ is a dimensionless factor introduced to ensure the modes are normalized: $[a_\ell, a_\ell^\dagger] = 1$.
In terms of these modes the Hamiltonian takes the concise form
\be \label{eq:Hint_simplified}
H_\text{int} = i\sum_\ell g_\ell \big( c_\ell^\dag a_\ell - c_\ell a_\ell^\dag \big), 
\qquad 
g_\ell = \ga B_0 \sqrt{\Omega_\ell}.
\ee

This shows that the interaction Hamiltonian of the axion-photon coupling can be written in the form of the toy model of Sec.~\ref{sec:toy}, up to two differences.
First, the expression involves a sum over cavity modes.
Experiments often isolate an individual mode of interest, so when relevant, we neglect all but one $\ell$; when doing so we also drop the subscripts for simplicity.
Second, the analogy presupposes that we can treat effective modes as conventional creation and annihilation mode operators.
We explore the subtleties of this analogy next.

\paragraph{Features of Effective Modes.}
%
Effective modes have several subtle features.
First, an excitation of an effective mode, generated by acting with $a_\ell^\dagger$, is \textit{not} an eigenstate of the free Hamiltonian, since it is formed by superposing axion plane-wave modes of different energies.
Equivalently, in the interaction picture, the operators $a_\bk(t) = e^{- i\omega_\bk t} a_\bk(0)$ in Eq.~\eqref{eq:effective_mode_def} do not rotate with the same angular frequency, so the identity of the effective mode $a_\ell(t)$ changes over time. 
However, since DM is nonrelativistic, we have $\omega_\bk \simeq m_a + k^2 / 2 m_a$, so that the range of \textit{occupied} angular frequencies scales as $\Delta \omega_a \sim m_a v_{\DM}^2 \sim 10^{-6} \, m_a$, corresponding to an axion coherence time $\tau_c \sim 1 / \Delta \omega_a$ (for a precise definition of $\tau_c$, see Ref.~\cite{Cheong:2024ose}). 
Thus, over measurement timescales shorter than the axion coherence time, $t_m \lesssim \tau_c$, we can consider the effective mode's state as fixed.

In addition, effective modes do \textit{not} commute with each other: $[a_\ell, a_m^\dag]$ can be nonzero for $\ell \neq m$.
We can see this explicitly by computing
\bea \label{eq:effective_commutator}
{[}a_\ell, a_m^\dag{]} &= \frac{V_c}{4\sqrt{\Omega_\ell \Omega_m}}\int \frac{d^3\bk}{(2\pi)^3} \, C_\ell(\bk) C^*_m(\bk) \\
&= \frac{\omega_\ell\omega_m}{\sqrt{\Omega_\ell \Omega_m}} \int_{V_c}\hspace{-0.1cm} d^3\bx \int_{V_c}\hspace{-0.1cm} d^3 \vb y\, u_\ell(\bx) u^*_m(\vb y) \int\! \frac{d^3 \bk}{(2\pi)^3} \frac{e^{i\bk \cdot (\bx -\vb y)}}{2\omega_\bk}.
\eea
If the denominator at the end of the expression had $m_a$ instead of $\omega_\bk$, then for the transverse magnetic (TM) modes of an azimuthally symmetric cavity, as relevant for an axion haloscope, the expression would vanish for $\ell \neq m$ by the orthogonality of the $u_\ell$; see Eq.~\eqref{eq:ul_norm}.
However, in reality the integral has support up to semi-relativistic axion modes, so the integral need not vanish even approximately.

The failure of orthogonality is unsurprising: precisely the same phenomenon occurs for a free relativistic scalar field in position space.
We can decompose the scalar field in Eq.~\eqref{eq:phi_x} into positive and negative frequency modes as $\phi(\bx) = \phi^+(\bx) + \phi^-(\bx)$ with
\be \label{eq:phi_pos}
\phi^+(\bx) = \int\! \frac{d^3 \bk}{(2\pi)^3}\, \frac{1}{\sqrt{2\omega_\bk}} \, a_\bk e^{i \bk \cdot \bx}
\ee
and $\phi^- = (\phi^+)^\dag$.
Then $[\phi^+(\bx),\phi^-(\vb y)] \neq 0$, even though $[\phi(\bx),\phi(\vb y)] = 0$ as required for causality.
The physical interpretation is that one cannot define perfectly localized excitations of a relativistic field.
Similarly, in our case where $a_\ell$ plays the role of $\phi^+(\bx)$, we learn that excitations of an effective mode generically couple to other cavity modes as well.\footnote{Even though $[a_\ell, a_m^\dag] \neq 0$, if we construct $\phi_\ell = a_\ell + a_\ell^\dag$, then for a generic cavity geometry $[\phi_\ell,\phi_m]=0$, similar to $\phi(\bx)$.
This can be seen most easily by noting it is always possible to choose a basis where $u_\ell(\bx) \in \mathbb{R}$.}
Correspondingly, we cannot construct independent $P$-functions for the different effective modes.
This subtlety will not impact the calculations below, as haloscope experiments typically read out a single cavity mode.

We can tighten the connection between $a_\ell$ and $\phi^+(\bx)$ as follows.
Although one cannot exactly localize excitations of a relativistic quantum field, the labeling of $\phi(\bx)$ by a position $\bx$ is tied to the fact that the Fourier transform of the mode functions $e^{i \bk \cdot \bx}$ is localized at $\bx$.
Writing out the effective mode using the explicit definition of $C_\ell$, we see
\be
a_\ell = \frac{\omega_\ell}{\sqrt{\Omega_\ell}} \int\! \frac{d^3\bk}{(2\pi)^3} \, \frac{1}{\sqrt{2\omega_\bk}} \,
a_\bk \left( \int_{V_c} \hspace{-0.1cm} d^3 \bx\, u_\ell(\bx) e^{i \bk \cdot \bx} \right)\!.
\ee
Up to an overall normalization, this is identical to Eq.~\eqref{eq:phi_pos}, except with $e^{i \bk \cdot \bx}$ replaced by the Fourier transform of $u_\ell(\bx)$.
Thus, to the extent $\phi(\bx)$ is associated with a position $\bx$, we can think of the effective modes as being associated with the cavity; loosely, they can be thought of as a wave packet of the axion field shaped to the cavity geometry.
Indeed, in the limit of zero cavity volume we schematically have $a_\ell \to \phi^+(\bx)$, whereas in the limit where the detector is as delocalized as a plane wave, $a_\ell \to a_\bk$.
How close $a_\ell$ is to the fundamental modes of the axion is a reflection of the overlap between the cavity and DM basis functions.
In general, there is a significant mismatch, as the natural scales for the DM and detector are galactic and terrestrial; this leads to the coarse-graining effects to be discussed in Sec.~\ref{sec:gaussian_clt}.

Finally, consider the expectation value of the number operator for the effective modes,
\be \label{eq:sec3_expected_number}
N_\eff^\ell = \la a_\ell^{\dag} a_\ell \ra
= \frac{1}{4} \frac{V_c}{\Omega_\ell} \int\! \frac{d^3\bk}{(2\pi)^3} \frac{d^3\bq}{(2\pi)^3}\,
C^*_\ell(\bk)\,
C_\ell(\bq)\,\tr \!\big[ \rho_a a_\bk^\dag a_{\bq} \big].
\ee
To evaluate the trace, we assume the axion state is homogeneous, so that $\tr \!\big[ \rho_a a_\bk^\dag a_{\bq} \big] = N_\bk\, (2\pi)^3 \delta^{(3)}(\bk-\bq)$.\footnote{Intuitively, homogeneity implies momentum conservation, and therefore $\la a_\bk^{\dag} a_{\bq} \ra \propto \delta^{(3)}(\bk-\bq)$.
More precisely, if we define the total momentum operator $\mathbf{P} = \int \tfrac{d^3\bk}{(2\pi)^3} \bk\, a^{\dag}_\bk a_\bk$ and unitary translation operator $T_{\bx} = e^{-i \bx \cdot \mathbf{P}}$, then we can express homogeneity of the system as $[T_{\bx},\rho_a]=0$ for all $\bx$.
As $T_{\bx} a_\bk T_{\bx}^{\dag} = e^{i \bk \cdot \bx} a_\bk$, homogeneity requires $\la a_\bk^{\dag} a_{\bq} \ra = e^{i(\bk-\bq) \cdot \bx} \,\la a_\bk^{\dag} a_{\bq} \ra$ for all $\bx$, so that the correlator must vanish unless $\bk = \bq$.
This can be compared with stationarity: it requires only $|\bk| = |\bq|$, which is insufficient to diagonalize the correlator.} 
Taking the momentum-space occupancy as $N_\bk = (2\pi)^3\, \bar{n} \,p(\bk)$~\cite{Cheong:2024ose}, with $\bar{n}$ the axion number density and $p(\bk)$ the unit-normalized momentum distribution, we have
\be \label{eq:nl_eff_def}
N_\eff^\ell = \frac{1}{4} \frac{\bar{n} V_c}{\Omega_\ell} \int\! d^3\bk\,
|C_\ell(\bk)|^2 \,p(\bk).
\ee
For non-relativistic DM it is more conventional to work with the velocity distribution $p(\mathbf{v})$.
As the DM speed distribution peaks around $v \sim 10^{-3}$, there is little support for large momentum, so that the integral is well-approximated as $\int\! d^3\bk\,
|C_\ell(\bk)|^2 \,p(\bk) \simeq |C_\ell(\vb{0})|^2 \int\! d^3\mathbf{v}
\,p(\mathbf{v}) = |C_\ell(\vb{0})|^2$.
The mean occupancy of effective DM modes is therefore
\be \label{eq:Neff-DM}
N_\eff^\ell \simeq \left(\frac{|C_\ell(\vb{0})|^2}{4 \Omega_\ell} \right) \bar{n} V_c 
\sim 10^{19} \, \left(\frac{|C_\ell(\vb{0})|^2}{4 \Omega_\ell} \right) \left( \frac{\rho_\DM}{0.4\, \mathrm{GeV}/\mathrm{cm}^3} \right) \left( \frac{3\, \mu\mathrm{eV}}{m_a} \right) \left( \frac{V_c}{0.1 \, \mathrm{m}^3} \right)\!.
\ee
As exhibited shortly, for the lowest-lying modes that are the focus of cavity haloscopes, the term $|C_\ell(\vb{0})|^2/ 4 \Omega_\ell$ is $\order{1}$.
This implies that for DM, the occupancy is $N_\eff^\ell \sim \bar{n} V_c$, the expected number of axions in the cavity volume.
That is, although cavity haloscopes target axion DM which is coherent over $\sim$km scales, the effective mode description reorganizes the field such that the detector only couples to the field contribution within the cavity volume.

\paragraph{Cylindrical Cavity Example.}
%
Although our results hold for general cavity geometries, we can also derive explicit results for the cylindrical cavity employed by ADMX.
We take the cylinder to have radius $R$ and length $L$ and expand in modes labeled by $\ell = mnp$ for the azimuthal, radial, and longitudinal quantum numbers.
For the $\bE \cdot \bB$ axion-photon interaction and a longitudinal magnetic field, only the TM modes are relevant, and we review their properties in App.~\ref{app:cavity_modes}.
These modes have a resonant frequency determined from $\omega_\ell^2 = (j_{|m|n}/R)^2 + (p\pi/L)^2$, where the $j_{|m|n}$ are Bessel zeros.
We can calculate the cylindrical form factor, finding
\be
C_\ell(\bk) = \frac{2\pi i^{|m|} R}{V_c} \sqrt{\frac{2^{1-\delta_{p0}}}{\omega_\ell \omega_\bk}} \left( \frac{k_x + ik_y}{k_\rho}\right)^{\!\! m} \left ( \frac{\omega_{\rho,\ell}^2 J_{|m|}(k_\rho R)}{\omega_{\rho,\ell}^2-k_\rho^2}\right)\left( \frac{ik_z (e^{i(k_zL+p\pi)}-1)}{\omega_{z,\ell}^2-k_z^2} \right)\!,
\ee
where $k_\rho^2 = k_x^2 + k_y^2$, and we have decomposed $\omega_\ell^2$ into its radial, $\omega_{\rho,\ell} = j_{|m|n}/R$, and longitudinal, $\omega_{z,\ell} = p\pi/L$, contributions.
For the lowest-lying and most commonly used $\text{TM}_{010}$ mode,
\be
C_{010}(\bk) = 2e^{i k_z L/2}\sqrt{\frac{\pi L}{V_c m_a \omega_\bk}} \, \text{sinc}\qty(\frac{k_z L}{2}) \, \frac{J_0(j_{01} k_\rho/m_a)}{1 - (k_\rho/m_a)^2}.
\ee
Here we have fixed the angular frequency to the mass, $m_a = \omega_{010} = j_{01}/R$.
If we focus on the most relevant contribution for non-relativistic DM (cf. the discussion above Eq.~\eqref{eq:Neff-DM}), we have $C_{010}(\vb{0}) = 2/j_{01}$, recovering the commonly used cylindrical form factor expression $|C_{010}(\vb{0})|^2 = 4/j_{01}^2 \simeq 0.69$.
Finally, the mode normalization is independent of the DM state, and calculating it requires considering arbitrary momentum $\bk$.
Focusing on the lowest-lying mode and setting $L = 5R$, as in the ADMX experiment, we find
\be
\Omega_{010} = \frac{5j_{01}}{4\pi} \int_0^\infty\! d x_\rho \, \frac{x_\rho J_0^2(j_{01}x_\rho)}{(1 - x_\rho^2)^2} \int_{-\infty}^{\infty}\! d x_z\, \frac{1}{\sqrt{1 + x_z^2 + x_\rho^2}} \, \text{sinc}^2\qty(\frac{5j_{01} x_z}{2}) \simeq 0.183.
\ee
This is in agreement with our argument in Ref.~\cite{Bao:2025nsd} that $\Omega_\ell$ is order-one for low-lying modes, and further we see that $|C_{010}(\vb{0})|^2/4 \Omega_{010} \simeq 0.943 = \mathcal{O}(1)$, as stated below Eq.~\eqref{eq:Neff-DM}.

\subsection{Gaussian States and the Quantum Central Limit Theorem}
\label{sec:gaussian_clt}

We now show that if the DM field is constructed from a set of many independent modes, then an effective mode which coarse grains over them has Gaussian statistics.
In particular, if the effective mode is stationary, its $P$-function takes the Gaussian form of Eq.~\eqref{eq:gaussian_state_def} with mean occupation $|\alpha_0|^2=N_\eff$, as given in Eq.~\eqref{eq:Neff-DM}.
When this holds, the effects of DM are the same as a classical Gaussian random field.

To show this, we must compute the $P$-function of the effective mode.
The state of the full DM field can be written in terms of a joint $P$-function for all the DM modes,
\be \label{eq:rhoDM}
\rho_a = \int d \bm{\alpha} \,P_a(\bm{\alpha})\, \ket{\bm{\alpha}} \bra{\bm{\alpha}}
\ee
where for simplicity we use discrete fundamental DM modes, enumerated by $\bm{\alpha} = \{\alpha_1,\alpha_2,\ldots\}$, in a quantization volume $\mathcal{V}$.
Within a DM coherence time, the detector only couples to DM through a fixed effective mode operator $a_\eff$ defined in Eq.~\eqref{eq:effective_mode_def}.
Thus, the effects of DM enter only through the effective mode's $P$-function, defined by tracing out all other DM modes~\cite{PhysRevA.87.033811},
\be \label{eq:Peff}
P_\eff(\beta) = \int d\bm{\alpha} \, P_a(\bm{\alpha}) \, \delta\big(\beta - \alpha_\eff(\bm{\alpha}) \big)
\ee
where we introduced the effective mode eigenvalue $a_\eff \ket{\bm{\alpha}} = \alpha_\eff(\bm{\alpha}) \ket{\bm{\alpha}}$.

We can show this more directly.
Experiments measure operators $\mathcal{O}$ constructed from $a_\eff$, its Hermitian conjugate, and cavity operators.
Assuming without loss of generality that $\mathcal{O}$ has been normal ordered, and suppressing dependence on the cavity operators, we have
\bea
\la \mathcal{O}(a_\eff^\dag, a_\eff) \ra 
&= \int d \bm{\alpha} \,P_a(\bm{\alpha})\, \bra{\bm{\alpha}} \mathcal{O}(a_\eff^\dag, a_\eff) \ket{\bm{\alpha}}
= \int d \bm{\alpha} \,P_a(\bm{\alpha})\, \mathcal{O}(\alpha_\eff^*, \alpha_\eff) \\
&= \int d\beta\,P_\eff(\beta)\, \mathcal{O}(\beta^*,\beta).
\eea
In other words, observable expectation values can be calculated by applying the optical equivalence theorem, discussed below Eq.~\eqref{eq:O_expt}, to the effective mode $P$-function alone.

\paragraph{The Quantum Central Limit Theorem.}
%
We now suppose the DM $P$-function factorizes mode-by-mode, that is $P_a(\bm{\alpha}) = \prod_\bk P_\bk(\alpha_\bk)$.
From Eq.~\eqref{eq:Peff}, this implies that $P_\eff$ is a weighted convolution of the $P_\bk$, so that one may invoke the quantum analog of the central limit theorem (CLT), as first discussed by Glauber~\cite{Glauber:1963tx}.
(Further aspects of the quantum CLT, such as its convergence rate and extension to qudits, were discussed in Refs.~\cite{Cushen:1971,Becker:2020myv,Bu:2023ssg}.)

As in the proof of the classical CLT, it is convenient to work not with $P(\alpha)$, but rather its Fourier transform, i.e.~the characteristic function
\begin{align} \label{eq:Glauber_P_characteristic_function}
\tilde{P}(\lambda) &= \int d\alpha \, P(\alpha) \, e^{\lambda \alpha^* - \lambda^* \alpha} = \tr\! \Big[ e^{-\lambda^* a}\rho \, e^{\lambda a^\dag} \Big], \\
P(\alpha) &= \int \frac{d\lambda}{\pi^2}\, \tilde{P}(\lambda) \, e^{-\lambda \alpha^* + \lambda^* \alpha}.
\end{align}
Here $\lambda$ is complex, and since $P(\alpha)$ is real, we have $\tilde{P}(\lambda)^* = \tilde{P}(-\lambda)$.
In addition, a Gaussian $P$-function corresponds to a Gaussian characteristic function.
Moreover, while the $P$-function can be a highly singular distribution for nonclassical states, the characteristic function $\tilde{P}$ is an ordinary function. 
Further properties of characteristic functions are discussed in App.~\ref{app:general_quasi_characteristic}.

To derive the quantum CLT, we compute the characteristic function of $P_\eff$.
For convenience, we define $\alpha_\eff = \sum_\bk c_\bk \alpha_\bk$, where the discrete version of Eq.~\eqref{eq:effective_mode_def} implies that the coefficients are $c_\bk = \tfrac{1}{2} C(\bk) \sqrt{V_c/\mathcal{V} \Omega}$ and obey $\sum_\bk |c_\bk|^2 = 1$.
Then from Eq.~\eqref{eq:Peff}, we have
\be
\tilde{P}_\eff(\lambda) = \int d\bm{\alpha} \, P_a(\bm{\alpha}) \, e^{\lambda \alpha_\eff^* - \lambda^* \alpha_\eff} = \int d\bm{\alpha} \, P_a(\bm{\alpha}) \prod_\bk e^{c_\bk^* \lambda \alpha_\bk^* - c_\bk \lambda^* \alpha_\bk}.
\ee
Assuming $P_a(\bm{\alpha})$ factorizes, this becomes a product over the $\tilde{P}_\bk$,
\be \label{eq:effectivemodecharacteristic}
\tilde{P}_\eff(\lambda) = \prod_\bk \int d\alpha_\bk P_\bk(\alpha_\bk) e^{c_\bk^* \lambda \alpha_\bk^* - c_\bk \lambda^* \alpha_\bk} = \prod_\bk \tilde{P}_\bk(c_\bk^*\lambda),
\ee
consistent with $P_\eff$ being a convolution over the $P_\bk$.

In the spirit of the classical CLT, we now imagine that a large number $\mathcal{N}$ of $\bk$-modes contribute comparably to $\alpha_\eff$.
The normalization condition then implies $|c_\bk| \sim 1/\sqrt{\mathcal{N}} \ll 1$.
Taylor expanding $\tilde{P}_\bk(c_\bk^*\lambda)$ in this limit yields
\bea \label{eq:Pp_expansion}
\tilde{P}_\bk(c_\bk^*\lambda) 
&\simeq \int d\alpha_\bk\, P_\bk(\alpha_\bk)\,
\Big[ 1 + (c_\bk^*\lambda \alpha_\bk^* - c_\bk \lambda^* \alpha_\bk) + \frac{1}{2} (c_\bk^* \lambda \alpha_\bk^* - c_\bk \lambda^* \alpha_\bk)^2 \Big] \\
&= 1 + c_\bk^* \lambda \la a_\bk^\dag\ra - c_\bk \lambda^* \la a_\bk\ra + \frac{1}{2}\Big[ (c_\bk^* \lambda)^2 \la a^{\dag2}_\bk \ra + (c_\bk \lambda^*)^2 \la a_\bk^2 \ra \Big] - |c_\bk \lambda|^2 \la a^\dag_\bk a_\bk \ra
\eea
up to $\mathcal{O}(|c_{\bk}|^3)$ corrections. 
Then taking the logarithm of Eq.~\eqref{eq:effectivemodecharacteristic} yields
\be \label{eq:tPeff-mid}
\log \tilde{P}_\eff(\lambda) 
= \sum_\bk \log \tilde{P}_\bk(c_\bk^*\lambda)
\simeq \lambda \mu^* - \lambda^* \mu + \frac{1}{2} \big( \lambda^2 \epsilon^* + \lambda^{*2} \epsilon \big) - |\lambda|^2 \sigma^2
\ee
where we have defined
\be
\mu = \sum_\bk c_\bk\la a_\bk \ra,
\hspace{0.5cm}
\epsilon = \sum_\bk c_\bk^2 \, \Big(\la a_\bk^2 \ra - \la a_\bk \ra^2\Big),
\hspace{0.5cm}
\sigma^2 = \sum_\bk |c_\bk|^2 \, \Big(\la a^\dag_\bk a_\bk \ra - |\la a_\bk \ra|^2\Big).
\ee
Using $|c_{\bk}| \sim 1/\sqrt{\mathcal{N}}$, these quantities scale at most as $\mu \propto \sqrt{\mathcal{N}}$, $\epsilon \propto \mathcal{N}^0$, and $\sigma^2 \propto \mathcal{N}^0$.
Just as in the classical CLT, cubic moments are at most $\propto \mathcal{N}^{-1/2}$ and thus become irrelevant in the limit $\mathcal{N} \gg 1$, and higher moments are even further suppressed.

Thus, in the limit of many contributing modes, the approximation in Eq.~\eqref{eq:tPeff-mid} becomes exact.
This corresponds to the log characteristic function for a Gaussian with mean $\mu$, variance $\sigma^2$, and an ellipticity controlled by $\epsilon$.
When $|\epsilon| < \sigma^2$, the corresponding $P$-function is a nonnegative multivariate Gaussian, 
\be \label{eq:multivariate_P}
P_\eff(\alpha) 
= \frac{e^{-\tfrac{1}{2}(\bm{\xi}_\alpha-\bm{\xi}_\mu)^T C^{-1} (\bm{\xi}_\alpha-\bm{\xi}_\mu)}}{\sqrt{\det(2\pi C)}},\hspace{0.5cm}
C = \frac{1}{2} \begin{pmatrix} \sigma^2 + \text{Re}(\epsilon) & \text{Im}(\epsilon) \\ \text{Im}(\epsilon) & \sigma^2 - \text{Re}(\epsilon)
\end{pmatrix},
\ee
where $\bm{\xi}_z = (\text{Re}(z),\,\text{Im}(z))$ and $C$ is the positive-definite covariance matrix.
However, this does not yet imply that the effective mode is classical: when $|\epsilon| > \sigma^2$, the Fourier transformation back to the $P$-function is not convergent (the equivalent covariance matrix above would have negative determinant), indicating the $P$-function is a singular distribution.
In this case the effective mode is in a squeezed state, as can be seen most directly using the Wigner characteristic function (see App.~\ref{app:general_quasi_characteristic}).

As a final simplification, we note that if the fundamental DM modes are stationary (see App.~\ref{app:stationarity}) we have $\la a_\bk \ra = \la a_\bk^2 \ra = 0$.
This sets $\mu=\epsilon=0$ and $\sigma^2 = N_\eff$, where the mean occupancy $N_\eff$ is as given in Eq.~\eqref{eq:Neff-DM}.
Then the effective mode is in a Gaussian state as defined in Eq.~\eqref{eq:gaussian_state_def}.\footnote{Light in a Gaussian state is sometimes called ``chaotic'' light, because it arises when one combines contributions from many independent emitters.
It arises here in a related but slightly different way, where we combine contributions from many independent axion modes.}
Even if the fundamental modes are not stationary, as long as their phases are roughly independent, we expect $|\epsilon| \ll \sigma^2$, which ensures a classical state.

The above analysis is almost identical to the proof of the classical CLT.
Nevertheless, it holds even for nonclassical states, where the $P_\bk$ can be negative, and even for those nonclassical states where the $P_\bk$ are highly singular, e.g.~as occurs for a Fock state, as we now show.

\paragraph{Example: Purity and Fock States.}

\begin{figure}[!t]
\centering
\includegraphics[width=0.8\linewidth]{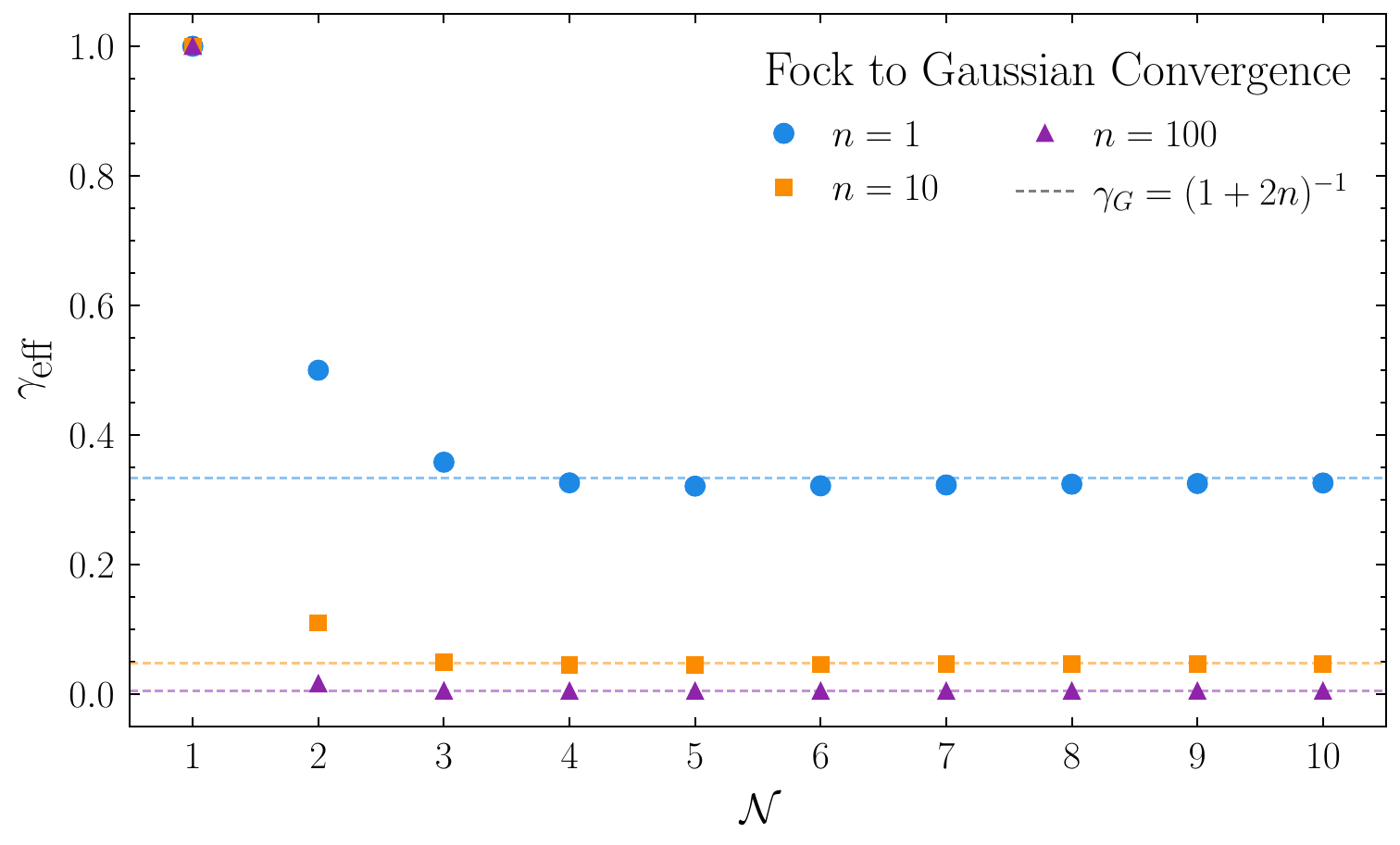}
\vspace{-0.3cm}
\caption{The purity $\gamma_\eff$ of a mode that coarse grains $\mathcal{N}$ identical and independent Fock states, with occupancy $n = 1$, $10$, and $100$. 
Convergence to the purity $\gamma_G = (1+2n)^{-1}$ of the Gaussian state predicted by the CLT is \textit{rapid}, whether $n$ is large or order-one.}
\label{fig:purity_fock_gaussian}
\end{figure}

The quantum CLT implies the effective mode is described by a Gaussian state, which is generally mixed, even if the fundamental modes are pure.
This is because coupling to the effective mode only samples a slice of the full DM density matrix; this coarse graining loses information about the fundamental modes.
To illustrate this fact, we consider the case where the fundamental modes are in Fock states, and exhibit the convergence of the effective mode's purity to that of the expected Gaussian state as $\mathcal{N} \to \infty$.

To begin, using the overlap of coherent states $|\la \alpha | \beta \ra|^2 = e^{-|\alpha - \beta|^2}$, we can derive a simple expression for the purity of a quantum state in terms of the characteristic function,
\be \label{eq:Gamma-characteristic}
\gamma = \tr (\rho^2 )
= \int\! d\alpha \, d\beta\, P(\alpha) P(\beta)\, e^{-|\alpha-\beta|^2}
= \int \frac{d\lambda}{\pi}\, |\tilde{P}(\lambda)|^2 \, e^{-|\lambda|^2}. 
\ee
This formula can be used, for instance, to derive Eq.~\eqref{eq:phase_averaged_purity}.
Recall that the purity $\gamma \in (0,1]$ achieves unity only for a pure state, with mixed states falling in the range $0 < \gamma < 1$.

We now imagine an effective mode constructed by equally weighting $\mathcal{N}$ independent fundamental modes, $c_k = 1/\sqrt{\mathcal{N}}$. 
Then by Eq.~\eqref{eq:effectivemodecharacteristic}, the characteristic function is
\be
\tilde{P}_\eff(\lambda) = \prod_{k=1}^\mathcal{N} \tilde{P}_k\Big(\lambda/\sqrt{\mathcal{N}}\Big).
\ee
We take each fundamental mode to be in an $n$-particle Fock state, so that $\tilde{P}_k (\lambda) = L_n (|\lambda|^2)$, where $L_n$ is a Laguerre polynomial. 
The Fock state is pure, as can be confirmed using Eq.~\eqref{eq:Gamma-characteristic} and the orthogonality condition for the Laguerre polynomials.
But the effective mode has
\be
\tilde{P}_{\mathrm{eff}}(\lambda)=\left[L_n\Big(|\lambda|^2 / \mathcal{N}\Big)\right]^\mathcal{N} = \left[ 1 - \frac{n}{\mathcal{N}} |\lambda|^2 + \order{\mathcal{N}^{-2}} \right]^\mathcal{N}\!,
\ee
and in the limit $\mathcal{N} \to \infty$, we see $\tilde{P}_{\mathrm{eff}} \to e^{-n|\lambda|^2}$, the Gaussian expected from the quantum CLT.\footnote{In App.~\ref{app:singular_P} we show that the same result can be reached working directly with $P$-functions, even though the $P$-function of a Fock state is highly singular.}
The purity has been reduced to
\be \label{eq:gamma_Fock}
\gamma_\eff = \int_0^{\infty} d |\lambda| \, 2 |\lambda| e^{-2n |\lambda|^2} e^{-|\lambda|^2 } = \frac{1}{1 + 2n}
\ee
corresponding to precisely that of the Gaussian state in Eq.~\eqref{eq:gaussian_state_def}. At finite $\mathcal{N}$, the leading deviation between $\gamma_\eff$ and the Gaussian state's purity $\gamma_G$ is 
\be
\frac{|\gamma_\eff - \gamma_G|}{\gamma_G} = \frac{n(n+1)}{\mathcal{N} (2n+1)^2} + \mathcal{O}\big(\mathcal{N}^{-2} \big),
\ee
which for $n \gtrsim 1$ scales as $1/\mathcal{N}$, depending only weakly on $n$, as is shown in Fig.~\ref{fig:purity_fock_gaussian}.
Consequently, convergence is rapid even for the $n \gg 1$ relevant for ultralight DM (although we are not suggesting this is a physical model for DM).
On the other hand, high mode occupancy $n$ is not the reason that ultralight DM appears classical, because convergence is also rapid for $n \sim 1$.

\paragraph{Evading the Quantum CLT.}
%
We have shown that if the DM occupancy is distributed across many independent modes, and the effective mode is stationary, then the CLT implies the effective mode has a thermal Gaussian $P$-function.
This process would erase any intrinsically quantum effects in the signal.
The assumption of independence is natural, as we expect that during virialization, the galaxy's DM is driven towards a higher entropy state. 
However, there are cases where this conclusion does not hold. 

In our derivation of the CLT, we assumed the first- and second-order moments in Eq.~\eqref{eq:Pp_expansion} are finite; for our case, this must hold for any finite-energy state.
We also implicitly assumed higher moments are finite.
Technically, this could be violated at finite energy, but we are unaware of any reasonable DM state that would achieve this.

There are three physically relevant ways to arrive at a nonclassical state.
First, the CLT requires a large number $\mathcal{N}$ of relevant occupied plane-wave modes.
For virialized DM, we can estimate $\mathcal{N} \sim (m_a v_\DM R)^3$ from the phase space volume, where $v_\DM \sim 10^{-3}$ is the DM speed, and $R$ is the DM halo radius.
This is extremely large in the cavity haloscope regime, $m_a \sim \mu\mathrm{eV}$, but only moderately large in the fuzzy DM regime $m_a \lesssim 10^{-20} \, \mathrm{eV}$.
Alternatively, it may be possible for interactions to drive the axion into a condensate where a single mode contains a significant fraction of the DM, so that we effectively have $\mathcal{N} = 1$.

Second, we have assumed a factorized $P$-function.
However, initial correlations between plane-wave modes with sufficiently similar momentum, $|\bk_i - \bk_j| \lesssim \Delta k$, could survive through virialization. 
In this case, we can perform a two-stage coarse graining, where one first combines the plane-wave modes into roughly uncorrelated mesoscopic modes, each with momentum spread $\Delta k$.
These mesoscopic modes can be combined into an effective mode as above, except that now the number of contributing independent modes is reduced to $\mathcal{N} \sim (m_a v_\DM / \Delta k)^3$, and can be reduced even further for frequency-selective detectors.
Of course, even for relatively modest values of $\mathcal{N}$ the Gaussian approximation can be an excellent one.

Alternatively, mode correlations can be introduced by axion self-interactions at late times.
For example, one can imagine that self-interactions within an axion star could drive the axion magnitude to a uniform value.
This is \textit{not} a nonclassical state, but it gives a simple example of how one might evade a thermal Gaussian.

Third, the quantum CLT only guarantees the characteristic function is a multivariate Gaussian; it can correspond to a squeezed state for $|\epsilon| > \sigma^2$.
This requires the fundamental modes to be squeezed, along roughly aligned axes.
However, if the relevant axions have propagated for a time $T$ with frequency spread $\Delta \omega \sim m_a v_\DM^2$, the squeezing axes will spread in phase by $T \, \Delta \omega \sim 10^6 \, (T/10^{10} \, \mathrm{yr}) (m_a / 10^{-20} \, \mathrm{eV})$, leading to a strong suppression of $|\epsilon|$.
Thus, we generically expect that ultralight DM of any mass yields an approximately stationary effective mode.
However, for highly monochromatic sources of axions or GWs at late times, it may be possible to have $T \, \Delta \omega \lesssim 1$, leading to a squeezed effective mode.

We discuss potential sources of nonclassical axions and GWs further in Sec.~\ref{sec:generate_nonclassical}.
In the following sections we allow the state of the effective mode to be arbitrarily nonclassical, but show that even under this maximally optimistic assumption, nonclassical effects remain strongly suppressed.

\section{Suppression of Nonclassical Effects}
\label{sec:projective}
We now consider a simple measurement model where the cavity is prepared in a given state at time $t = 0$, interacts with the axion on resonance, then is projectively measured at time $t_m$.
We optimistically take $t_m \sim Q_c/m_a$, the maximum timescale over which cavity dissipation can be neglected.
Since axion haloscopes operate at $Q_c \lesssim 1/v_\DM^2 \sim 10^6$, this implies $t_m \lesssim 1/\Delta \omega_a \sim 1/(m_a v_\DM^2)$, so the effective mode's state evolution can be neglected as well.

Our goal is to exhibit the difficulty of observing nonclassical effects, even under optimistic assumptions.
We thus allow an arbitrary initial axion effective mode $P$-function $P_a$, and suppose the measurement process can be repeated independently arbitrarily many times. 
(We revisit these assumptions in Sec.~\ref{sec:discussion}.)
To build intuition, in the first parts of this section, we assume the cavity is perfectly prepared in the vacuum state. 
In this case, the final cavity $P$-function in Eq.~\eqref{eq:approx_cavity_P} is simply a rescaled version of $P_a$,
\be \label{eq:simple_final_P}
P_c^f(\alpha) = \frac{P_a(\alpha / \sqrt{\eta})}{\eta},
\ee
which shows that any negativity in $P_a$ is automatically transferred to $P_c^f$. 
However, we show that the observability of the associated nonclassical effects is suppressed by additional powers of the extremely small conversion efficiency $\eta$, where
\be \label{eq:eta_value}
\eta = \sin^2(g t_m) \sim 10^{-21} \left( \frac{\ga}{10^{-15} \, \mathrm{GeV}^{-1}} \frac{B_0}{8 \, \mathrm{T}^{\vrule height 4.5pt width 0pt}} \frac{Q_c}{10^5} \frac{3\, \mu\mathrm{eV}}{m_a} \right)^2\!,
\ee
where we recalled from Eq.~\eqref{eq:Hint_simplified} that $g = \ga B_0 \sqrt{\Omega}$ and $\Omega \sim 1$.

Specifically, in Sec.~\ref{sec:number_meas} we consider projective measurements of the cavity number $n = c^\dagger c$.
The distinctive effects of nonclassical states can be parametrized by nonclassicality measures such as Mandel $Q = \var(n) / \la n \ra - 1$, which can only be negative for nonclassical states.
We show that even if the axion state is highly nonclassical, nonclassicality measures for the cavity statistics are suppressed by powers of $\eta$.
In particular, detecting negative $Q$ requires a prohibitively long integration time $t_{\text{int}} \sim t_m / \eta^2$, \textit{independent of the axion occupation number}. 
In Sec.~\ref{sec:quad_meas}, we consider projective measurements of the cavity quadrature $X = (c + c^\dagger) / \sqrt{2}$.
Here the simplest nonclassicality measure is the squeezing parameter $S = \var(X) - 1/2$, and we again find it is highly suppressed.

In Sec.~\ref{sec:general_suppression} we show that \textit{all} nonclassical effects, for any initial cavity state, are suppressed by an additional power of $\eta$. 
This suppression can only be canceled by preparing the cavity in an initial state with quantum resources (e.g. squeezing) scaling as $1/\eta$, which is far beyond the reach of existing experiments.
We also extend the analysis to multi-mode haloscopes, where we find that axion-induced mode entanglement is again suppressed by $\eta$.

\subsection{Number Measurements}
\label{sec:number_meas}

Here we consider measuring the number of photons in the final cavity state, relevant for the DM searches in Refs.~\cite{Dixit:2020ymh,Agrawal:2023umy,Gu:2025pms,Zheng:2025qgv}.
As we have seen in Eq.~\eqref{eq:nc-na-calc}, the expected number of photons in the cavity is related to the expected number of axions in the effective mode by $\la n_c \ra = \eta \la n_a \ra = \eta N_\eff$.
Further information about the DM state comes from higher moments of the cavity number distribution.
For example, the argument leading to $\la n_c \ra = \eta \la n_a \ra$ readily generalizes to any normally ordered operator, and considering $\normord{n^2} = n(n-1)$ yields
\be \label{eq:falling_moment_result}
\la n_c (n_c - 1) \ra = \eta^2 \la n_a (n_a - 1) \ra
\ee
along with similar results for other falling factorial moments.

Now, the simplest nonclassicality criterion involving moments of the number distribution is Mandel $Q$, which can only be negative for nonclassical states.\footnote{This follows as $\var(n)-\la n \ra = \int d\alpha\, P(\alpha)\, \big(|\alpha|^2-\la n \ra \big)^2$, which can be negative only when $P(\alpha) < 0$.}
It is defined by
\be \label{eq:MandelQ_def}
Q = \frac{\var(n)}{\la n \ra} - 1 = \frac{\la n(n-1) \ra}{\la n \ra} - \la n \ra.
\ee
From the first form of $Q$, it is straightforward to show that $Q = 0$ for a coherent state (which has Poisson number distribution), $Q = \la n \ra$ for a Gaussian state (which has geometric number distribution), and the minimum value of $Q = -1$ is achieved for a Fock state, as it has zero number variance.
From the second form of $Q$ and Eq.~\eqref{eq:falling_moment_result}, we have
\be \label{eq:suppressed_mandel_Q}
Q_c = \eta Q_a \geq - \eta
\ee
so that negative values of Mandel $Q$ in the cavity are always extremely small in magnitude.

\paragraph{The Photon Number Distribution.}
%
We can gain intuition for Eq.~\eqref{eq:suppressed_mandel_Q} by explicitly computing the probability $p_n$ to find $n$ photons in the cavity.
From Eq.~\eqref{eq:simple_final_P}, we have
\bea \label{eq:num_dist}
p_n &= \int d\alpha \, P_c^f(\alpha) \, | \la n | \alpha \ra |^2
= \int d\alpha \, P_a(\alpha) | \la n | \sqrt{\eta} \, \alpha \ra |^2 \\
&= \frac{1}{n!} \int d\alpha \, P_a(\alpha) \, e^{- \eta |\alpha|^2} \big(\eta |\alpha|^2\big)^n.
\eea
We can already heuristically see how nonclassical effects are suppressed: $P_a(\alpha)$ is integrated against the very slowly varying function $e^{-\eta |\alpha|^2} \big(\eta |\alpha|^2\big)^n$, but since $P_a$ can only be negative over regions of size at most ${\cal O}(1)$, its sign must oscillate, suppressing the observed nonclassicality.

To connect this with the axion number distribution $p_n^\DM$, we can expand out the exponential in Eq.~\eqref{eq:num_dist} and apply the optical equivalence theorem Eq.~\eqref{eq:O_expt} to write it as a normal ordered expectation value in the axion state,
\be
p_n = \frac{\eta^n}{n!} \sum_{m=0}^\infty \frac{(-\eta)^m}{m!} \la (a^\dagger)^{n+m} a^{n+m} \ra_\DM.
\ee
The expectation value can then be evaluated in terms of $p_n^\DM$, and applying standard combinatoric identities yields
\be \label{eq:num_dist_v2}
p_n = \sum_{k=n}^\infty \binom{k}{n} \, p_k^\DM \, \eta^n (1-\eta)^{k-n}.
\ee
That is, the cavity photon number distribution can be found by giving each axion an independent chance $\eta$ to be converted to a photon.
Thus, for small $\eta$ and large $N_\eff$, the conversion process adds approximately Poisson fluctuations, suppressing negative Mandel $Q_c$.
For the examples below, in the weak-signal regime $\langle n_c \rangle \ll 1$, we have $p_1 \sim \langle n_c \rangle$ and $p_2 \sim \langle n_c \rangle^2$.

For example, if the axion is in a Fock state $\ket{N_\eff}$, the final cavity number distribution is binomial.
If the axion is in a Gaussian state, as motivated by the quantum CLT, then both the axion and the cavity number distribution are geometric, with
\be \label{eq:thermal_geom}
p_n = \frac{\la n_c \ra^n}{(1+\la n_c \ra)^{n+1}}.
\ee
This holds for $t_m \lesssim 1/\Delta \omega_a$, while in Ref.~\cite{continuous_paper} we present a result valid for $t_m \gtrsim 1/\Delta \omega_a$.

Note that Eq.~\eqref{eq:num_dist} takes the same form as the Kelley--Kleiner formula for photodetection with efficiency $\eta$, first derived in Ref.~\cite{PhysRev.136.A316}.\footnote{We also note that Eq.~\eqref{eq:num_dist} can be derived without $P$-functions by directly computing the overlap $\la n | e^{- i H t_m} | 0 \ra$, though this route is more involved; one must use the ``disentangling'' identities derived in the appendix of Ref.~\cite{PhysRevA.6.2211}, applied to the $\mathfrak{su}(2)$ operators defined in Eq.~\eqref{eq:J-13-K-twomode}.}
However, Ref.~\cite{PhysRev.136.A316} considered light continually falling on a photodetector which clicks when it detects individual photons, while here we are considering coherent conversion of axions to photons in a cavity.
The results are the same in this simple case, where both detectors can only absorb energy.
But if we had not taken the cavity to start in the vacuum state, then the $p_n$ here would be completely different in form.

\paragraph{Measuring Negative Mandel $Q$.}
%
Suppose one performs $N_{\text{shot}}$ independent measurements.
Here we show that detecting the negative $Q_c = - \eta$ induced by an axion Fock state requires $N_{\text{shot}} \sim 1/\eta^2$, corresponding to an impractically long integration time $t_{\text{int}} = N_{\text{shot}} t_m$.

First, consider the weak-signal regime $\la n_c \ra \ll 1$, corresponding to a haloscope operating near the edge of its sensitivity.
In this regime, we have $p_0 \simeq 1$, and discovering the axion corresponds to measuring nonzero $p_1 \simeq \la n_c \ra$. 
However, since $\la n_c \ra$ scales with the unknown axion density and coupling, it does not yield nontrivial information about the number distribution.
The leading information regarding the axion state comes from $p_2 \sim \la n_c \ra^2$, and in the limit where we can neglect events with more than $2$ photons, we have
\be \label{eq:p2_stats}
p_2 = \frac{1}{2} \la n_c \ra \Big(\la n_c \ra + Q_c \Big) 
= \begin{cases} \frac{1}{2} \la n_c \ra^2 & \text{coherent DM}, Q_c = 0 \\ 
\la n_c \ra^2 & \text{Gaussian DM}, Q_c = \la n_c \ra \\ 
\frac{1}{2} \la n_c \ra \big(\la n_c \ra - \eta \big) & \text{Fock DM}, Q_c = - \eta \end{cases}
\ee
where the special cases on the right-hand side can also be derived by using the fact that the number distributions are Poisson, geometric, and binomial, respectively.

Measuring negative $Q$ thus requires measuring $p_2$ to a fractional uncertainty $\sim\!\eta/\la n_c \ra$, sufficient to distinguish the Fock and coherent states.
Since we are assuming the measurements are independent, the fractional uncertainty of $p_n$ scales as $1/\sqrt{p_n N_{\text{shot}}}$, so that the estimation of $p_2$ dominates the error over any uncertainty in $p_1$ and $\la n_c \ra$.
Since both the Fock and coherent states have $p_2 \sim \la n_c \ra^2$, we require at least $N_{\text{shot}} \sim 1/\eta^2$ measurements.

The above argument assumed $\la n_c \ra \ll 1$, but the same conclusion holds in the strong-signal regime $\la n_c \ra \gg 1$, relevant if an axion was discovered at couplings far above the edge of sensitivity.
In this case, $\la n_c \ra$ is set by the unknown axion density and coupling, and Mandel $Q$ is best inferred directly from its definition, $Q_c = \var(n_c) / \la n_c \ra - 1$.
To distinguish Fock and coherent axion states, one must estimate $Q_c$ to a precision of $\eta$.
Since estimating the variance is parametrically harder than estimating the mean, this corresponds to estimating $\var(n_c)$ to a fractional precision of $\eta$.
Since both distributions are close to Poisson, the fractional uncertainty on the variance is $\sim 1 / \sqrt{N_{\text{shot}}}$, again implying $N_{\text{shot}} \sim 1/\eta^2$.

In either case, the required integration time scales as
\be \label{eq:integration_time_quantum}
t_{\text{int}} \sim \frac{t_m}{\eta^2} 
\gtrsim 10^{17} \, \mathrm{yr} \left( \frac{10^{-12} \, \mathrm{GeV}^{-1}}{\ga} \frac{8 \, \mathrm{T}^{\vrule height 4.5pt width 0pt}}{B_0} \right)^4 \left( \frac{10^5}{Q_c} \frac{m_a}{3\, \mu\mathrm{eV}} \right)^3
\ee
where we note that since $\eta \propto t_m^2$, we have $t_{\text{int}} \propto 1/t_m^3$, so that it is advantageous for $t_m$ to be as large as possible.
This is impractically long, even for the optimistic coupling value taken above, and it is \emph{not at all} enhanced by the large occupancy $N_\eff$. Furthermore, this result follows from maximally optimistic assumptions.
As we discuss in Sec.~\ref{sec:discussion}, lifting these assumptions makes nonclassical effects even harder to observe.

\paragraph{Distinguishing Coherent and Gaussian States.} 
%
By contrast, it is feasible to distinguish sufficiently different classical axion states, and in some cases, doing so is not significantly more difficult than discovering the axion.
For example, if the axion is in a Gaussian state, then number fluctuations are significantly higher than for a coherent state,
\be
\var(n_c) = \begin{cases} \la n_c \ra & \text{coherent DM}, \\ \la n_c \ra \big( \la n_c \ra + 1 \big) & \text{Gaussian DM}. \end{cases}
\ee
In the strong-signal regime $\la n_c \ra \gg 1$, one can distinguish these scenarios with only $\mathcal{O}(1)$ measurements.
In other words, if one discovered an axion in this regime, one would immediately be able to distinguish these states.

In the weak-signal regime $\la n_c \ra \ll 1$, the states can be distinguished by measuring $p_2$, which from Eq.~\eqref{eq:p2_stats} is twice as large for the Gaussian state.
Measuring $p_2$ to $\mathcal{O}(1)$ fractional precision requires $N_{\text{shot}} \sim 1 / \la n_c \ra^2$ measurements.
This is larger than the $N_{\text{shot}} \sim 1 / \la n_c \ra$ measurements needed to initially discover the axion through $p_1$.
However, this is compensated by the fact that an axion search conventionally scans over many candidate axion masses, while a post-discovery measurement can sit directly at the known axion mass.

Explicitly, suppose a haloscope experiment allots a time $t_e$ to scan an $e$-fold of axion masses.
Then the time spent at each axion mass is $t_{\text{int}}^{\text{scan}} \sim t_e / Q_c$, and for optimal sensitivity one should take the maximum possible $t_m \sim Q_c / m_a$.
Now suppose an axion is discovered at the smallest coupling to which the experiment is sensitive, $t_{\text{int}}^{\text{scan}} \sim t_m / \la n_c \ra$.
Then the time $t_{\text{int}}^{\text{post}} \sim t_m/\la n_c \ra^2$ to determine the axion state in a post-discovery measurement is
\be \label{eq:distinguish_time}
t_{\text{int}}^{\text{post}} \sim \frac{(t_{\text{int}}^{\text{scan}})^2}{t_m} 
\sim \frac{m_a t_e^2}{Q_c^3} 
\sim 1 \, \mathrm{yr}\, \bigg(\frac{m_a}{3 \, \mu\mathrm{eV}} \bigg) \bigg( \frac{t_e}{1 \, \mathrm{month}} \bigg)^2 \bigg( \frac{10^5}{Q_c} \bigg)^3.
\ee
This is a reasonable time, and it rapidly decreases away from the edge of sensitivity, since $t_{\text{int}}^{\text{post}} \propto 1/g_{a\gamma\gamma}^4$.
For instance, an experiment with scanning sensitivity to the DFSZ axion should also be able to distinguish between coherent and Gaussian states for a KSVZ axion.

The estimate in Eq.~\eqref{eq:distinguish_time} can be refined.
First, we have taken the measurements to be independent over timescales $\sim\!\,t_m$, but the Gaussian state has intensity fluctuations over the timescale $1/\Delta \omega_a$, so that measurements within this window are correlated.
We show how to account for this in Ref.~\cite{continuous_paper}, but it does not qualitatively affect the conclusion here.
Second, we have neglected errors in preparing and reading out the cavity state, and for a fixed conversion efficiency, these effects can make it more difficult to distinguish states.

\paragraph{More Nonclassicality Measures.}
%
Mandel $Q$ is the simplest nonclassicality measure involving number statistics, but many nonclassical states have $Q \geq 0$, such as the cat state proportional to $\ket{\alpha} + \ket{-\alpha}$.
There are more refined measures which can identify broader families of nonclassical states, but they are no more practical to measure than Mandel $Q$.

For example, for a general mode, define the normally ordered moments $G_n = \la (a^\dagger)^n a^n \ra$, related to Mandel $Q$ by $Q = (G_2 - G_1^2) / G_1$.
More general nonclassicality measures can be constructed from higher $G_n$~\cite{PhysRevA.46.485}.
These higher-order measures also have a bounded negative range for the axion, but using $G_n^c = \eta^n G_n^a$, one can show their in-cavity values are suppressed by more powers of $\eta$ than $Q_c = \eta Q_a$, making them even less practical to observe.

The fact that the higher moments do not yield substantial additional information can be shown rigorously by computing the Kullback--Leibler (KL) divergence~\cite{Kullback:1951zyt} of the cavity number distributions, which for the Fock and coherent DM states is $D_{\rm KL}({\rm Fock} || {\rm coherent}) \sim \eta^2$. 
Then by Stein's lemma~\cite{Cover:2005lom}, the number of measurements required to distinguish the two states scales as $N_{\text{shot}} \sim 1/D_{\rm KL} \sim 1/\eta^2$. 
On the other hand, in the weak-signal regime we have $D_{\rm KL}({\rm Gaussian}|| {\rm coherent}) \sim \la n_c \ra^2$, so that distinguishing Gaussian and coherent DM states requires $N_{\text{shot}} \sim 1/\la n_c \ra^2$.
These results confirm the estimates made above.

Other nonclassicality measures can be written in terms of the generating function
\be
M(\mu) = \sum_{n=0}^{\infty}p_{n} (1-\mu)^n
\ee
which is defined for $\mu \in [0, 2]$.
This quantity can be used to compute factorial moments.
For example, the Mandel $Q$ parameter is 
\be
Q = M^{\prime} (0) - \frac{M^{\prime\prime}(0) }{M^{\prime}(0)}.
\ee
The generating function can be written in terms of the $P$-function, 
\be \label{eq:mgf_P_function}
M(\mu) = \left\la (1 - \mu)^{a^{\dagger} a} \right\ra = \la \normord{e^{-\mu a^{\dagger} a}} \ra = \int d \alpha \, P(\alpha) e^{-\mu|\alpha|^2} 
\ee
where we used a standard identity (see Eqs.~(4.31) and (4.35) of Ref.~\cite{PhysRev.177.1857}), then applied the optical equivalence theorem Eq.~\eqref{eq:O_expt}.
Therefore, any negativity of $M(\mu)$ indicates a nonclassical state.
For example, only nonclassical states can be more likely to have an odd number of photons than an even number~\cite{PhysRevA.31.338}, which is equivalent to $M(2) < 0$.

Again, the low efficiency $\eta$ suppresses these signatures, as we have
\be
M_c(\mu) = \int d\alpha \, \frac{P_a(\alpha/\sqrt{\eta})}{\eta} \, e^{- \mu |\alpha|^2} = M_a(\eta \mu)
\ee
so that negativity in $P_a$ is washed out by integration against the smooth exponential $e^{- \mu |\alpha|^2}$.
For example, for an axion Fock state we have $M_a(\mu) = (1 - \mu)^{N_{\eff}}$, but in the cavity one has $M_c(\mu) = (1 - \eta \mu)^{N_{\eff}} $, which given $\mu \in [0, 2]$ is always nonnegative if $\eta < 1/2$. 
(In addition, we see that $M_c(\mu) \simeq e^{- \mu \eta N_{\eff}}$ up to $\mathcal{O}(\eta^2 N_{\eff})$ error terms.
This is precisely the generating function for a Poisson distribution, providing another way to see that low efficiency drives Fock number statistics to Poisson.)

Finally, there are signatures involving tails of the number distribution.
For example, Ref.~\cite{mandel1995optical} shows that for any state besides the vacuum state, if $p_n$ vanishes for any $n$, then the state is nonclassical.
Accordingly, for an axion Fock state, it is impossible to observe more than $N_{\eff}$ photons in the cavity, while it is possible for an axion coherent state.
But the probability of this occurring is exponentially suppressed with a very large exponent, $p_{N_\eff+1} \propto \eta^{N_\eff+1}$.

\subsection{Quadrature Measurements}
\label{sec:quad_meas}

We next consider cavity quadrature measurements, where many of the qualitative conclusions we found for number measurements hold.
Though there are no current axion experiments that perform projective quadrature measurements, many perform continuous quadrature measurements.
The two are roughly related by identifying the measurement rate with $1/t_m$, so the results here serve as a toy model for the continuous measurement results in Ref.~\cite{continuous_paper}.

We work with the dimensionless quadratures defined in Eq.~\eqref{eq:quad_defs}, so that $[X, Y] = i$ and the vacuum state has $\var(X) = \var(Y) = 1/2$.
For an initial cavity vacuum state, Eq.~\eqref{eq:simple_final_P} implies $\la X_c \ra = \sqrt{\eta} \, \la X_a \ra$.
By similar logic to that employed in Sec.~\ref{sec:number_meas}, higher normally ordered moments are also scaled by powers of $\sqrt{\eta}$, and in particular we have
\be
\la \normord{X_c^2} \ra = \eta \, \la \normord{X_a^2} \ra.
\ee
The simplest nonclassicality measure involving quadratures is the squeezing parameter,
\be \label{eq:Sparam_def}
S = \var(X) - \frac{1}{2} = \la \normord{X^2} \ra - \la X \ra^2 \geq -\frac12.
\ee
Classical states satisfy $S \geq 0$.\footnote{Defining $x_{\alpha} = (\alpha+\alpha^*)/\sqrt{2}$, we have $S = \int d\alpha\,P(\alpha)\, \big(x_{\alpha}-\la X \ra\big)^2$, which is nonnegative when $P(\alpha) \geq 0$.
Not all nonclassical states have negative $S$: a Fock state has $S = \la n \ra$, identical to the Gaussian.}
For example, from the first form of $S$ we see that $S = 0$ for a coherent state and $S = \la n \ra$ for a Gaussian state. 
An example of a state that can enter the nonclassical range is the squeezed vacuum, $\ket{z} = \hat{\mathcal{S}}(z) \ket{0}$, where the squeezing operator is
\be \label{eq:squeeze-op}
\hat{\mathcal{S}}(z) = \exp(\frac{1}{2} \big(z^* a^2 - z a^{\dag 2} \big)),
\ee
and $z = r e^{i\theta}$, with $r,\,\theta$ the squeezing amplitude and angle.
The squeezed vacuum has $S = (\cosh 2r - \cos \theta \sinh 2r-1)/2$.
For $\theta=0$, we have $S = (e^{-2r}-1)/2$, which achieves $S = -1/2$ in the limit of infinite squeezing, $r \to \infty$.
From the second form of $S$ in Eq.~\eqref{eq:Sparam_def}, 
\be \label{eq:Scav_relation}
S_c = \eta S_a \geq - \frac{\eta}{2}
\ee
which already indicates that it is very difficult to observe axion-induced squeezing.

More generally, defining $q_n = \la \normord{X^n} \ra$, the squeezing parameter is $S = q_2 - q_1^2$, and further measures can be constructed from higher $q_n$~\cite{AGARWAL1993109}, but their observability is suppressed by more powers of $\eta$.
The same applies for appropriately normalized nonclassicality measures involving higher-order moments of both quadratures~\cite{PhysRevA.71.011802}.

\paragraph{The Quadrature Distribution.}
%
We can gain intuition for these suppressions by examining the probability distribution of the $X$ quadrature in the cavity,
\bea \label{eq:quad_dist}
p(x) &= \int d\alpha \, P_c^f(\alpha) \, | \la x | \alpha \ra |^2
= \int d\alpha \, P_a(\alpha) | \la x | \sqrt{\eta} \, \alpha \ra |^2 \\
&= \frac{1}{\sqrt{\pi}} \int d\alpha \, P_a(\alpha) \, \exp(- \big(x - \sqrt{\eta}\, x_{\alpha}\big)^2),
\eea
where $x_{\alpha} = (\alpha+\alpha^*)/\sqrt{2} = \sqrt{2}\, \text{Re}(\alpha)$.
Similar to Eq.~\eqref{eq:num_dist}, the quadrature distribution involves an integral of $P_a$ against a slowly varying function, suppressing nonclassical effects.

We can also directly relate $p(x)$ to the axion quadrature distribution $p_\DM(x)$.
The simplest route is to work with the Wigner characteristic function introduced in App.~\ref{app:general_quasi_characteristic}, since the marginalized Wigner function is the quadrature distribution.\footnote{In detail, defining $y_{\alpha} = \sqrt{2}\,\text{Im}(\alpha)$, $p(x) = \frac{1}{2} \int d y_{\alpha}\, W(\alpha)$.}
From Eq.~\eqref{eq:simple_final_P}, we have $\tilde{P}^f_c(\lambda) = \tilde{P}_a(\sqrt{\eta} \, \lambda)$, so that the Wigner characteristic function obeys
\be
\tilde{W}^f_c(\lambda) = e^{(\eta-1) |\lambda|^2/2} \, \tilde{W}_a(\sqrt{\eta} \, \lambda).
\ee
Fourier transforming back to the Wigner function and marginalizing, we conclude
\be \label{eq:convolved_quad_dist}
p(x) = \int dx' \, \frac{p_\DM(x'/\sqrt{\eta})}{\sqrt{\eta}} \, \frac{e^{-(x - x')^2 / (1 - \eta)}}{\sqrt{(1 - \eta) \pi}}.
\ee
That is, the axion quadrature distribution is scaled down and convolved against a Gaussian with variance $(1-\eta)/2$.
For small $\eta$, this variance is only very slightly smaller than that of the vacuum state, suppressing the effect of even highly squeezed axion states.

\paragraph{Distinguishing Quadrature Distributions.}
%
As for number measurements, we find that with quadrature measurements, it is extremely difficult to detect nonclassical effects, such as those due to squeezed states, while it is feasible to distinguish different classical states.

First, consider the ideal case of an axion state squeezed with arbitrarily high strength, so that $p_\DM(x) \simeq \delta(x)$.
This state drives the cavity into a very slightly squeezed state, whose quadrature distribution is a Gaussian with variance $\var(X_c) = (1-\eta) / 2$.
Detecting this effect requires differentiating this scenario from a coherent state with variance $1/2$ and therefore measuring $\var(X_c)$ to a fractional precision of $\eta$.
By the same logic as in Sec.~\ref{sec:number_meas}, this requires $N_{\text{shot}} \sim 1/\eta^2$ measurements, corresponding to the infeasibly long integration time in Eq.~\eqref{eq:integration_time_quantum}.
As for number measurements, information from higher moments does not help, as can be confirmed by computing the KL divergence, $D_{\rm KL}({\rm squeezed} || {\rm coherent}) \sim \eta^2$.

In fact, this estimate is highly optimistic, because a squeezed state is not stationary; in the Schr\"odinger picture it continually rotates in phase space, so that $\var(X_a)$ oscillates between a very small value, and a large value of order $N_\eff$.
To detect the nonclassical squeezing in one quadrature, one must track the candidate axion's phase between measurements to an extreme precision $\sim\! 1/\sqrt{N_\eff}$ to avoid overwhelming the signal with noise from the other quadrature. 
This treatment also assumes identical independent measurements, which neglects both the nontrivial time evolution of the effective mode's state, and the backreaction that our measurements induce on the state of the DM.

Again, it is readily possible to distinguish between different classical axion states.
This is clearly possible in the strong-signal regime, so we focus on the weak-signal case $\eta N_\eff \ll 1$.
As a simple example, a coherent axion state aligned along the $X$ quadrature would yield $\la X_c \ra = \sqrt{2 \eta N_\eff}$.
The final quadrature distribution is a slightly shifted Gaussian with variance $\var(X_c) = 1/2$, so detecting this shift and thereby discovering the axion requires $N_{\text{shot}} \sim 1/(\eta N_\eff)$. 
Furthermore, this time is also sufficient to distinguish a coherent state from any stationary state, such as a Gaussian state, as such states yield $\la X_c \ra = 0$.

Since a coherent state is not stationary, this strategy requires tracking the phase of the axion oscillation with order-one precision across measurements.
If this is not possible, we should instead consider distinguishing the phase-averaged coherent state Eq.~\eqref{eq:phaseMixtureP} from, e.g.~a Gaussian state.
In both cases, one would discover the axion by detecting an increase in the quadrature variance to $\var(X_c) = 1/2 + \eta N_\eff$, which requires $N_{\text{shot}} \sim 1 / (\eta N_\eff)^2$.
The leading moment that differs between the states is $\la \normord{X_c^4} \ra$, which contains a contribution $\sim \eta^2 \la X_a^4 \ra \sim (\eta N_\eff)^2$, representing the strength of axion intensity fluctuations.
The coefficient of this term differs by order-one between the two states.
Thus, one must measure the raw fourth moment $\la X_c^4 \ra$, which contains an order-one vacuum contribution, to absolute precision $\sim (\eta N_\eff)^2$, corresponding to $N_{\text{shot}} \sim 1 / (\eta N_\eff)^4$.
Since $\eta N_\eff$ can only be mildly small for the axion to be discoverable, distinguishing these states after discovery is thus only moderately more difficult than discovery itself.

\subsection{Entanglement and General Observables}
\label{sec:general_suppression}

We have shown explicitly that two simple nonclassicality measures -- Mandel $Q$ and the squeezing parameter $S$ -- are extremely suppressed, and for each resolving nonclassical effects requires $N_{\text{shot}} \sim 1/\eta^2$ independent measurements.
We further argued that higher moments of quadrature or number do not provide a loophole to this argument.

Nonetheless, one could go on to consider even more general higher-order moments of the number distribution~\cite{PhysRevA.41.1721,PhysRevA.41.1569,KLYSHKO19967}, or mixed number-quadrature moments~\cite{PhysRevA.72.043808}.
Yet other nonclassicality criteria can be written in terms of the Wigner function~\cite{PhysRevLett.124.133601,PhysRevResearch.3.043116} or the Husimi $Q$-function~\cite{PhysRevA.51.3340}.
In particular, nonclassical states which achieve a negative value of the Wigner function possess ``quantum magic,''~\cite{Bravyi:2004isx} which encodes the degree to which a system cannot be efficiently simulated classically~\cite{Ferrie:2011rce,Veitch:2012ttw,Mari:2012ypq,Wang:2019nve}.

In lieu of considering all of these cases, here we instead discuss two broader points.
First, we present a general argument that the visibility of any nonclassical effect is always suppressed by at least one additional power of $\eta$.
Second, we show that our conclusions generalize to multi-detector observables such as entanglement.

\paragraph{A General Argument.}
%
The DM $P$-function is scaled down by $\sqrt{\eta}$ when mapped to the cavity, and heuristically this means that nonclassical effects are generically suppressed by rapid sign oscillations in $P_c^f(\alpha)$.
To make this precise, including for states with singular $P$-functions, here we show that for every nonclassical axion state, one can construct a smeared, \textit{classical} axion state which averages out these sign fluctuations. 
This state yields almost identical statistics in the cavity, up to corrections suppressed by $\eta$.

In the previous subsections we assumed an initial cavity vacuum state, but here we let the initial cavity state be arbitrary, so the final cavity $P$-function $P_c^f$ is given by Eq.~\eqref{eq:approx_cavity_P}.
Now, the probability of any cavity measurement outcome $\mathcal{O}$ can be written in the form
\be \label{eq:prob_def}
p_\mathcal{O} = \int d \alpha \, P_c^f(\alpha) \, f_\mathcal{O}(\alpha).
\ee
The function $f_\mathcal{O}(\alpha)$ is nonnegative, because if the projection operator for $\mathcal{O}$ is $\Pi_\mathcal{O}$, then $f_\mathcal{O}(\alpha) = \la \alpha | \Pi_\mathcal{O} | \alpha \ra \geq 0$.
For example, for observing $n$ photons we have $f_n(\alpha) = | \la n | \alpha \ra |^2$, for observing an even number of photons we have $f_{\text{even}}(\alpha) = \sum_{m=0}^\infty | \la 2m | \alpha \ra |^2$, and for observing a value of $x$ for the $X$ quadrature we have $f_x(\alpha) = | \la x | \alpha \ra |^2$.

As we discuss in App.~\ref{app:general_quasi_characteristic},
convolving a $P$-function with an order-one width Gaussian $e^{-|\alpha|^2}/\pi$ yields the corresponding Husimi $Q$-function, which is nonnegative.
Thus, we consider an alternative classical axion state whose $P$-function is the $Q$-function of the original axion state.
Changing to this state is equivalent to convolving the original $P_c^f(\alpha)$ in Eq.~\eqref{eq:prob_def} with a Gaussian $g(\alpha) = e^{-|\alpha|^2/\eta}/(\pi \eta)$ of order-$\sqrt{\eta}$ width, so that the probability of outcome $\mathcal{O}$ is
\be \label{eq:smearing_construction}
p_\mathcal{O}^{\text{cl}} = \int d\alpha \, d\beta \, P_c^f(\beta) g(\alpha - \beta) f_\mathcal{O}(\alpha).
\ee
We can equivalently regard this convolution as acting on $f_\mathcal{O}$, so that 
\be \label{eq:f_X_def}
p_\mathcal{O}^{\text{cl}} = \int d\alpha \, P_c^f(\alpha) f_\mathcal{O}^{\text{cl}}(\alpha), \qquad f_\mathcal{O}^{\text{cl}}(\alpha) = \int d\beta \, f_\mathcal{O}(\alpha-\beta) g(\beta).
\ee

For smooth $f_\mathcal{O}$, this narrow Gaussian smoothing has a very small effect. 
To see this heuristically, we let $r$ and $i$ subscripts denote real and imaginary parts, and Taylor expand
\bea
f_\mathcal{O}(\alpha-\beta) &= f_\mathcal{O}(\alpha) - \beta_r \partial_{\alpha_r} f_\mathcal{O}(\alpha) - \beta_i \partial_{\alpha_i} f_\mathcal{O}(\alpha) \\
&\qquad + \frac{1}{2} \beta_r^2 \partial_{\alpha_r}^2 f_\mathcal{O}(\alpha) + \beta_r \beta_i \partial_{\alpha_r} \partial_{\alpha_i} f_\mathcal{O}(\alpha) + \frac{1}{2} \beta_i^2 \partial_{\alpha_i}^2 f_\mathcal{O}(\alpha) + \ldots.
\eea
Plugging this back into Eq.~\eqref{eq:f_X_def}, all terms odd in $\beta_i$ or $\beta_r$ integrate to zero, leaving
\be
f_\mathcal{O}^{\text{cl}}(\alpha) = f_\mathcal{O}(\alpha) + \frac{\eta}{4} \, \nabla^2_\alpha f_\mathcal{O}(\alpha) + \mathcal{O}(\eta^2),
\ee
where $\nabla_\alpha^2 = \partial_{\alpha_r}^2 + \partial_{\alpha_i}^2$, and the $\mathcal{O}(\eta^2)$ contribution comes from the fourth-order term in the Taylor expansion.
Therefore, the change in the probability of outcome $\mathcal{O}$ is
\be \label{eq:pcl_difference}
p_\mathcal{O}^{\text{cl}} - p_\mathcal{O} = \frac{\eta}{4} \int d\alpha \, P_c^f(\alpha) \nabla_\alpha^2 f_\mathcal{O}(\alpha) + \mathcal{O}(\eta^2).
\ee
For simple measurement outcomes, such as those below Eq.~\eqref{eq:prob_def}, the function $\nabla_\alpha^2 f_\mathcal{O}$ is not much larger in magnitude than $f_\mathcal{O}$ itself, so that $p_\mathcal{O}^{\text{cl}} - p_\mathcal{O}$ is always suppressed by a power of $\eta$, with no enhancement by the axion occupancy $N_\eff$. This implies that one needs at least $\sim\! 1/\eta$ measurements to detect any nonclassical effect.

We discussed a related, but distinct issue in Ref.~\cite{Bao:2025nsd}.
There we showed that if the cavity began in a thermal state with occupancy $n_t$, then $P_c^f(\alpha)$ is convolved with a Gaussian, erasing its negativity if $n_t > \eta$.
Achieving $n_t \lesssim \eta$ requires $T \lesssim m_a / \log(1/\eta)$, which is difficult, but plausible in a dilution fridge for $m_a \gtrsim \text{few} \times 10^{-5} \, \mathrm{eV}$.
By contrast, the argument here shows that even in the absence of noise, the effects of negativity are always suppressed.

\paragraph{Evading the Suppression.}
To cancel the suppression in Eq.~\eqref{eq:pcl_difference}, the integral must scale as $1/\eta$.
However, this requires a highly excited nonclassical cavity state, for any choice of $\mathcal{O}$.
To see this, let $\rho^f$ and $\rho^f_{\text{cl}}$ be the final cavity states due to interaction with a nonclassical DM state, and the nearby classical DM state constructed above. 
For any measurement outcome $\mathcal{O}$ with associated projector $\Pi_{\mathcal{O}}$,
\be
\big|p_{\mathcal{O}} - p_{\mathcal{O}}^{\text{cl}}\big| = \bigg| \tr((\rho^f - \rho^f_{\text{cl}}) \Pi_{\mathcal{O}}) \bigg| \leq D \big(\rho^f, \rho^f_{\text{cl}} \big) = \frac{1}{2} \tr \big|\rho^f - \rho^f_{\text{cl}}\big|
\ee
where $D$ is the trace distance and $\tr |A| = \tr \sqrt{A^\dagger A}$ is the trace norm.
(For a pedagogical discussion of results used here, see Sec.~9.2 of Ref.~\cite{nielsen2010quantum}.
The result above holds unchanged for an arbitrary POVM.)
Next, as shown above Eq.~\eqref{eq:smearing_construction}, $\rho_{\text{cl}}^f$ can be constructed by Gaussian smearing of $\rho^f$, which can be realized by a Lindbladian,
\be
\rho_{\text{cl}}^f = e^{\eta \mathcal{L}} \rho^f, \hspace{0.5cm}
\mathcal{L}[\rho] = c \rho c^\dagger + c^\dagger \rho c - \frac12 \{c^\dagger c + c c^\dagger, \rho\}.
\ee
We can then write $\rho^f - \rho^f_{\text{cl}}$ as an integral, giving
\be
\tr \big|\rho^f - \rho^f_{\text{cl}}\big| = \tr \bigg| \int_0^\eta ds\, e^{s \mathcal{L}} \mathcal{L}[\rho^f] \bigg| \leq \int_0^\eta ds\, \tr \big| e^{s \mathcal{L}} \mathcal{L}[\rho^f] \big| \leq \eta \tr |\mathcal{L}[\rho^f]|
\ee
where we used the triangle inequality and then the fact that the trace-preserving quantum operation $e^{s \mathcal{L}}$ cannot increase the trace norm.

Thus, the trace distance is penalized by a factor of $\eta$; avoiding this penalty requires $\tr |\mathcal{L}[\rho^f]|$ to be very large.
Since $\tr |\mathcal{L}[\rho^f]| \leq 2 \langle n_c \rangle + 1 + \sqrt{\langle (2 n_c + 1)^2 \rangle}$, we would need the final cavity state to have $\langle n_c \rangle \sim 1/\eta$ or $\sqrt{\langle n_c^2 \rangle} \sim 1/\eta$.
However, if the cavity began in the vacuum state, then Eq.~\eqref{eq:acHP} implies that final-state cavity moments scale as powers of $\eta N_\eff$, and therefore cannot produce a bare factor of $N_\eff$ that might compensate the penalizing $\eta$.
Instead, a suitably large $\tr |\mathcal{L}[\rho^f]|$ can only arise from the cavity initial state.

Setting up an initial state with $\la n_c \ra \sim 1/\eta$ alone is not sufficient and indeed not difficult to achieve; for instance, loading the cavity with a coherent oscillating magnetic field for the parameters in Eq.~\eqref{eq:eta_value} and $V_c = 0.1\,$m$^3$, only requires a field amplitude of $B \sim 10^{-4} \, \mathrm{T}$. 
However, starting with a cavity coherent state simply translates the final cavity $P$-function, which provides no advantage over starting in the vacuum state. 
Furthermore, starting in a classical mixture of coherent states would simply add additional noise. 
Instead, we must initialize the cavity in a highly excited nonclassical state, which is very difficult.
Even if this is achieved, one would need to overcome the further challenge of performing precise measurements on top of such a state in the presence of the noise associated with realistic state preparation and readout.

For example, suppose both the DM effective mode and the cavity are initially in uncorrelated squeezed vacuum states.\footnote{We thank Liang Jiang for suggesting this example.} 
Then the final quadrature variance is 
\be
\var(X_c) = (1 - \eta) \var(X_{c, 0}) + \eta \var(X_a).
\ee
For an initial vacuum cavity state, $\var(X_{c, 0}) = 1/2$, we saw in Sec.~\ref{sec:quad_meas} that it would take $\sim\! 1/\eta^2$ measurements to detect the small fractional decrease in $\var(X_c)$ due to a DM state squeezed along the $X$ quadrature, compared to a coherent DM state. 
However, if $\var(X_{c, 0}) \sim \eta$, it would only take $\mathcal{O}(1)$ measurements to make this distinction.

Such a level of squeezing is many orders of magnitude beyond current experimental capabilities.
Consistent with our general argument above, such a state requires $\var(Y_{c, 0}) \sim 1/\eta$ and hence an occupancy $\langle n_c \rangle \sim 1/\eta$. 
Crucially, the advantage of such a state is due to $\var(X_{c, 0})$ being small; there is no advantage just from having large $\var(Y_{c, 0})$, which would be easy to achieve classically.
Finally, even given perfect squeezed state preparation, one still has the problem of phase alignment noted in Sec.~\ref{sec:quad_meas}.
Here the problem is even more severe, as one must precisely align the squeezing axes of the cavity and DM to avoid noise from $Y_{c,0}$.

\paragraph{Mode Entanglement.}
%
The preceding argument applies to general measurements in a single-mode cavity, but the effective mode formalism also applies to multiple cavity modes.
Therefore, as a final example we show that nonclassical signatures unique to multiple detector modes, such as entanglement, are suppressed by $\eta$ like the single-mode observables considered above.
This is relevant to proposed DM searches using multiple qubits, possibly prepared in entangled states~\cite{Chen:2022quj,Chen:2023swh,Ito:2023zhp,Chen:2024aya,Fukuda:2025zcf,Bodas:2025vff}.

We consider joint measurements of the $\text{TM}_{010}$ mode in two identical cavities.
For simplicity we assume the cavities are widely separated, so that the commutator in Eq.~\eqref{eq:effective_commutator} is negligible.
Then for an initial joint cavity vacuum state, the final joint cavity $P$-function is 
\be
P_c^f(\alpha_1, \alpha_2) = \frac{1}{\eta^2} \, P_a\!\left( \frac{\alpha_1}{\sqrt{\eta}}, \frac{\alpha_2}{\sqrt{\eta}} \right)\!.
\ee
As a result, entanglement between the two axion effective modes can be transduced into entanglement between the two cavity modes.\footnote{This differs from the example we considered in Ref.~\cite{Bao:2025nsd}, which involved an axion effective mode in a cat state becoming entangled with a single cavity mode.
The observability of that effect is even more suppressed, because one can only read out the cavity mode, while here one can read out both entangled cavity modes.}

Entanglement is an inherently quantum correlation between two systems, and for pure states it can always be identified by computing the von Neumann entropy, $-\tr(\rho \log \rho)$, of the reduced density matrix of either mode.
As for quadrature and number statistics, there exist a variety of simpler entanglement measures that are more amenable to computation and measurement~\cite{PhysRevA.60.2752,PhysRevLett.84.2726,PhysRevLett.96.050503}.
Here we consider the simple entanglement witness
\be
E = \frac{\var(X_1 - X_2)}{2} + \frac{\var(Y_1 + Y_2)}{2} - 1
\ee
which is a special case of the one defined in Ref.~\cite{PhysRevLett.84.2722}.
For joint coherent states we have $E = 0$, and for a joint thermal Gaussian state with mean occupancy $\langle n \rangle$ in each mode, we have $E = 2 \langle n \rangle$.
Again, negative values of $E$ are intrinsically quantum,\footnote{To see this, note that $2E = \la \normord{(X_1 - X_2)^2} \ra - \la X_1 - X_2 \ra^2 + \la \normord{(Y_1 + Y_2)^2} \ra - \la Y_1 + Y_2 \ra^2$, cf. Eq.~\eqref{eq:Sparam_def}.
Combined with the optical equivalence theorem, this shows that classical states must have $E \geq 0$.} but have a restricted range.
For example, the two-mode squeezed state
\be
\ket{\lambda} = \exp( r ( a_1^\dag a_2^\dag - a_1 a_2)) \ket{0} \ket{0} = \sqrt{1- \lambda^2} \sum_{n=0}^\infty \lambda^n \ket{n} \ket{n},
\ee
has $E = -2\lambda/(1+\lambda)$, where $\lambda = \tanh r$.
The minimum possible value of $E = -1$ is achieved in the limit of infinite squeezing ($r \to \infty$, $\lambda \to 1$), where the two quadratures become perfectly correlated and anticorrelated, $\la X_1, X_2 | \lambda \ra \propto \delta(X_1 - X_2)$ and $\la Y_1, Y_2 | \lambda \ra \propto \delta(Y_1 + Y_2)$.

As usual, negative values of $E_c$ that the axion can imprint in the cavities are extremely suppressed.
To see this, note that in Heisenberg picture (with the free evolution factored out), using Eq.~\eqref{eq:acHP} the final $X$ quadrature operator for cavity $i=1,2$ can be written in terms of the initial cavity and axion $X$ quadrature operators by
\be
X_i = \sqrt{1-\eta} \, X_{i,0} + \sqrt{\eta} \, X_{i,a}.
\ee
Since the initial axion and cavity states are uncorrelated, $\text{covar}(X_{1, 0}-X_{2,0},\, X_{1,a}-X_{2,a}) = 0$, we have 
\be
\var(X_1 - X_2) = (1 - \eta) \var(X_{1,0} - X_{2, 0}) + \eta \var(X_{1,a} - X_{2,a})
\ee
and if the cavities start in their joint ground state, we have $\var(X_{1, 0} - X_{2, 0}) = 1$.
Adding the analogous result for the other quadrature yields 
\be
E_c = (1-\eta) + \eta (E_{a} + 1) - 1 = \eta E_{a}
\ee
which is precisely the same scaling we have seen for Mandel $Q$ and squeezing $S$.
In all three cases observing a nonclassical effect requires measuring an effect that is suppressed by $\eta$ but not enhanced by $N_\eff$.

Accordingly, entanglement behaves similarly to number or quadrature.
Given the tiny magnitude of $\eta$, it is impractical to observe axion-induced entanglement between cavity modes, which would be intrinsically quantum.
No such penalty applies for observing axion-induced \textit{classical} correlations, and indeed one can use this to perform interferometry on the axion wave as outlined in Ref.~\cite{Foster:2020fln}, which can also be implemented using quantum sensors~\cite{Fukuda:2025zcf}.

\section{Decoherence From Classical and Quantum Dark Matter} 
\label{sec:decoherence}
One could search for DM via the decoherence it induces upon interacting with a detector prepared in a suitable state. 
For example, Refs.~\cite{Riedel:2012ur,Riedel:2016acj} proposed to detect light particle DM through the decoherence of a test mass prepared in a macroscopic superposition of positions.
(In addition, Refs.~\cite{Du:2022ceh,Badurina:2024nge,Badurina:2026owr} discuss decoherence as a DM signal in atom interferometers, and Ref.~\cite{Badurina:2025idj} mentions the $P$-function in the context of trapped-ion interferometry.)
For very soft DM scatterings, the energy deposited would be too small to detect, but the partial measurement of the test mass's state performed by each passing DM particle could induce observable decoherence.
This detection scheme is inherently quantum mechanical, and its signal and background rates can differ qualitatively from more traditional measurements.

Thus, it is interesting to consider decoherence induced by axion DM.
In this section, we take the first steps in this direction by computing the decoherence a general DM state induces in a cavity haloscope.
Our focus is on understanding whether nonclassical DM states can leave a distinctive, potentially detectable imprint through decoherence, given the quantum mechanical nature of the process.
Ultimately, however, we show that they cannot.

There are multiple ways to quantify decoherence.
Suppose one prepares a detector in a pure, equal superposition of very distinct states, $\rho^c \propto (\ket{\alpha} + \ket{\beta}) (\bra{\alpha} + \bra{\beta})$.
In the context of Refs.~\cite{Riedel:2012ur,Riedel:2016acj}, DM scatterings could not cause transitions between the states $\ket{\alpha}$ and $\ket{\beta}$, so decoherence was quantified in terms of the decay of the off-diagonal elements $\bra{\alpha} \rho^c \ket{\beta}$.
However, axion DM interacts inelastically; absorption and emission of axion quanta can significantly change the detector state.
Thus, we instead quantify decoherence in terms of the decrease of $\tr((\rho^c)^2)$, the purity of the cavity state, over a timescale $t$.
This quantity serves as a generic estimate of the size of decoherence effects.

In Sec.~\ref{sec:dec_perturbative}, we calculate the purity loss perturbatively, to second order in the coupling.
We show that for fixed DM occupancy and a given cavity state, nonclassical DM states cannot yield larger purity loss than classical states.
It is possible for nonclassical DM to induce a smaller purity loss than any classical DM state, but just as for the observables considered in Sec.~\ref{sec:projective}, this difference is suppressed by $\eta$ but not enhanced by $N_\eff$, rendering it extremely difficult to observe. 
We also discuss when decoherence might yield competitive axion sensitivity.
In Sec.~\ref{sec:dec_general}, we derive a more general expression for the purity loss in terms of the DM and cavity $P$-functions, which elucidates the origin of certain features of the perturbative result.

\subsection{Perturbative Calculation}
\label{sec:dec_perturbative}

The DM-cavity density matrix can be expanded order by order in $g$ as $\rho = \rho_0 + \rho_1 + \rho_2 + \mathcal{O}(g^3)$, where $\rho_0 = \rho_0^a \otimes \rho_0^c$ is the initial state, and the equation of motion $\dot{\rho} = - i [H_{\text{int}}, \rho]$ gives
\be
\rho_1 = - i t [H_{\text{int}}, \rho_0], \qquad \rho_2 = - \frac{t^2}{2} [H_{\text{int}}, [H_{\text{int}}, \rho_0]],
\ee
where $H_{\text{int}} = i g (c^\dag a - c a^\dag)$ is the toy model Hamiltonian of Eq.~\eqref{eq:H-toy}.
Letting $\tr_c$ and $\tr_a$ be the partial traces over the cavity and DM modes respectively, the reduced density matrix for the cavity can be expanded as
\be
\rho^c = \tr_a \rho = \rho_0^c + \tr_a(\rho_1) + \tr_a(\rho_2) + \mathcal{O}(g^3) = \rho_0^c + \rho_1^c + \rho_2^c + \mathcal{O}(g^3).
\ee
For simplicity, we assume in this perturbative calculation that the initial cavity state is pure, $\tr_c \!\big((\rho_0^c)^2\big) = 1$, so the purity loss is
\be \label{eq:purity_loss}
\delta \equiv 1 - \tr_c((\rho^c)^2) = - \tr_c \!\left(\{\rho_0^c, \rho_1^c\} + (\rho_1^c)^2 + \{\rho_0^c, \rho_2^c\}\right) + \mathcal{O}(g^4).
\ee

\paragraph{No Purity Loss at First Order.}
The first term in Eq.~\eqref{eq:purity_loss} is $\mathcal{O}(g)$, but it always vanishes.
To see this, we note that
\be
\rho_1^c = g t \, [\la a \ra c^\dag - \la a^\dag \ra c,\, \rho_0^c]
\ee
where $\la \cdot \ra$ denotes an expectation value with respect to the DM state. Then we have 
\be
\tr_c (\rho_0^c \, \rho_1^c) = gt \la a \ra \tr_c ( \rho_0^c (c^\dag \rho_0^c - \rho_0^c c^\dag) ) - gt \la a^\dag \ra \tr_c ( \rho_0^c (c \rho_0^c - \rho_0^c c) ) = 0
\ee
by the cyclic property of the trace.
In other words, purity loss begins at $\mathcal{O}(g^2)$ even though the axion's effect on the cavity state starts at $\mathcal{O}(g)$.
This is not fundamentally different from how the axion signal power scales as $\mathcal{O}(g^2)$ even though the signal amplitude is $\mathcal{O}(g)$.

\paragraph{Purity Loss at Second Order.}
We write the two second-order terms in Eq.~\eqref{eq:purity_loss} as
\be
\delta = -\eta (z_1 + z_2)
\ee
where the conversion efficiency, defined in Eq.~\eqref{eq:efficiency}, is $\eta = \sin^2(gt) \simeq g^2 t^2$.
For $z_1$, we have
\bea \label{eq:c1_terms}
z_1 &= \tr_c \!\Big( [\la a \ra c^\dag - \la a^\dag \ra c,\, \rho_0^c]^2 \Big) \\
&= \tr_c \!\Big( \la a \ra^2 [c^\dag, \rho_0^c]^2 + \la a^\dag \ra^2 [c, \rho_0^c]^2 - \la a \ra \la a^\dag \ra \big\{[c^\dag, \rho_0^c],\,[c, \rho_0^c] \big\} \Big).
\eea
All of these terms have two powers of $\rho_0^c$ and can be simplified with the assumption that the initial detector state is pure, so that $\rho_0^c = \ket{\psi} \bra{\psi}$ for some $\ket{\psi}$.
This implies, for example, that 
\bea
\tr_c (c \rho_0^c c \rho_0^c) &= \tr_c (c \ket{\psi} \bra{\psi} c \ket{\psi} \bra{\psi}) = \bra{\psi} c \ket{\psi} \, \bra{\psi} c \ket{\psi} = \la c \ra^2 \\
\tr_c \!\big(c^2 (\rho_0^c)^2 \big) &= \tr_c \!\big(c^2 \rho_0^c \big) = \la c^2 \ra.
\eea
Applying similar manipulations to each term in Eq.~\eqref{eq:c1_terms}, we find
\be
\frac{z_1}{2} = \la a \ra^2 \left(\la c^\dag \ra^2 - \la {c^\dag}^2 \ra \right) + \la a^\dag \ra^2 \left(\la c \ra^2 - \la c^2 \ra \right) + \la a \ra \la a^\dag \ra \left(\la c^\dag c + c c^\dag \ra - 2 \la c^\dag \ra \la c \ra \right)\!.
\ee
To evaluate $z_2$, we first note that
\be
\rho_2^c = \frac{\eta}{2} \tr_a \Bigl([a c^\dag - a^\dag c,\, [a c^\dag - a^\dag c,\, \rho_0]]\Bigr).
\ee
From here, expanding and simplifying assuming an initial pure detector state gives
\be
\frac{z_2}{2} = \la a^2 \ra (\la {c^\dag}^2 \ra - \la c^\dag \ra^2) + \la {a^\dag}^2 \ra (\la c^2 \ra - \la c \ra^2) - \la a^\dag a \ra (\la c c^\dag \ra - \la c \ra \la c^\dag \ra) - \la a a^\dag \ra (\la c^\dag c \ra - \la c^\dag \ra \la c \ra).
\ee
Collecting our results, we conclude that
\bea \label{eq:final_dec}
\delta = 2 \eta \Bigg( 2 &\left(\la a^\dag a \ra - |\la a \ra|^2 + \frac12 \right) \left(\la c^\dag c \ra - |\la c \ra|^2 + \frac12 \right) - \frac12 \\
&-\Big(\la {a^\dag}^2 \ra - \la a^\dag \ra^2 \Big) \Big(\la c^2 \ra - \la c \ra^2 \Big) - \Big(\la a^2 \ra - \la a \ra^2 \Big) \Big(\la {c^\dag}^2 \ra - \la c^\dag \ra^2 \Big) \Bigg).
\eea
This gives the purity loss at second order in terms of first and second moments of the initial DM and cavity states. 
(A similar expression was derived, in a different context, in Ref.~\cite{Goldberg:2021syi}.)
Such a structure was inevitable, since we were working perturbatively, but it demonstrates that decoherence is not qualitatively different from more mundane observables such as signal power; both can be computed in terms of low-order moments.
As such, we should not expect decoherence to be parametrically more sensitive to nonclassical states than any of the observables considered in Sec.~\ref{sec:projective}.
We confirm this intuition below.

\paragraph{Examples of Purity Loss.}
We can read off special cases from the general result Eq.~\eqref{eq:final_dec}.
First, for a coherent DM state we have
\be \label{eq:coherent_dec}
\delta = 2 \eta \left( \la c^\dag c \ra - |\la c \ra|^2 \right)\!.
\ee
This contribution arises solely from the axion's vacuum fluctuations.
While decoherence from electromagnetic vacuum fluctuations has been considered as an observable effect~\cite{Gundhi:2025bwj}, the analogous effect for the axion is extremely weak, since it is suppressed by $\eta$ but not enhanced by the axion's large occupancy.\footnote{There is another very small contribution due to the terms dropped in the rotating wave approximation.} In the exact calculation below, we will see that coherent DM states give the minimum purity loss among all classical DM states.

Second, for any stationary DM state, such as a thermal Gaussian state or Fock state, the leading-order purity loss is
\be \label{eq:stationary_delta}
\delta_{\text{st}} = 2 \eta \Bigg( 2 \left(\la a^\dag a \ra + \frac12 \right) \left(\la c^\dag c \ra - |\la c \ra|^2 + \frac12 \right) - \frac12 \Bigg)
\ee
which only depends on the mean occupancy of the DM, and not on more detailed properties of the DM state.
The purity loss can be somewhat altered for a nonstationary state, with $\la a^2 \ra \neq \la a \ra^2$. 
For example, if the DM state is an equal classical mixture of $\ket{\alpha}$ and $\ket{-\alpha}$, then the purity loss becomes 
\be
\delta = \delta_{\text{st}} - 4 \eta \, \text{Re} \big( \alpha^2 (\la c^2 \ra - \la c \ra^2 )^* \big).
\ee
The additional term has the same scaling with DM occupancy as $\delta_{\text{st}}$, so the purity loss can be altered by an order-one factor.
However, to see this effect, one would have to prepare the cavity itself in a nonstationary state, with $\la c^2 \ra \neq \la c \ra^2$.

Nonclassical states can yield slightly lower purity loss than any classical state.
For example, if both the cavity and DM are in identical squeezed states, then the purity loss would be zero,\footnote{To see this, use the fact that squeezed vacuum states obey $a \ket{\psi} = \nu a^\dagger \ket{\psi}$ for a coefficient $\nu$. When $\nu_a = \nu_c$, the second line of Eq.~\eqref{eq:final_dec} cancels the first.} smaller than the classical minimum given by Eq.~\eqref{eq:coherent_dec}. 
However, like Mandel $Q$ or squeezing $S$, the difference in purity loss is suppressed by $\eta$ but not enhanced by $N_\eff$.

For fixed $N_\eff$, nonclassical DM states do not yield significantly higher purity loss.
At $\mathcal{O}(\eta)$, the extreme case of a DM Fock state yields the same purity loss as a thermal Gaussian state, since both are stationary. 
The DM cat state proportional to $\ket{\alpha} + \ket{-\alpha}$, highlighted in Ref.~\cite{Allali:2021puy}, yields almost exactly the same decoherence as a classical mixture of $\ket{\alpha}$ and $\ket{-\alpha}$, up to exponentially suppressed corrections proportional to $\braket{\alpha}{-\alpha} \sim e^{-2 |\alpha|^2}$.
Moreover, for a stationary detector state it yields the same decoherence as a thermal Gaussian.
A very slightly higher purity loss can be achieved if the cavity and DM are squeezed along orthogonal axes, but the purity loss is only increased by $\sim \eta$, with no enhancement with $N_\eff$.

\paragraph{Decoherence as a Signal.}
%
Our results show that it is extremely difficult in practice to use decoherence to test whether the axion is quantized, though it can be used to distinguish between different classical states.
In addition, for large $n_a$ and $n_c$ the maximum purity loss scales as $\delta \sim \eta\, n_a n_c$, which is parametrically equal to the expected number of quanta exchanged between the axion DM and cavity.
Thus, if one is working in the regime where single-quantum exchange can be detected directly, as is true for the superconducting cavity setups in Refs.~\cite{Agrawal:2023umy,Zheng:2025qgv}, then decoherence does not provide an advantage to the signal rate.
On the other hand, it may provide an advantage in the regime $m_a \ll 10^{-6} \, \mathrm{eV}$, for setups where individual signal quanta cannot be detected, provided that background contributions to decoherence are sufficiently low.
For example, it would be interesting to consider nuclear spin ensembles, which could potentially be prepared in highly exotic states that enhance the decoherence rate~\cite{Arvanitaki:2024taq}.

\subsection{General Calculation}
\label{sec:dec_general}

When both $n_a$ and $n_c$ are large, the perturbative expansion Eq.~\eqref{eq:final_dec} of the purity loss is parametrically an expansion in $\eta n_a n_c$, so it only applies when the expected number of transitions is small.
However, it is also possible to derive exact expressions.

To do this, we recall from Eq.~\eqref{eq:approx_cavity_P} that the final cavity $P$-function $P_c^f$ can be written as a scaled convolution of the effective mode's $P$-function $P_a$ and the initial cavity $P$-function $P_c^0$.
Thus, the final characteristic function is a scaled product, 
\be
\tilde{P}_c^f(\lambda)
= \tilde{P}_a(\sqrt{\eta} \, \lambda)\, \tilde{P}_c^0(\sqrt{1-\eta}\, \lambda).
\ee
Next, Eq.~\eqref{eq:Gamma-characteristic} gives the purity in terms of the characteristic function.
Applying it gives an exact expression for the purity loss, valid for any initial DM and cavity state,\footnote{Since we are now allowing the initial cavity state to be mixed, it is possible for the purity loss to be \textit{negative}.
This always occurs, for instance, if the cavity starts in a thermal Gaussian and the axion is in a coherent state.
However, this does not qualitatively affect any of our conclusions.}
\be \label{eq:final_dec_P}
\delta = \tr_c\!\big((\rho_0^c)^2\big) - \tr_c\!\big((\rho^c)^2\big) 
= \tr_c\!\big((\rho_0^c)^2\big) - \int \frac{d\lambda}{\pi} \, |\tilde{P}_a(\sqrt{\eta} \, \lambda)|^2 \, |\tilde{P}_c^0(\sqrt{1-\eta} \, \lambda)|^2 \, e^{-|\lambda|^2}.
\ee

\paragraph{Simple Examples.}
%
In App.~\ref{app:uniqueness} we show that classical states always have $|\tilde{P}(\lambda)| \leq 1$, and only coherent states saturate this inequality for all $\lambda$.
This implies that coherent DM states give the minimum purity loss among all classical DM states, equal to that due to the vacuum.

As another example, suppose the cavity is prepared in a coherent state (such as the vacuum state) so that $|\tilde{P}_c^0(\lambda)|^2 = \tr_c \!\big((\rho_0^c)^2\big) = 1$.
Then taking the DM to have mean occupancy $n_a$, the exact purity loss is
\be \label{eq:purity_loss_examples}
\delta = \begin{cases} 1 - \dfrac{1}{1+2\eta n_a} & \text{thermal\ Gaussian\ DM}, \tilde{P}_a(\lambda) = e^{- n_a |\lambda|^2} \\
1 - (1-2\eta)^{n_a} P_{n_a}\!\left(\dfrac{1+(1-2\eta)^2}{2(1-2\eta)} \right) & \textrm{Fock\ DM}, \tilde{P}_a(\lambda) = L_{n_a}(|\lambda|^2)
\end{cases}
\ee
where $P_n$ denotes the $n^{\text{th}}$ Legendre polynomial.
The two expressions above are manifestly distinct, but they match at $\mathcal{O}(\eta)$.
This is consistent with the prediction from Sec.~\ref{sec:dec_perturbative} that for an initially pure cavity state and a stationary DM state, the purity loss at $\mathcal{O}(\eta)$ depends only on the mean occupancy.

\paragraph{Connection to Perturbative Results.}
%
We can obtain another perspective on the results of Sec.~\ref{sec:dec_perturbative} by expanding Eq.~\eqref{eq:final_dec_P} perturbatively in $\eta$. 
Defining $\lambda_r \equiv \text{Re}\, \lambda$ and $\lambda_i \equiv \text{Im}\, \lambda$, 
\be
\tilde{P}_a(\sqrt{\eta} \, \lambda) 
= 1 + \sqrt{\eta} \big[\lambda_r d_{10} + \lambda_i d_{01}\big] + \eta \left[\frac12 \lambda_r^2 d_{20} + \lambda_r \lambda_i d_{11} + \frac12 \lambda_i^2 d_{02}\right] + \ldots
\ee
where the coefficients are moments of $P_a(\alpha)$, 
\be
d_{pq} = \partial_{\lambda_r}^p \partial_{\lambda_i}^q \tilde{P}_a(\lambda) \bigg|_{\lambda = 0} = 2^{p+q} (-i)^p i^q \int d\alpha\, \alpha_i^p \alpha_r^q\,P_a(\alpha).
\ee
Since $d_{10}$ and $d_{01}$ are pure imaginary, and characteristic functions always appear in Eq.~\eqref{eq:final_dec_P} via $|\tilde{P}(\lambda)|^2$, there is no $\mathcal{O}(\sqrt{\eta} \sim g)$ contribution to the purity loss.
Working to $\mathcal{O}(\eta)$, we have
\bea \label{eq:pert_exp_delta}
\delta &= \eta \int \frac{d\lambda}{\pi} |\tilde{P}_c^0(\lambda)|^2 e^{-|\lambda|^2} \big[ |\lambda|^2 - 1 \big]\\
& \quad - \eta \int \frac{d\lambda}{\pi} 
|\tilde{P}_c^0(\lambda)|^2 e^{-|\lambda|^2} \big[ 
\lambda_r^2 d_{20} + \lambda_i^2 d_{02} + 2 \lambda_r \lambda_i d_{11} - (\lambda_r d_{10} + \lambda_i d_{01})^2
\big].
\eea
The first line here, independent of $P_a(\alpha)$, is the vacuum contribution to decoherence.
Again, we see that the leading effect of the DM state only enters through low moments of $P_a(\alpha)$.

If the DM is in a stationary state, then Eq.~\eqref{eq:pert_exp_delta} greatly simplifies. The only surviving moments at this order are $d_{20} = d_{02} = - 2 n_a$, leaving
\be
\delta_{\text{st}} = 2 \eta \int \frac{d\lambda}{\pi} |\tilde{P}_c^0(\lambda)|^2 e^{-|\lambda|^2} \Bigg[ \left(n_a+\frac12 \right)|\lambda|^2 - \frac12 \Bigg].
\ee
This expression generalizes Eq.~\eqref{eq:stationary_delta} to a cavity that need not initially be pure. 
We again see that at $\mathcal{O}(\eta)$, the purity loss only depends on the mean DM occupancy $n_a$.

To recover Eq.~\eqref{eq:stationary_delta} for a pure initial cavity state, note that for pure states, 
\be
\int \frac{d\lambda}{\pi}\, |\tilde{P}(\lambda)|^2 \, e^{-|\lambda|^2} = 1
\ee
by Eq.~\eqref{eq:Gamma-characteristic}.
In addition, observe that in general, we have
\bea
\tr_c \!\big(\rho c^\dag \rho c - \rho^2 c^\dag c\big)
&= \int d\alpha \, d\beta\,P(\alpha)P(\beta) \left[\la \alpha \vt c^\dag \vt \beta \ra \la \beta \vt c \vt \alpha \ra - \la \alpha \vt \beta \ra \la \beta \vt c^\dag c \vt \alpha \ra \right] \\
&= \frac{1}{2} \int d\alpha \, d\beta\,P(\alpha)P(\beta)\, |\alpha-\beta|^2\, e^{-|\alpha-\beta|^2} \\
&= \frac{1}{2} \int \frac{d\lambda}{\pi}\, |\tilde{P}(\lambda)|^2 \,e^{-|\lambda|^2} \, (1-|\lambda|^2),
\eea
where in the second line we symmetrized over $\alpha$ and $\beta$.
Then for pure states, we have
\be
\int \frac{d\lambda}{\pi}\, |\tilde{P}(\lambda)|^2 \,e^{-|\lambda|^2} \, |\lambda|^2
= 2 \left[ \la c^\dag c \ra - |\la c \ra|^2 + \frac{1}{2} \right].
\ee
Applying these results recovers Eq.~\eqref{eq:stationary_delta}.

\section{Discussion}
\label{sec:discussion}
Weakly coupled waves have very low conversion efficiency $\eta$, but are detectable due to a compensating high effective mode occupancy $N_\eff$.
(For a cavity haloscope targeting axion DM, Eqs.~\eqref{eq:Neff-DM} and~\eqref{eq:eta_value} give $N_\eff \sim 10^{19}$ and $\eta \sim 10^{-21}$.)
The classical signatures of such waves depend on $\eta N_\eff$, but their intrinsically quantum effects are suppressed by an \textit{additional} power of $\eta$, not compensated by $N_\eff$.
We have shown this holds for a broad range of possible measurement procedures, involving number, quadrature, entanglement, and decoherence.

We have also shown in general that it arises fundamentally from the smallness of $\eta$, \textit{not} the high mode occupancy. 
To overcome this obstacle, one requires integration times scaling as powers of $1/\eta$, or cavity states prepared with quantum resources scaling as $1/\eta$, both of which are far beyond present experiment capabilities.
Our results significantly sharpen the claim of our previous work~\cite{Bao:2025nsd}: \textit{intrinsically quantum effects of weakly coupled waves are strongly suppressed.}

We now place our results in a broader context.
In Sec.~\ref{sec:other_fields}, we first discuss how our conclusions extend to more general detectors, other types of DM, and GWs.
In Sec.~\ref{sec:generate_nonclassical}, we explore how nonclassical states for these waves could be generated in practice, evading the obstruction due to the quantum central limit theorem. 
In Sec.~\ref{sec:detect_nonclassical}, we critically examine claims in the literature that quantum effects of axions or GWs are, in fact, detectable.
Finally, in Sec.~\ref{sec:conclusion} we present our conclusions.

\subsection{Gravitational Waves and Ultralight Fields}
\label{sec:other_fields}

Throughout we have focused on axion DM, because its deep motivations have made it the driver of significant efforts to build experiments operating at and beyond the standard quantum limit.
Even within that space, we further restricted our attention to resonant cavity haloscopes to provide a concrete setting for all calculations.
Here, we argue that our conclusions extend to a very wide variety of signals and detectors.

\paragraph{Continuous Measurements.}
%
Our measurement procedure was to repeatedly prepare and projectively measure the cavity state.
In many cases, however, the cavity is instead continuously weakly measured, e.g.~by measuring the fields propagating out along a waveguide coupled to the cavity.
Heuristically, the information from continuous measurement with coupling rate $\kappa$ is comparable to that from performing discrete projective measurements at rate $\kappa$, so we expect our conclusions to be qualitatively unchanged.
We confirm this in Ref.~\cite{continuous_paper}, which generalizes the description of continuous measurement via input-output theory to compute the full statistical distributions induced by nonclassical axion states.

One important difference is that continuous measurement is typically performed for a long time $t_{\text{int}} \gg \tau_c \sim 1/\Delta \omega_a$.
Thus, we would either have to account for the evolution of the effective mode, or further divide the effective mode into frequency bins of width $\sim\!1/t_{\text{int}}$.
Taking the latter route, for realistic $t_{\text{int}}$, the number of occupied independent DM modes within each bin (in the cavity haloscope regime $m_a \sim \mu\mathrm{eV}$) would still be very large, so the quantum CLT logic in Sec.~\ref{sec:gaussian_clt} should still apply.
If so, then we expect the axion to behave like a classical Gaussian random field, e.g.~as described in Ref.~\cite{Foster:2017hbq,Cheong:2024ose}.
On the other hand, for sufficiently long $t_{\text{int}}$ or small $m_a$, the bins may become so fine that we begin to resolve the scale over which the DM is correlated in momentum space and the CLT breaks down.

\paragraph{Dark Photon Detection.}
%
Kinetically mixed dark photon DM $A^{\prime}_{\mu}$ can be detected in many of the same instruments designed to detect axion DM~\cite{Holdom:1985ag,ADMX:2010ubl,Chaudhuri:2014dla,Caputo:2021eaa,Cervantes:2022epl,BREAD:2023xhc,Beadle:2025dgy}, without requiring a background magnetic field. 
The kinetic mixing $\mathcal{L} \supset \tfrac{\epsilon}{2} F^{\mu \nu} F_{\mu \nu}'$ is equivalent to a mass mixing with interaction Hamiltonian 
\be 
H_{\text{int}} = - \epsilon m_{A'}^2 \int_{V_c} \hspace{-0.1cm} d^3 \bx \, A^{\mu}(\bx)\, A'_{\mu}(\bx).
\ee
From here, we can repeat the argument from Sec.~\ref{sec:effective_intro} regarding DM effective modes: we expand $A'_\mu(\bx)$ in plane-wave modes, obtaining new form factors $C_{\ell, s}(\bk)$ where $s$ denotes the polarization index of the dark photon; unlike the single degree of freedom for an axion, a massive dark photon has three polarization states $a^\dagger_s(\bk)$ for each momentum $\bk$.
The presence of polarizations introduces no conceptual changes to our arguments, so we conclude that nonclassical effects in dark photon DM should be no easier to detect than for an axion.

\paragraph{Other Ultralight DM Signals.}
%
Our conclusions apply immediately to any DM interaction which reduces to the linear-mixing interaction $ig (c^\dag a - c a^\dag)$ of Eq.~\eqref{eq:H-toy}. 
This includes lumped element detection~\cite{Kahn:2016aff}, provided we replace the cavity mode profile $\mathbf{u}_\ell(\mathbf{x})$ with the profile of the LC circuit mode, and heterodyne detection~\cite{Berlin:2019ahk}, provided we replace the static background field $\mathbf{B}_0$ with the oscillating $\mathbf{B}_0(\mathbf{x}, t)$.
For setups with broadband sensitivity, such as Refs.~\cite{Kahn:2016aff,Berlin:2020vrk}, the rotating wave approximation does not apply, but we expect that similar conclusions hold, because the DM ladder operators are still very weakly mixed with those of the detector.

Ultralight DM can also be absorbed into collective excitations such as phonons and magnons. In a translationally-invariant medium (which, e.g.~includes those in Refs.~\cite{Knapen:2017ekk,Mitridate:2020kly} but not Ref.~\cite{Bloch:2024qqo}), each mode of the DM field mixes with one mode of the medium.
In such cases there is no need for effective modes, but the $\eta$ suppression discussed in Sec.~\ref{sec:projective} still applies.

Finally, we could consider DM absorption into electronic excitations~\cite{Hochberg:2016sqx,Mitridate:2021ctr}, or DM interactions with fermion spins~\cite{Graham:2017ivz,Berlin:2023ubt}.
This case is different because for each transition, only two states in the detector are relevant (e.g.~excited and unexcited, or spin up and down), rather than the infinite tower of a bosonic mode.
Still, in the weak coupling case, only the lowest levels of a bosonic mode are relevant, so we expect our conclusions to hold.

\paragraph{Gravitational Wave Detection.}
%
Though we focused on DM in this work, our analysis also directly applies to relativistic signals.
In particular, GWs interact via the coupling $\mathcal{L} \supset \tfrac{1}{2} h_{\mu \nu} T^{\mu \nu}$.
For electromagnetic detection ($T^{\mu\nu} = T^{\mu \nu}_{\scriptscriptstyle \textrm{EM}}$) this interaction takes the linear mixing form $ig (c^\dag a - c a^\dag)$, as was shown in Ref.~\cite{Carney:2023nzz} for photon-graviton conversion in a magnetic field, and Refs.~\cite{Pang:2018eec,Carney:2024dsj} for laser interferometers (after linearizing about the optical carrier).
A similar conclusion applies to the mechanical coupling of a GW to the matter density of a Weber bar~\cite{Tobar:2023ksi} or its magnetic variant~\cite{Domcke:2024mfu}.

The appropriate definition of the effective mode depends on the detector, and the value of $N_\eff$ depends on both the source and detector.
As in the axion case, the efficiency $\eta$ will be extremely small, suppressing the visibility of nonclassical effects, which do not enter with a compensating $N_\eff$.
For example, a graviton propagating through a constant orthogonal magnetic field $B_0$ for a distance $L$ converts to photons with efficiency~\cite{Carney:2023nzz}
\be
\eta = \frac{B_0^2 L^2}{2 M_{\mathrm{pl}}^2} \sim 10^{-33} \, \left( \frac{B_0}{10 \, \mathrm{T}} \right)^2 \left( \frac{L}{10 \, \mathrm{m}} \right)^2\!.
\ee
Here $M_{\mathrm{pl}}$ is the reduced Planck mass and the result holds at leading order in $B_0$.

As another example, consider a Weber bar with length scale $L$, mass density $\rho$, and mechanical quality factor $Q_m$, sensitive to GWs of angular frequency $\omega_0$.
In the presence of GWs with strain $h_0$, the expected number of phonons excited in the bar is $\sim M L^2 \omega_0^3 h_0^2 T^2$~\cite{Tobar:2023ksi}, where $M \sim \rho L^3$ and $T \sim Q_m / \omega_0$ is the ringup time.
Setting this equal to $\eta N_\eff$, the highest value of $\eta$ corresponds to the lowest value of $N_\eff$, which corresponds to a minimally-sized wavepacket with cross-section $1/\omega_0^2$ and length $T$ which is precisely shaped to ring up the Weber bar. 
Since the energy density in gravitational radiation is $\sim \omega_0^2 h_0^2 M_{\mathrm{pl}}^2$, we have $N_\eff \sim \omega_0^2 h_0^2 M_{\mathrm{pl}}^2 (T / \omega_0^2) / \omega_0$, which implies
\be
\eta \sim \frac{\rho \, Q_m v_s^5}{\omega_0^2 M_{\mathrm{pl}}^2} \sim 10^{-32} \, \left( \frac{\rho}{3 \, \mathrm{g}/\mathrm{cm}^3} \right) \left( \frac{Q_m}{10^6} \right) \bigg( \frac{v_s}{10^{-5}} \bigg)^5 \left( \frac{2 \pi \cdot \mathrm{kHz}}{\omega_0} \right)^2
\ee
where $v_s \sim \omega_0 L$ is the sound speed in the material.\footnote{Equivalently, the efficiency $\eta$ is roughly the probability that a phonon in the bar decays into a GW rather than to heat, which we can compute using the quadrupole formula.}
A further challenge for Weber bar detectors is that, as discussed in Sec.~\ref{sec:general_suppression}, detector thermal noise already washes out quantum effects unless the thermal occupancy obeys $n_t < \eta$. For a kHz signal, this corresponds to a stringent requirement $T \lesssim 10^{-9} \, \mathrm{K}$ on the detector temperature.

\paragraph{Parametric Interactions.}
%
Weakly coupled waves can also act through a parametric interaction, rather than a linear mixing.
For example, dilaton DM~\cite{Damour:1994zq,Damour:2010rp,Arvanitaki:2014faa,Stadnik:2015kia} shifts the frequency of a resonator proportionally to its value, corresponding to interactions such as $(a + a^\dagger)\, (c + c^\dagger)^2$.
The interaction of a GW with a photon propagating in vacuum has a similar form~\cite{Guerreiro:2019vbq}.\footnote{This is distinct from the signal in a laser interferometer, discussed above, which involves reading out a phase difference between two beams.}
In addition, quadratically coupled scalar DM~\cite{Hees:2018fpg,Banerjee:2022sqg,Kim:2022ype,Kim:2023pvt,Beadle:2023flm} could yield an interaction of the form $(a + a^\dagger)^2 \, (c + c^\dagger)^2$.

In these cases, even a classical wave can drive the detector to a nonclassical state, by implementing a squeezing interaction.\footnote{It was claimed in Ref.~\cite{Guerreiro:2019vbq} that only quantum GWs can lead to ``revivals'' of squeezing, citing Ref.~\cite{Ma:2018bxa}, which only considered a classical state with Gaussian noise.
Again, to establish that an effect is intrinsically quantum, one must show that it cannot occur under \textit{any} classical ensemble.}
It was shown in Ref.~\cite{Howl:2020isj} that for a known classical wave, interactions of this form preserve Gaussianity (in the more general sense of having a Gaussian characteristic function, discussed below Eq.~\eqref{eq:multivariate_P}, which includes squeezed states). Intrinsically quantum states could yield non-Gaussianity, though this would also be possible for classical ensembles of coherent states.

It would be interesting to identify the relevant nonclassicality measures in this case, and work out their observability. 
However, based on the physical picture discussed in Sec.~\ref{sec:projective}, where a nonclassical DM state can only imprint a rapidly oscillating $P$-function onto the detector state, one could expect they will also be suppressed by an extra power of the low efficiency $\eta$.

\subsection{The Origins of Nonclassical States}
\label{sec:generate_nonclassical}

We demonstrated in Sec.~\ref{sec:gaussian_clt} that the coarse-grained effective mode seen by a realistic detector is often driven to a thermal Gaussian state by the quantum CLT, erasing nonclassical effects.
Here we discuss how potential loopholes to this argument could arise.

\paragraph{Axion States.}
%
The quantum state of axion DM is highly uncertain today.
Fully quantum simulations are very computationally expensive, but Refs.~\cite{Eberhardt:2022rcp,Eberhardt:2023axk} indicate that quantum corrections grow rapidly during halo collapse, though this process competes with decoherence.
Moreover, Refs.~\cite{Allali:2020shm,Allali:2021puy} showed that DM states involving superpositions of the axion phase with the same axion magnitude, such as $\ket{\alpha} + \ket{-\alpha}$, experience negligible gravitational decoherence because the Newtonian gravitational potential is independent of the phase.

Nonetheless, for standard virialized DM, away from the lowest ``fuzzy'' DM masses, the CLT argument is highly robust and indicates the detector cannot see nonclassical effects.
Furthermore, throughout this work we optimistically assumed that each measurement was independent, neglecting the effect of these measurements on the axion state.
(We discuss this further in Ref.~\cite{continuous_paper}.)
But if, e.g.~the effective mode were in a state such as $\ket{\alpha} + \ket{-\alpha}$, phase-sensitive quadrature measurements from a haloscope would decohere it to a mixture of $\ket{\alpha}$ and $\ket{-\alpha}$. In the regime $\eta N_\eff \gtrsim 1$, this would occur in a single measurement, leaving no nonclassical effects for later measurements.

As we mentioned in Sec.~\ref{sec:gaussian_clt}, this pair of issues might be avoided for DM models with significant substructure and self-interactions. For example, axion DM might condense into a large number of axion stars which individually pass through the detector, with each axion star driving itself to a nonclassical state.
They could also be avoided by continuously sourcing highly monochromatic axions at late times from compact objects.

In a light-shining-through-a-wall experiment~\cite{Anselm:1985obz,VanBibber:1987rq,Redondo:2010dp} such as ALPS~\cite{ALPS:2009des,Ehret:2010mh,ALPSII:2025eri}, the axion is sourced by a laser beam propagating through a static magnetic field.
For laser light in a coherent state, the axion is produced in a coherent state of the appropriate effective mode of the source.
It then propagates out of the source cavity, and the signal strength depends on its overlap with the effective mode of the detector cavity. 
Unlike axion DM, there is no mismatch of scales leading to coarse graining, so the CLT does not apply.
That said, a source beam in a coherent state yields an axion in a coherent state, which is not nonclassical.
One could generate a nonclassical axion using a light beam in a nonclassical state, though it would be difficult to realize such a light beam with suitably large amplitude.

Finally, inflation applies squeezing to modes $(a_{\mathbf{k}} \pm i a_{-\mathbf{k}})/\sqrt{2}$, which corresponds to driving the pair of plane wave modes $(a_{\mathbf{k}}, a_{-\mathbf{k}})$ into a two-mode squeezed state~\cite{Grishchuk:1990bj}.
The CLT argument does allow for the effective mode to be in a squeezed state, as long as the squeezing axes of the plane wave modes are sufficiently aligned today.
However, observing nonclassical effects in this case would still be very difficult, because the nonclassical features are associated with the quadrature that is squeezed, not the one that is amplified.\footnote{In the inflation literature, squeezed states are described as effectively classical because the nonclassicality is associated with a small ``decaying mode''~\cite{Albrecht:1992kf,Polarski:1995jg,Kiefer:2008ku}.
This is essentially equivalent to our perspective here.
Nonclassical effects are also further suppressed due to decoherence.}

\paragraph{Gravitational Wave States.}
%
Similar conclusions can be drawn for the quantum state of a GW.
As recently reiterated in Ref.~\cite{Laga:2026vwm}, if the GW is produced by a known classical source, then within linearized gravity we expect it to be in a coherent state, whose amplitude matches that computed in the classical theory.
However, a coherent state is not the only option.
For instance, inflation produces primordial GWs in squeezed states, and a stochastic background of GWs produced by many overlapping, unresolved sources would appear as a thermal Gaussian state.

There are also a variety of potential astrophysical sources of squeezed GWs.
Reference~\cite{Kanno:2025how} estimated that nonlinear gravitational interactions, relevant at the end of a binary merger, generate a small squeezing parameter $r \sim 10^{-4}$.
Recently, Refs.~\cite{Dorlis:2025zzz,Dorlis:2025amf} claimed that squeezed GWs can be sourced from superradiant axion clouds.

\subsection{Prior Claims on Detecting Nonclassical Effects}
\label{sec:detect_nonclassical}

A wide variety of papers have argued that existing or near-future experiments can detect the quantization of the gravitational field (or the axion field) using observations of GWs or axion DM.
If true, such claims would directly contradict our central thesis.
Nevertheless, in all cases we have examined, these claims rely on signatures that can either be realized by a classical field, or are suppressed by the low efficiency $\eta$.
Here, we exemplify this point by discussing a selection of recent examples.

\paragraph{Number Measurements.}
%
Reference~\cite{Tobar:2023ksi} (with follow-up work in Ref.~\cite{Loughlin:2025rih}) claimed that performing number measurements on a Weber bar excited by GWs would show that gravity is quantum.
However, as noted in Refs.~\cite{Carney:2023nzz,Carney:2024dsj}, the same behavior could occur with a quantum detector and a classical gravitational field, in analogy to the semiclassical explanation of the photoelectric effect.
(Recently, Ref.~\cite{Gouin:2026mdz} claimed that in a fully quantum mechanical calculation of the Weber bar excitation, coherent and squeezed GWs can yield ``intermittent conversion'' in the detector, which does not occur classically.
This arises from unphysically projecting the joint final state onto a given GW state, when in reality the final GW mode is unmeasured and must be traced over.)

As we demonstrated in Sec.~\ref{sec:number_meas}, number measurements can be used to distinguish between coherent and thermal Gaussian states.
This same observation was made in Refs.~\cite{Manikandan:2024fmf,Manikandan:2025hlz}, but these works claimed that measuring any deviation from a coherent state would show that gravity is quantum.
However, a thermal Gaussian state is also classical; moreover, as we have discussed, in many contexts one expects to observe this state.

We showed that intrinsically quantum features of the number distribution, such as negative Mandel $Q$, are presently undetectable.
A number of works have claimed otherwise.
For example, Ref.~\cite{Arani:2026jyz} considered coupling GWs to cavity photons, highlighting quantum effects in the limit where $t_m$ was so large that $\eta \sim 1$. 
However, this would require both a cavity with a quality factor many orders of magnitude higher than those of existing cavities (cf. Eq.~\eqref{eq:eta_value}) and a signal that remains coherent over the significantly lengthened $t_m$. 
References~\cite{Kanno:2018cuk,Kanno:2019gqw} proposed detecting negative $Q$ in GWs using interferometry, but appear to have assumed the GW profile could be read out directly, dropping the $\eta$ penalty.
Reference~\cite{Toccacelo:2026hcz} claimed that nonclassical gravity effects were detectable via $g^{(2)}(0) = Q/\langle n \rangle + 1$ because the factors of $\eta$ cancel out, $g^{(2)}_c = g^{(2)}_\GW$.
However, simply rescaling Mandel $Q$ does not improve its detectability; our discussion involving the KL divergence shows that one needs at least $\sim 1/\eta^2$ measurements to detect nonclassicality in number statistics.

The general arguments in Sec.~\ref{sec:general_suppression} show that one cannot evade this problem by considering correlations of number measurements across detectors.
Nonetheless, Refs.~\cite{Manikandan:2025lfx,Athulya:2026wpl} proposed coupling two identical detectors to the same GW and measuring $\kappa_2 \equiv \langle n_1 n_2 \rangle - \langle n_1 \rangle \langle n_2 \rangle = \eta^2 Q_\GW N_\GW$.
It was heuristically argued in Ref.~\cite{Athulya:2026wpl} that observing quantum effects becomes \textit{easier} in the limit $\eta \to 0$ and $N_\GW \to \infty$ with $\eta N_\GW$ fixed, because the contribution of thermal noise to the mean value of $\kappa_2$ becomes suppressed.
However, only computing a mean value is insufficient to determine observability.
Moreover, our general argument shows that observing nonclassical effects is infeasible even in the complete absence of thermal noise, as they are suppressed by $\eta$ without a compensating factor of $N_\GW$.

\paragraph{Quadrature Variance.}
%
We saw in Sec.~\ref{sec:quad_meas} that quadrature fluctuations of DM or GWs can be imprinted into quadrature fluctuations in the detector.
For many states, the quadrature variance can be much larger than that of a coherent state, which is equal to that of vacuum.

In an influential proposal, Ref.~\cite{Parikh:2020kfh} (see also Refs.~\cite{Parikh:2020nrd,Parikh:2020fhy,Kanno:2020usf,Guerreiro:2021qgk,Manikandan:2025qgv,Dorlis:2026gth}) showed that squeezed states and thermal Gaussian states can impart large, potentially detectable quadrature fluctuations in a GW detector.
Follow-up work has considered more general states~\cite{Cho:2021gvg} and pairs of detectors~\cite{Parikh:2023zat}, and applied the concept to LIGO~\cite{Hertzberg:2021rbl} and a BEC detector~\cite{Sen:2024nhb}.

Many of these works argued that additional quadrature noise would be a signal of quantum gravity.
However, the potentially observable effects in these works involved quadrature variance greater than the vacuum value, which is not intrinsically quantum, because it can also be realized by classical ensembles.
On the other hand, variance \textit{smaller} than the vacuum value is intrinsically quantum but impractical to detect, as was emphasized in Ref.~\cite{Carney:2024dsj}.
Finally, quadrature noise due to vacuum fluctuations is classical under our definition, but as we discuss below, one could still argue that it is intrinsically quantum.
However, in perturbative quantum gravity it corresponds to an unobservably small $\Delta L \sim \ell_{\mathrm{pl}}$~\cite{Carney:2024wnp}.

\paragraph{Decoherence and Entanglement.}
%
Decoherence is often described as an intrinsically quantum effect, associated with entanglement between the detector and its environment; however, it can also result from noise due to a classical environment. 
Accordingly, in Sec.~\ref{sec:decoherence} we computed cavity decoherence rates for a general effective mode state, and showed that intrinsically quantum states do not yield significantly higher or lower decoherence rates.

Counter to this, several works have claimed that one could establish the quantization of gravity by decoherence measurements.
For example, Ref.~\cite{Schutzhold:2025vti} considered interferometry with light in a cat-like state, and GWs in a Fock or coherent state, and Ref.~\cite{Sen:2024rot} studied decoherence in a BEC due to GWs in a squeezed state.
The common issue among these proposals is that they only consider a handful of states, rather than the full range of states.
However, intrinsically quantum states only extend the range of decoherence rates by an amount suppressed by $\eta$.
Simply measuring a high decoherence rate cannot show that gravity is quantum; instead one has to detect very small changes in the decoherence rate.
Finally, Ref.~\cite{Nandi:2026sww} considered decoherence structure in mechanical resonators, from GWs in either a phase-averaged coherent state or a thermal Gaussian state.
These states are indeed different, but they identified the thermal Gaussian state as quantum, when it is actually classical.

For detectors coupled locally to a mediator field, only a quantum mediator field can generate entanglement between the detectors~\cite{Marshman:2019sne}.
However, in Sec.~\ref{sec:general_suppression} we showed that signals of entanglement are also suppressed by $\eta$.
In particular, the entanglement of primordial GWs in two-mode squeezed states was explored in Refs.~\cite{Maity:2021zng,Kanno:2021vwu,Ikeda:2025uae}. 
However, these works did not consider a realistic measurement protocol, and thus missed the $\eta$ suppression.
Other works considered quantities which are not suppressed by $\eta$, but do not indicate entanglement.
For example, Ref.~\cite{Lentz:2025lkg} considered measuring $\langle X_1 X_2 \rangle - \langle X_1 \rangle \langle X_2 \rangle$ in a pair of cavity haloscopes, and Ref.~\cite{Mavromatos:2026nin} stated that only quantum GW states can have $\langle a_i a_j \rangle \neq 0$, but both quantities can be generated by ordinary classical correlation.

It may be possible for entanglement prepared within the detector to serve as a quantum resource that enhances the visibility of nonclassical effects, analogous to our squeezing example in Sec.~\ref{sec:general_suppression}. 
This is an interesting question which, to our knowledge, has not been studied.

\paragraph{Other Definitions of Nonclassicality.}
%
In Sec.~\ref{sec:toy} we showed that the effects of quantum states with nonnegative $P(\alpha)$ are identical, at leading order in $\eta$, to a classical ensemble with probability distribution $P(\alpha)$.
We have thus identified nonclassicality with negative $P(\alpha)$, which is the standard choice in the quantum optics community.
However, a variety of alternative definitions are often invoked, and we comment on these here.

First, stimulated emission is often called a signature of quantum mechanics.
However, by comparing Eqs.~\eqref{eq:approx_cavity_P} and~\eqref{eq:classical_cavity_P}, one sees the exact same enhancement is also present in a semiclassical model, with a quantum detector and classical field.
The quantum-mechanical Bose enhancement is simply the equivalent of the classical fact that the effects of a field scale with its magnitude.
One could rule out exotic theories by testing if stimulated emission occurs, but this would not distinguish between the standard classical and quantum theories.

Second, spontaneous emission of a particle can be ascribed to intrinsically quantum vacuum fluctuations of the corresponding field.
This is debatable, as one can heuristically derive similar phenomenology in the semiclassical theory by adding $\hbar \omega / 2$ of classical noise energy to each field mode~\cite{milonni1976semiclassical}.
Regardless, spontaneous emission obeys our general power counting because it is suppressed by $\eta$ but not enhanced by $N_\eff$, and is thus impractical to observe for axions or gravitons.
The same holds for other signatures of vacuum fluctuations.

Third, classical effects are often identified as those which survive in the limit $\hbar \to 0$. 
Using this criterion, Refs.~\cite{Britto:2021pud,Cristofoli:2021jas} found that the amplitude for two masses to scatter while emitting one graviton was classical, while the amplitude to emit two gravitons was quantum.
It is not clear if this is equivalent to our definition of nonclassicality.
However, exponentiating the one-graviton amplitude yields a coherent state, as expected from linearized gravity, which is indeed classical.
Moreover, higher-order gravitational interactions can generate squeezed states, which are nonclassical, and this behavior might be captured in the two-graviton amplitude.
It would be interesting to investigate the link between these perspectives further.

\subsection{Conclusion}
\label{sec:conclusion}

We have shown that the intrinsically quantum effects of GWs and axion DM are impractical to detect.
We emphasize that there is no obstruction to distinguishing between different \textit{classical} states.
In particular, many of the observables we consider can be used to distinguish between classical models of axion DM, and would yield additional information in a post-discovery scenario.
Furthermore, it remains interesting to search for exotic GWs produced in nonclassical states.
In this case experiments can focus on the most effective classical observables to detect these waves, rather than constructing more complicated observables to attempt to establish the quantization of gravity, which we argue is not presently feasible.

In the context of gravity, our results shed new light on the classic question of whether gravitons are detectable~\cite{Dyson:2013hbl,Rothman:2006fp}.
They lend further motivation to ongoing efforts to test if gravity is quantum by seeing if gravity can mediate entanglement~\cite{Marletto:2017kzi,Bose:2017nin,Carney:2018ofe}.\footnote{These experiments are of course challenging, since their signal is also suppressed by the weak gravitational coupling.
Our point is that one cannot circumvent this difficulty using GW detectors, because the nonclassical signal in a GW detector is not enhanced by $N_\eff$.}
In the context of axions, we have justified the conventional wisdom that the axion appears classical in direct detection experiments, and found that classical behavior emerges not because of high occupancy, but rather due to coarse graining and weak coupling.

\begin{acknowledgments}
We thank Abhishek Banerjee, Itay Bloch, Zachary Bogorad, Daniel Carney, Andrew Eberhardt, Sebastian Ellis, James Gardner, Anson Hook, Liang Jiang, Giacomo Marocco, David Marsh, Gilad Perez, Ryan Plestid, and Ritoban Basu Thakur for discussions.
We acknowledge the use of GPT‑5.6 Sol for proofreading the final draft.
YB and LTW are supported by the Department of Energy grant DE-SC0009924.
DYC is supported by the Enrico Fermi and KICP fellowship from the Enrico Fermi Institute and the Kavli Institute for Cosmological Physics at the University of Chicago and the Kavli Foundation.
The research of NLR, JT, and KZ was supported by the Office of High Energy Physics of the U.S. Department of Energy under contract DE-AC02-05CH11231.
The work of JT was further supported by the NSF Graduate Research Fellowship Program under Grant DGE2146752.
\end{acknowledgments}

\appendix
\addtocontents{toc}{\protect\setcounter{tocdepth}{1}}
\section{Exact Solution of a Multimode Toy Model}
\label{app:exact_solution}

If we define $\vb{a} = (c, a)^T$, the evolution in Eq.~\eqref{eq:acHP} takes the form $\vb{a}(t) = A(t)\, \vb{a}(0)$ for some matrix $A(t)$.
This is possible because the Hamiltonian is quadratic in the ladder operators and number-conserving in the rotating wave approximation, so that $c$ and $a$ can only evolve into each other.
Since the time evolution preserves the commutation relations of the ladder operators (e.g.~$[a(t), a^\dagger(t)] = 1$), $A(t)$ is a unitary matrix.

The preceding fact can be used to derive a compact form for the time evolution, even when the cavity couples to multiple axion modes.
For the two-mode case, any number-conserving Hamiltonian quadratic in the ladder operators can be written as
\be
H = 2\bar{\omega}K + 2\vb{g} \cdot \vb{J},
\ee
in terms of the four operators
\be \label{eq:J-13-K-twomode}
J_1 = \frac{1}{2}(c^\dagger a + c a^\dagger),\hspace{0.25cm}
J_2 = -\frac{i}{2}(c^\dagger a - c a^\dagger),\hspace{0.25cm}
J_3 = \frac{1}{2}(c^\dagger c - a^\dagger a),\hspace{0.25cm}
K = \frac{1}{2}(c^\dagger c + a^\dagger a).
\ee
These operators satisfy the commutation relations of $\mathfrak{u}(2) = \mathfrak{su}(2) \oplus \mathfrak{u}(1)$, 
\be
[J_i, J_j] = i \varepsilon_{ijk}J_k, \hspace{0.5cm} [K, J_i]=0,
\ee
so that the Hamiltonian is an element of $\mathfrak{u}(2)$.
Under commutation with these operators, $\vb{a}$ transforms in the fundamental representation of $\mathfrak{u}(2)$: $[K,\vb{a}] = - \vb{a}/2$ and $[J_i,\vb{a}]=-\sigma_i\, \vb{a}/2$, with $\sigma_i$ the Pauli matrices.
Accordingly, the Heisenberg equation of motion simplifies to
\be
\dot{\vb{a}} = 2i\left(\bar\omega [K, \vb{a}] + \sum_i g_i [J_i, \vb{a}] \right)
= -i\big( \bar{\omega}\, \mathds{1} + \vb{g}\cdot \boldsymbol{\sigma} \big) \vb{a}.
\ee
The general solution is then simply a $U(2)$ transformation, where
\be
\vb{a}(t) = \exp(-i ( \bar\omega \mathds{1} + \vb{g}\cdot\boldsymbol{\sigma}) t)\, \vb{a}(0)
\ee
for time-independent couplings.

Though this language was unnecessary for two modes, it generalizes directly to $\mathcal{N}$ total modes $\vb{a} = (a_1, \ldots, a_{\mathcal{N}})^T$ coupled through quadratic, number-conserving interactions.
For multiple modes, the general Hamiltonian can be written as $H = 2\bar\omega K + 2 \sum_i g_i J_i$ up to an irrelevant additive constant, where the generators
\bea
&J^1_{mn} = \frac{1}{2}(a_m^\dagger a_n + a_ma^\dagger_n),\quad
J^2_{mn} = -\frac{i}{2}(a_m^\dagger a_n - a_m a_n^\dagger), \\
&J^3_{n} = \frac{1}{2}(a_1^\dagger a_1 - a^\dagger_n a_n),\quad\quad\quad\,
K = \frac{1}{2}\sum_n a^\dagger_n a_n,
\eea
satisfy the $\mathfrak{u}(\mathcal{N})$ commutation relations.
Here, the $J^{1,2}_{mn}$ are defined for $1 \leq m < n \leq \mathcal{N}$, and $J^3_n$ is defined for $1 < n \leq \mathcal{N}$.
By similar logic, the Heisenberg equation of motion is 
\be
\dot{\vb{a}} = -i \Bigl( \bar\omega \mathds{1} + \sum_i g_i \lambda_{i} \Bigr) \vb{a},
\ee
where the $\lambda_i$ are the generators of $\mathfrak{su}(\mathcal{N})$ in the fundamental representation, and 
\be
\vb{a}(t) = \exp(-i\left( \bar\omega \mathds{1} + \sum_i g_i \lambda_i \right) t)\,\vb{a}(0).
\ee

In fact, one can generalize even further, dropping the rotating wave approximation.
In this case, creation and annihilation operators can mix, but an exact solution continues to exist~\cite{COLPA1978327}, where the set of Hamiltonians corresponds to the algebra $\mathfrak{sp}(2\mathcal{N}, \mathbb{R})$.

\section{TM Modes of a Cylindrical Cavity}
\label{app:cavity_modes}

Here we review the explicit construction of cavity modes in a cylindrical cavity of radius $R$, centered on the $z$-axis, and bounded by $z = 0$ and $z = L$.
We use Coulomb-temporal gauge, $A^0 = 0$ and $\nabla \cdot \mathbf{A} = 0$, and focus on TM modes, as these are relevant for cavity haloscopes with a longitudinal background magnetic field. 
The general solution for the TM modes is 
\be
\vb{A}(\bx) = \nabla \times \qty( \nabla \times F(\bx, t)\, \hat{\vb{z}} )
\ee
for a nonsingular function $F$ which vanishes at $\rho = R$, and obeys $\del_z F = 0$ at $z = 0$ and $z = L$.
The function $F$ can be expanded in modes as 
\be
F(\bx, t) = \sum_{\ell} q_\ell(t) N_{\ell}\, \xi_{\ell}(\bx),
\ee
where $N_\ell$ is a mode normalization constant, to be fixed below.
For a cylindrical cavity the modes are indexed by $\ell = mnp$ and $\xi_\ell$ can be written using separation of variables as 
\be
\xi_\ell(\bx) = R_{mn}(\rho)\, \Phi_m(\phi)\, Z_p(z), 
\ee
where the basis functions are 
\be
\Phi_{m}(\phi) = \frac{e^{i m \phi}}{\sqrt{2\pi}}, \hspace{0.5cm} 
R_{mn}(\rho) = \frac{\sqrt{2} \, J_{\abs{m}}(\omega_{\rho, mn} \rho)}{R J_{\abs{m}+1}(j_{|m|n})}, \hspace{0.5cm} 
Z_{p}(z) = \sqrt{\frac{2^{1 - \delta_{p0}}}{L}} \cos(\omega_{z, p} z).
\ee
Above, $m$ is an integer, $n$ is a positive integer, $p$ is a nonnegative integer, and $j_{|m|n}$ is the $n^{\text{th}}$ zero of $J_{\abs{m}}$, the Bessel function of order $\abs{m}$.
The prefactors are set so that the basis functions are complete and orthonormal over the cavity.

Decomposing the Laplacian as $\nabla^2 = \nabla_\parallel^2 + \nabla_\perp^2$, with $\nabla_\parallel^2 = \del_z^2$, these basis functions satisfy the eigenvalue equations
\bea
\nabla_\parallel^2 Z_p(z) &= -\omega_{z, p}^2 Z_p(z), \\
\nabla_\perp^2 \qty[R_{mn}(\rho) \Phi_m(\phi)] &= -\omega_{\rho, mn}^2 \qty[ R_{mn}(\rho)\Phi_m(\phi) ],
\eea
where $\omega_{\rho, mn} = j_{|m|n}/R$ is the transverse wavenumber and $\omega_{z, p} = p \pi/ L$ is the longitudinal wavenumber.
This implies that the $\text{TM}_{mnp}$ mode obeys $\nabla^2 \xi_\ell = - \omega_\ell^2 \xi_\ell$, with angular frequency
\be
\omega_{\ell} = \sqrt{\omega_{z, p}^2 + \omega_{\rho, mn}^2}. 
\ee

To quantize the theory in the Coulomb-temporal gauge, we write the mode expansion of $\vb{A}$ and its conjugate momentum $\bm{\Pi} = \dot{\vb{A}} = -\bE$ as 
\be
\vb{A}(\bx) = \sum_{\ell} \bu_{\ell}(\bx) c^\dagger_{\ell} + \hc,\hspace{0.5cm}
\bm{\Pi}(\bx) = \sum_{\ell} i \omega_{\ell} \bu_\ell(\bx) c_\ell^\dagger + \hc
\ee
where the mode functions which give the vector potential's profile are
\be
\bu_{\ell}(\bx) = \nabla \times \qty( \nabla \times N_{\ell} \xi_{\ell}(\bx) \hat{\vb{z}} ).
\ee
These fields must obey the canonical commutation relation
\be \label{appeqn:CavityCCA}
\comm{A_i(\bx)}{\Pi_j(\vb{y})} = i \qty( \delta_{ij} - \frac{\del_i \del_j}{\nabla^2}) \delta^{(3)}(\bx - \vb{y}), 
\ee
where the transverse projector $\delta_{ij} - \del_i \del_j / \nabla^2$ is used because the gauge condition enforces a transversality condition on $\vb{A}$, and the derivatives all act on $\bx$.

The mode operators $c_\ell$ must also satisfy the standard commutation relations
\be
\comm{c_{\ell}}{c_{\ell'}} = \comm{c^\dagger_{\ell}}{c^\dagger_{\ell'}} = 0, \hspace{0.5cm}
\comm{c_\ell}{c^\dagger_{\ell'}} = \delta_{\ell' \ell} = \delta_{m'm} \delta_{n'n} \delta_{p'p}. 
\ee
Because only TM modes contribute to $A_z$ and $\Pi_z$, we can determine their mode normalization $N_\ell$ by evaluating the component
\bea
\,[A_z(\bx), \Pi_z(\vb{y})] &= 2 i \sum_\ell |N_\ell|^2 \omega_\ell \, (\nabla_\perp^2 \xi_\ell(\bx)) (\nabla_\perp^2 \xi_\ell^*(\vb{y}) ) \\
&= 2 i \, \frac{\nabla_\perp^2}{\nabla^2} \sum_\ell |N_\ell|^2 \omega_\ell^3 \omega_{\rho, mn}^2 \, \xi_\ell(\bx) \xi_\ell^*(\vb{y}) 
\eea
where the derivatives outside act on $\bx$, and we applied the eigenvalue equations.
This is compatible with Eq.~\eqref{appeqn:CavityCCA} if we take 
\be \label{eq:N_ell_def}
N_{\ell} = \frac{1}{\sqrt{2} \, \omega_{\ell}^{3/2} \omega_{\rho, mn}},
\ee
so that the sum collapses to $\delta^{(3)}(\bx - \vb{y})$ by the completeness of the $\xi_\ell$.

The normalization in Eq.~\eqref{eq:N_ell_def} implies the mode functions $\bu_\ell$ obey the orthogonality relation Eq.~\eqref{eq:orthog_relation}.
If the external magnetic field points along the $\hat{\vb{z}}$ direction, it is useful to define $u_{\ell}(\bx) = \bu_\ell(\bx) \cdot \hat{\vb{z}}$, which obeys the relation
\be \label{eq:ul_norm}
u_{\ell}(\bx) = - N_{\ell} \nabla_\perp^2 \xi_\ell(\bx) = \frac{\omega_{\rho, mn}}{\sqrt{2} \, \omega_{\ell}^{3/2}} \, \xi_{\ell}(\bx).
\ee
Since the functions $\xi_\ell$ are orthogonal over the cavity, the $u_\ell$ are as well.

\section{Quasiprobability Distributions}
\label{app:quasiprobability}

The Glauber--Sudarshan $P$-function, which is a quasiprobability distribution representation of the density matrix, is the primary tool used in this work to identify genuinely nonclassical phenomena.
In this appendix we collect a number of useful properties of the $P$-function and related distributions used in the main text.

\subsection{General Quasiprobability Distributions and Characteristic Functions}
\label{app:general_quasi_characteristic}

The characteristic function introduced in Eq.~\eqref{eq:Glauber_P_characteristic_function} can be generalized to~\cite{PhysRev.177.1857}
\be \label{appeq:s_ordered_characteristic_function}
\tilde{P}^{(s)} (\lambda) = e^{s|\lambda|^2/2} \tr\!\Big[ \rho \, e^{\lambda a^{\dagger}-\lambda^* a} \Big] 
= \begin{cases} 
\tr\!\big[ e^{- \lambda^* a} \, \rho \, e^{\lambda a^\dagger} \big] & s = 1 \\ 
\tr\!\big[ \rho \, e^{\lambda a^{\dagger}-\lambda^* a} \big] & s = 0 \\ 
\tr\!\big[ e^{\lambda a^\dagger} \rho \, e^{- \lambda^* a} \big] & s = -1  \end{cases}
\ee
for $s \in [-1, 1]$.
They are related to quasiprobability distributions by a Fourier transform,
\be \label{appeq:s_ordered_quasi}
P^{(s)}(\alpha) = \int \frac{d{\lambda}}{\pi^2} \, \tilde{P}^{(s)} (\lambda) \, e^{-\lambda \alpha^{*} + \lambda^{*} \alpha} 
= \begin{cases} \text{Glauber--Sudarshan} \ P(\alpha) & s = 1 \\ 
\mathrm{Wigner} \ W(\alpha) & s = 0 \\ 
\mathrm{Husimi} \ Q(\alpha) & s = -1 
\end{cases}
\ee
As discussed below Eq.~\eqref{eq:O_expt}, expectation values for normally ordered operators can be computed by integrating against $P(\alpha)$ ($s = 1$), while the $s = 0$ and $s = -1$ cases fulfill the same role for symmetric and antinormal ordering.
(Marginalizing the Wigner function over the real or imaginary part of $\alpha$ yields, up to a factor of $\sqrt{2}$, the probability distribution of the other quadrature.)
Table~\ref{apptab:example_characteristic_functions} provides several examples of these functions; for a full introduction, see Ref.~\cite{barnett2002methods}.
Next we derive several results needed in the main text.

\begin{table}
\centering
\renewcommand{\arraystretch}{1.1}
\resizebox{\linewidth}{!}{%
\begin{tabular}{c|c|c|c|c}
\hline
\textbf{State} & $\tilde{P}^{(s)} (\lambda)$ & $P(\alpha) $  & $W(\alpha) $  & $Q(\alpha) $ \\
\hline& & & & \\[-9.5pt]
Coherent $\ket{\alpha_0}$ & $\displaystyle e^{(s-1)|\lambda|^2 / 2} e^{\lambda \alpha_{0}^{*} - \lambda^{*}\alpha_{0}}$ & $\displaystyle \delta ( \alpha - \alpha_{0}) $    & $\displaystyle \frac{2}{\pi} e^{-2 | \alpha - \alpha_{0}|^2}$  & $\displaystyle \frac{1}{\pi} e^{-|\alpha - \alpha_{0}|^2}$ \\[10pt]
Gaussian, $\langle n \rangle = N$ & $\displaystyle e^{-[N+(1-s) / 2]|\lambda|^2}$ & $\displaystyle \frac{1}{\pi N} e^{-|\alpha|^2 / N}$  & $\displaystyle \frac{1}{\pi ( N + 1/2)} e^{-|\alpha|^2 / (N+1/2)}$ & $\displaystyle \frac{1}{\pi(N+1)}e^{-|\alpha|^2 / (N+1)}$  \\[10pt]
Fock $\ket{n}$ & $\displaystyle e^{(s-1)|\lambda|^2 / 2} L_n\big(|\lambda|^2\big)$ & $\displaystyle L_n(-\del_\alpha \del_{\alpha^*}) \delta(\alpha) $ & $\displaystyle \frac{2}{\pi}(-1)^n L_n \big(4|\alpha|^2\big) e^{-2|\alpha|^2} $  & $\displaystyle \frac{1}{\pi} \frac{|\alpha|^{2 n}}{n!} e^{-|\alpha|^2} $\\[10pt]
\hline
\end{tabular}}
\caption{Examples of $\tilde{P}^{(s)} (\lambda)$ and their corresponding quasiprobability distributions.
Here, $L_n$ is the $n^{\text{th}}$ Laguerre polynomial.}
\label{apptab:example_characteristic_functions}
\end{table}

First, Eq.~\eqref{appeq:s_ordered_characteristic_function} encodes that the different characteristic functions are related by multiplication by a Gaussian.
A corollary is that quasiprobability distributions are related by convolution with a Gaussian.
Explicitly, we have 
\be \label{eq:gaussian_convs}
W(\alpha) = \frac{2}{\pi} \int d{\beta}  \, P(\beta) \, e^{-2|\alpha-\beta|^2 }, \hspace{0.5cm}
Q(\alpha) = \frac{1}{\pi} \int d{\beta}  \, P(\beta) \, e^{-|\alpha-\beta|^2 }.
\ee
As the Fock state example in Tab.~\ref{apptab:example_characteristic_functions} highlights, the $P$-function can be highly singular, but this Gaussian smearing generally removes such singularities from $W$ and $Q$.
However, while negativity of $P$ captures \textit{all} nonclassical effects, the same does not hold for $W$ and $Q$.
For instance, $W(\alpha)$ is nonnegative for squeezed states, which are nonclassical.

\begin{figure}[t!]
\centering
\includegraphics[width=0.75\linewidth]{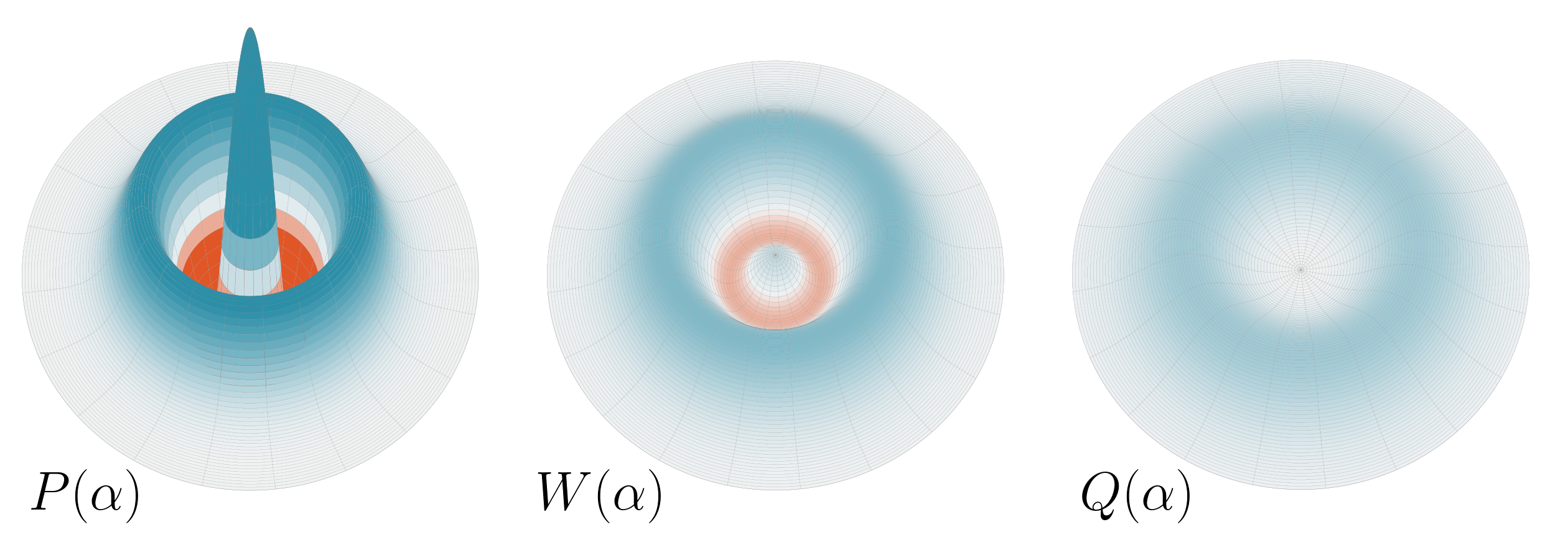}
\vspace{-0.3cm}
\caption{Quasiprobability distributions for the state obtained by adding two photons to a thermal Gaussian state with mean occupancy $N = 1$~\cite{PhysRevA.46.485}.
Negative regions are shown in red.
Nonclassicality always corresponds to negative $P$, though squeezed states achieve negative $P$ but have nonnegative $W$.}
\label{appfig:PQW_2photon_added_thermal}
\end{figure}

In fact, $Q(\alpha)$ is never negative.
To see this, observe that from Eq.~\eqref{appeq:s_ordered_characteristic_function} we have
\be
\tilde{P}^{(-1)}(\lambda) = \int \frac{d \alpha}{\pi} \, \bra{\alpha} e^{\lambda a^\dagger} \rho e^{- \lambda^* a} \ket{\alpha} = \int \frac{d\alpha}{\pi} \, e^{\lambda \alpha^* - \lambda^* \alpha} \bra{\alpha} \rho \ket{\alpha}
\ee
from which we can read off 
\be
Q(\alpha) = \frac{1}{\pi} \langle \alpha| \rho|\alpha \rangle \geq 0.
\ee
Thus, the $Q$-function can be regarded as a genuine probability distribution.
Furthermore, we see that convolving any $P$-function with a Gaussian of order-one width erases its negativity, since this yields the $Q$-function; heuristically, this means negative regions in $P$ cannot be larger than order-one in size.
These remarks are illustrated by Fig.~\ref{appfig:PQW_2photon_added_thermal}.

\subsection{Singular and Nonsingular $P$-Functions}
\label{app:singular_P}

The $P$-functions of simple nonclassical states are often highly singular distributions~\cite{PhysRevA.94.013814}.
Here we explain the origin of this behavior, and why it does not affect our conclusions.

As an example, we consider a Fock state $\ket{n}$.
From Eq.~\eqref{appeq:s_ordered_characteristic_function}, the normal-ordered characteristic function $\tilde{P}^{(1)}(\lambda)$ is simply an $n^{\text{th}}$ order polynomial in $|\lambda|^2$.
Then the Fourier transform that defines $P(\alpha)$ in Eq.~\eqref{appeq:s_ordered_quasi} is not convergent, and it is sometimes said that a Fock state does not have a $P$-function at all.
More commonly, one can formally construct one by identifying the Fourier transform of $1$ with $\delta(\alpha)$, and each power of $|\lambda|^2$ with $-\del_\alpha \del_{\alpha^*}$.
This yields a highly singular $P$-function involving derivatives of a $\delta$-function, and the same behavior arises for, e.g.~squeezed states or superpositions of coherent states.
This makes it impossible to plot the $P$-function for these simple idealized states.

In some cases, these singularities can be removed by adding a small amount of noise, which would always exist in practice.
For instance, an ideal Fock state $\ket{n}$ can be prepared by adding $n$ quanta to the vacuum state; however, if one instead starts from a thermal Gaussian state of arbitrarily small mean $N$, the characteristic function will decay rapidly at large $|\lambda|$.
This renders the Fourier transform convergent, so that the $P$-function is the ordinary function~\cite{PhysRevA.46.485}
\be
P(\alpha) = \frac{1}{\pi N (-N)^n} \, L_n\!\left( \left( 1 + \frac{1}{N} \right) |\alpha|^2 \right) e^{-|\alpha|^2/N} 
\ee
which is smooth for any nonzero $N$, and achieves negative values for any $n > 0$.

Even if we allow for states with singular $P$-functions, the arguments in the main text still hold.
For instance, the quantum central limit theorem is proven in Sec.~\ref{sec:gaussian_clt} by directly considering the characteristic function, not the $P$-function, and the general arguments in Sec.~\ref{sec:general_suppression} only rely on the general identities in Eq.~\eqref{eq:gaussian_convs}.
In some cases one can even work directly with singular $P$-functions, as shown in the following example.

\paragraph{Central Limit Theorem with Fock States.}
%
Consider $\mathcal{N}$ modes $b_i$ all in an $n = 1$ Fock state.
Then the $P$-function corresponding to the mode $a_{\text{eff}} = (b_1 + \ldots + b_{\mathcal{N}}) / \sqrt{\mathcal{N}}$ is
\bea
P_{\text{eff}}(\alpha) &= \int \prod_{i=1}^\mathcal{N} \left( d \beta_i \left( 1 + \del_{\beta_i} \del_{\beta_i^*} \right) \delta(\beta_i) \right) \, \delta\!\left( \alpha - \sum_{i=1}^{\mathcal{N}} \frac{\beta_i}{\sqrt{\mathcal{N}}} \right) \\
&= \int \prod_{i=1}^\mathcal{N} \left( d \beta_i \, \delta(\beta_i) \left( 1 + \frac{1}{\mathcal{N}} \, \del_\alpha \del_{\alpha^*} \right) \right) \, \delta\!\left( \alpha - \sum_{i=1}^\mathcal{N} \frac{\beta_i}{\sqrt{\mathcal{N}}} \right) \\
&= \left( 1 + \frac{1}{\mathcal{N}} \, \del_\alpha \del_{\alpha^*} \right)^\mathcal{N} \delta(\alpha) 
\eea
where we integrated by parts and then performed the integrals.
In the limit $\mathcal{N} \to \infty$ this approaches $P_{\text{eff}}(\alpha) = \exp(\del_\alpha \del_{\alpha^*}) \, \delta(\alpha)$, which appears to be a highly singular distribution, while the central limit theorem would predict the thermal Gaussian $P_{\text{eff}}(\alpha) = e^{-|\alpha|^2} / \pi$.
However, one can straightforwardly check that these expressions are equivalent, e.g.~they give the same result when integrated against smooth test functions.
They both correspond to a Gaussian characteristic function $e^{-|\lambda|^2}$, with the former generated by mapping $|\lambda|^2$ to $-\del_\alpha \del_{\alpha^*}$.

\subsection{Criterion for Stationarity}
\label{app:stationarity}

The DM field is typically assumed to be in a stationary state, i.e.~to have a time-independent density matrix.
Here we clarify what this implies for the DM $P$-function.

First, suppose only a single DM mode is occupied, with angular frequency $\omega$.
To compute the time evolution of the $P$-function, we use the Mehta formula~\cite{mehta1967diagonal}, 
\be \label{eq:Mehta_formula}
P(\alpha) = \int \frac{d \beta}{\pi^2} \bra{-\beta} \rho \ket{\beta}\, e^{|\alpha|^2+|\beta|^2+\alpha \beta^*-\alpha^* \beta}.
\ee
The time-evolved density matrix is $\rho(t) = U(t) \rho U^\dagger(t)$, and we know that $U^\dagger(t) \ket{\beta} = \ket{e^{i \omega t} \beta}$.
Changing variables to $\beta' = e^{i \omega t} \beta$, the time-evolved $P$-function is
\be
P(\alpha, t) = \int \frac{d\beta'}{\pi^2} \bra{-\beta'} \rho \ket{\beta'}\, e^{|\alpha|^2+|\beta'|^2+\alpha e^{i \omega t} \beta'^*- (\alpha e^{i \omega t})^* \beta'} = P(\alpha e^{i \omega t}, 0).
\ee
If $\rho$ is stationary, then we must have $P(\alpha, t) = P(\alpha)$, and this can be satisfied if and only if $P(\alpha) = P(|\alpha|)$. 
This is the standard, textbook stationarity criterion~\cite{mandel1995optical}.
It implies, for instance, that all phase-dependent expectation values vanish, e.g.~$\langle (a^\dagger)^m a^n \rangle = 0$ for $m \neq n$.

In reality the DM field has many modes, and in the multi-mode case the criterion can be more subtle.
It is still true that if $P$ depends on $|\alpha_{\bk}|$ for each mode, then the system is stationary, but the converse is technically false.
For example, consider two modes with angular frequencies $\omega_1$ and $\omega_2$.
An argument analogous to the above yields the stationarity criterion 
\be
P(\alpha_1 e^{i \omega_1 t}, \alpha_2 e^{i \omega_2 t}) = P(\alpha_1, \alpha_2)
\ee
for the two-mode joint $P$-function.
In the generic case, this is equivalent to requiring $P(\alpha_1, \alpha_2) = P(|\alpha_1|, |\alpha_2|)$, but when $\omega_1 = \omega_2$, the state with 
\be
P(\alpha_1, \alpha_2) = \frac{e^{-|\alpha_1|^2 - |\alpha_2|^2}}{\pi^2} \, \left( 1 + \frac{\alpha_1 \alpha_2^* + \alpha_1^* \alpha_2}{2 \, |\alpha_1| \, |\alpha_2|} \right)
\ee
is stationary.
One can construct similar examples whenever $\omega_1 / \omega_2$ is rational.
More generally, for any number of modes, exceptions to the usual stationarity condition can occur whenever there are degenerate energy eigenstates. 
This is because within a subspace of degenerate states, the density matrix of a stationary state can be completely arbitrary.

However, in this work we focus on the case where the DM can be treated as having a single effective mode, for which the simple stationarity criterion $P(\alpha) = P(|\alpha|)$ can be used.

\subsection{Uniqueness of Coherent States}
\label{app:uniqueness}

In Sec.~\ref{sec:dec_general}, we showed that the purity loss is independent of the DM state if $|\tilde{P}_{a} (\lambda)|^2 = 1$.
Here we show that this condition is satisfied if and only if the DM is in a coherent state.  
Working in terms of the $P$-function is somewhat delicate, because $P$-functions can be singular distributions.
Instead, we convert this to an equivalent question in terms of the $Q$-function, which is a legitimate probability distribution as discussed in App.~\ref{app:general_quasi_characteristic}.

Using Eq.~\eqref{appeq:s_ordered_characteristic_function}, the condition $|\tilde{P}_{a} (\lambda)|^2 = 1$ is equivalent to 
\be
e^{-2 |\lambda|^2} = \tilde{Q}(\lambda)\tilde{Q}(-\lambda)
\ee
where $\tilde{Q}(\lambda) \equiv \tilde{P}^{(-1)}(\lambda)$ is the characteristic function of the $Q$-function.
Taking the inverse complex Fourier transform of both sides implies
\be \label{eq:Qcalc_midpoint}
\frac{1}{2\pi} \, e^{-|\alpha|^2/2} = \int d\beta \, Q(\beta) Q(\beta+\alpha).
\ee
That is, up to a sign, the convolution of $Q$ with itself is a Gaussian.
However, Cram{\'e}r's decomposition theorem~\cite{Cramer_1970} states that if the sum of two independent random variables is Gaussian distributed, then both random variables must be Gaussian distributed.
Then $Q(\alpha)$ must be a Gaussian with variance fixed by Eq.~\eqref{eq:Qcalc_midpoint}.
This gives the general solution 
\be
Q(\alpha) = \frac{1}{\pi} \, e^{- |\alpha-\alpha_{0}|^2}
\ee
for any $\alpha_0$.
By Eq.~\eqref{eq:gaussian_convs}, this corresponds precisely to coherent states, $P(\alpha) = \delta(\alpha- \alpha_{0})$.

As a corollary, we can show that the only pure classical states are coherent states~\cite{HILLERY1985409}.
This is because the characteristic function of a classical state satisfies
\be \label{eq:char_func_bound}
|\tilde{P}(\lambda)|=\left|\int d \alpha \, P(\alpha) \, e^{\lambda \alpha^*-\lambda^* \alpha}\right| \leq \int d \alpha \, P(\alpha)\left|e^{\lambda \alpha^*-\lambda^* \alpha}\right|=\int d \alpha \, P(\alpha)=1.
\ee
However, Eq.~\eqref{eq:Gamma-characteristic} implies that a pure state must satisfy
\be
\int \frac{d\lambda}{\pi}\, |\tilde{P}(\lambda)|^2 \, e^{-|\lambda|^2} = 1.
\ee
Thus, to have a classical pure state we must saturate Eq.~\eqref{eq:char_func_bound} by having $|\tilde{P}(\lambda)|^2 = 1$, which we have just shown is satisfied only for coherent states.

\bibliographystyle{utphys3}
{\scriptsize
\bibliography{refs}}

\providecommand{\href}[2]{#2}\begingroup\raggedright\begin{thebibliography}{100}

\bibitem{LIGOScientific:2014pky}
{\bfseries LIGO Scientific} Collaboration, J.~Aasi {\em et~al.}, ``{Advanced
  LIGO},'' \href{https://dx.doi.org/10.1088/0264-9381/32/7/074001}{{\em Class.
  Quant. Grav.} {\bfseries 32} (2015) 074001},
  \href{https://arxiv.org/abs/1411.4547}{{\ttfamily arXiv:1411.4547 [gr-qc]}}.

\bibitem{LIGOScientific:2016aoc}
{\bfseries LIGO Scientific, Virgo} Collaboration, B.~P. Abbott {\em et~al.},
  ``{Observation of Gravitational Waves from a Binary Black Hole Merger},''
  \href{https://dx.doi.org/10.1103/PhysRevLett.116.061102}{{\em Phys. Rev.
  Lett.} {\bfseries 116} no.~6, (2016) 061102},
  \href{https://arxiv.org/abs/1602.03837}{{\ttfamily arXiv:1602.03837
  [gr-qc]}}.

\bibitem{ADMX:2001dbg}
{\bfseries ADMX} Collaboration, S.~J. Asztalos {\em et~al.}, ``{Large scale
  microwave cavity search for dark matter axions},''
  \href{https://dx.doi.org/10.1103/PhysRevD.64.092003}{{\em Phys. Rev. D}
  {\bfseries 64} (2001) 092003}.

\bibitem{ADMX:2003rdr}
{\bfseries ADMX} Collaboration, S.~J. Asztalos {\em et~al.}, ``{An Improved RF
  cavity search for halo axions},''
  \href{https://dx.doi.org/10.1103/PhysRevD.69.011101}{{\em Phys. Rev. D}
  {\bfseries 69} (2004) 011101},
  \href{https://arxiv.org/abs/astro-ph/0310042}{{\ttfamily
  arXiv:astro-ph/0310042}}.

\bibitem{ADMX:2009iij}
{\bfseries ADMX} Collaboration, S.~J. Asztalos {\em et~al.}, ``{A SQUID-based
  microwave cavity search for dark-matter axions},''
  \href{https://dx.doi.org/10.1103/PhysRevLett.104.041301}{{\em Phys. Rev.
  Lett.} {\bfseries 104} (2010) 041301},
  \href{https://arxiv.org/abs/0910.5914}{{\ttfamily arXiv:0910.5914
  [astro-ph.CO]}}.

\bibitem{ADMX:2018gho}
{\bfseries ADMX} Collaboration, N.~Du {\em et~al.}, ``{A Search for Invisible
  Axion Dark Matter with the Axion Dark Matter Experiment},''
  \href{https://dx.doi.org/10.1103/PhysRevLett.120.151301}{{\em Phys. Rev.
  Lett.} {\bfseries 120} no.~15, (2018) 151301},
  \href{https://arxiv.org/abs/1804.05750}{{\ttfamily arXiv:1804.05750
  [hep-ex]}}.

\bibitem{ADMX:2019uok}
{\bfseries ADMX} Collaboration, T.~Braine {\em et~al.}, ``{Extended Search for
  the Invisible Axion with the Axion Dark Matter Experiment},''
  \href{https://dx.doi.org/10.1103/PhysRevLett.124.101303}{{\em Phys. Rev.
  Lett.} {\bfseries 124} no.~10, (2020) 101303},
  \href{https://arxiv.org/abs/1910.08638}{{\ttfamily arXiv:1910.08638
  [hep-ex]}}.

\bibitem{ADMX:2021nhd}
{\bfseries ADMX} Collaboration, C.~Bartram {\em et~al.}, ``{Search for
  Invisible Axion Dark Matter in the
  3.3{\textendash}4.2{\,}{\,}{\ensuremath{\mu}}eV Mass Range},''
  \href{https://dx.doi.org/10.1103/PhysRevLett.127.261803}{{\em Phys. Rev.
  Lett.} {\bfseries 127} no.~26, (2021) 261803},
  \href{https://arxiv.org/abs/2110.06096}{{\ttfamily arXiv:2110.06096
  [hep-ex]}}.

\bibitem{Fang:2024ple}
Y.~Fang, C.~Gao, Y.-Y. Li, J.~Shu, Y.~Wu, H.~Xing, B.~Xu, L.~Xu, and C.~Zhou,
  ``{Quantum frontiers in high energy physics},''
  \href{https://dx.doi.org/10.1007/s11433-024-2635-4}{{\em Sci. China Phys.
  Mech. Astron.} {\bfseries 68} no.~6, (2025) 260301},
  \href{https://arxiv.org/abs/2411.11294}{{\ttfamily arXiv:2411.11294
  [hep-ph]}}.

\bibitem{LIGOScientific:2013pcc}
{\bfseries LIGO Scientific} Collaboration, J.~Aasi {\em et~al.}, ``{Enhancing
  the sensitivity of the LIGO gravitational wave detector by using squeezed
  states of light},'' \href{https://dx.doi.org/10.1038/nphoton.2013.177}{{\em
  Nature Photon.} {\bfseries 7} (2013) 613--619},
  \href{https://arxiv.org/abs/1310.0383}{{\ttfamily arXiv:1310.0383
  [quant-ph]}}.

\bibitem{HAYSTAC:2020kwv}
{\bfseries HAYSTAC} Collaboration, K.~M. Backes {\em et~al.}, ``{A
  quantum-enhanced search for dark matter axions},''
  \href{https://dx.doi.org/10.1038/s41586-021-03226-7}{{\em Nature} {\bfseries
  590} no.~7845, (2021) 238--242},
  \href{https://arxiv.org/abs/2008.01853}{{\ttfamily arXiv:2008.01853
  [quant-ph]}}.

\bibitem{Jia:2024iqe}
W.~Jia {\em et~al.}, ``{Squeezing the quantum noise of a gravitational-wave
  detector below the standard quantum limit},''
  \href{https://dx.doi.org/10.1126/science.ado8069}{{\em Science} {\bfseries
  385} no.~6715, (2024) 1318},
  \href{https://arxiv.org/abs/2404.14569}{{\ttfamily arXiv:2404.14569
  [gr-qc]}}.

\bibitem{Abbott:1982af}
L.~F. Abbott and P.~Sikivie, ``{A Cosmological Bound on the Invisible Axion},''
  \href{https://dx.doi.org/10.1016/0370-2693(83)90638-X}{{\em Phys. Lett. B}
  {\bfseries 120} (1983) 133--136}.

\bibitem{Preskill:1982cy}
J.~Preskill, M.~B. Wise, and F.~Wilczek, ``{Cosmology of the Invisible
  Axion},'' \href{https://dx.doi.org/10.1016/0370-2693(83)90637-8}{{\em Phys.
  Lett. B} {\bfseries 120} (1983) 127--132}.

\bibitem{Jaeckel:2010ni}
J.~Jaeckel and A.~Ringwald, ``{The Low-Energy Frontier of Particle Physics},''
  \href{https://dx.doi.org/10.1146/annurev.nucl.012809.104433}{{\em Ann. Rev.
  Nucl. Part. Sci.} {\bfseries 60} (2010) 405--437},
  \href{https://arxiv.org/abs/1002.0329}{{\ttfamily arXiv:1002.0329 [hep-ph]}}.

\bibitem{Marsh:2015xka}
D.~J.~E. Marsh, ``{Axion Cosmology},''
  \href{https://dx.doi.org/10.1016/j.physrep.2016.06.005}{{\em Phys. Rept.}
  {\bfseries 643} (2016) 1--79},
  \href{https://arxiv.org/abs/1510.07633}{{\ttfamily arXiv:1510.07633
  [astro-ph.CO]}}.

\bibitem{Irastorza:2018dyq}
I.~G. Irastorza and J.~Redondo, ``{New experimental approaches in the search
  for axion-like particles},''
  \href{https://dx.doi.org/10.1016/j.ppnp.2018.05.003}{{\em Prog. Part. Nucl.
  Phys.} {\bfseries 102} (2018) 89--159},
  \href{https://arxiv.org/abs/1801.08127}{{\ttfamily arXiv:1801.08127
  [hep-ph]}}.

\bibitem{Parikh:2020kfh}
M.~Parikh, F.~Wilczek, and G.~Zahariade, ``{Quantum Mechanics of Gravitational
  Waves},'' \href{https://dx.doi.org/10.1103/PhysRevLett.127.081602}{{\em Phys.
  Rev. Lett.} {\bfseries 127} no.~8, (2021) 081602},
  \href{https://arxiv.org/abs/2010.08205}{{\ttfamily arXiv:2010.08205
  [hep-th]}}.

\bibitem{Tobar:2023ksi}
G.~Tobar, S.~K. Manikandan, T.~Beitel, and I.~Pikovski, ``{Detecting single
  gravitons with quantum sensing},''
  \href{https://dx.doi.org/10.1038/s41467-024-51420-8}{{\em Nature Commun.}
  {\bfseries 15} no.~1, (2024) 7229},
  \href{https://arxiv.org/abs/2308.15440}{{\ttfamily arXiv:2308.15440
  [quant-ph]}}.

\bibitem{Schutzhold:2025vti}
R.~Sch{\"u}tzhold, ``{Stimulated Emission or Absorption of Gravitons by
  Light},'' \href{https://dx.doi.org/10.1103/xd97-c6d7}{{\em Phys. Rev. Lett.}
  {\bfseries 135} no.~17, (2025) 171501},
  \href{https://arxiv.org/abs/2502.10221}{{\ttfamily arXiv:2502.10221
  [gr-qc]}}.

\bibitem{mandel1995optical}
L.~Mandel and E.~Wolf, {\em Optical Coherence and Quantum Optics}.
\newblock Cambridge University Press, 1995.

\bibitem{loudon2000quantum}
R.~Loudon, {\em The Quantum Theory of Light}.
\newblock OUP Oxford, 2000.

\bibitem{barnett2002methods}
S.~Barnett and P.~M. Radmore, {\em Methods in Theoretical Quantum Optics},
  vol.~15 of {\em Oxford Series in Optical and Imaging Sciences}.
\newblock Oxford University Press, 2002.

\bibitem{Cheong:2024ose}
D.~Y. Cheong, N.~L. Rodd, and L.-T. Wang, ``{Quantum description of wave dark
  matter},'' \href{https://dx.doi.org/10.1103/PhysRevD.111.015028}{{\em Phys.
  Rev. D} {\bfseries 111} no.~1, (2025) 015028},
  \href{https://arxiv.org/abs/2408.04696}{{\ttfamily arXiv:2408.04696
  [hep-ph]}}.

\bibitem{HAYSTAC:2023cam}
{\bfseries HAYSTAC} Collaboration, M.~J. Jewell {\em et~al.}, ``{New results
  from HAYSTAC{\textquoteright}s phase II operation with a squeezed state
  receiver},'' \href{https://dx.doi.org/10.1103/PhysRevD.107.072007}{{\em Phys.
  Rev. D} {\bfseries 107} no.~7, (2023) 072007},
  \href{https://arxiv.org/abs/2301.09721}{{\ttfamily arXiv:2301.09721
  [hep-ex]}}.

\bibitem{HAYSTAC:2024jch}
{\bfseries HAYSTAC} Collaboration, X.~Bai {\em et~al.}, ``{Dark Matter Axion
  Search with HAYSTAC Phase II},''
  \href{https://dx.doi.org/10.1103/PhysRevLett.134.151006}{{\em Phys. Rev.
  Lett.} {\bfseries 134} no.~15, (2025) 151006},
  \href{https://arxiv.org/abs/2409.08998}{{\ttfamily arXiv:2409.08998
  [hep-ex]}}.

\bibitem{CAPP:2020utb}
{\bfseries CAPP} Collaboration, O.~Kwon {\em et~al.}, ``{First Results from an
  Axion Haloscope at CAPP around 10.7 $\mu$eV},''
  \href{https://dx.doi.org/10.1103/PhysRevLett.126.191802}{{\em Phys. Rev.
  Lett.} {\bfseries 126} no.~19, (2021) 191802},
  \href{https://arxiv.org/abs/2012.10764}{{\ttfamily arXiv:2012.10764
  [hep-ex]}}.

\bibitem{CAPP:2024dtx}
{\bfseries CAPP} Collaboration, S.~Ahn {\em et~al.}, ``{Extensive Search for
  Axion Dark Matter over 1~GHz with CAPP{\textquoteright}S Main Axion
  Experiment},'' \href{https://dx.doi.org/10.1103/PhysRevX.14.031023}{{\em
  Phys. Rev. X} {\bfseries 14} no.~3, (2024) 031023},
  \href{https://arxiv.org/abs/2402.12892}{{\ttfamily arXiv:2402.12892
  [hep-ex]}}.

\bibitem{Ahn:2026ssw}
S.~Ahn {\em et~al.}, ``{Extended Haloscope Search and Exclusion of a Candidate
  Signal near 1.036~GHz},'' \href{https://dx.doi.org/10.1103/2sn2-h97m}{{\em
  Phys. Rev. Lett.} {\bfseries 137} no.~2, (2026) 021803},
  \href{https://arxiv.org/abs/2602.05388}{{\ttfamily arXiv:2602.05388
  [hep-ex]}}.

\bibitem{Glauber:1963tx}
R.~J. Glauber, ``{Coherent and incoherent states of the radiation field},''
  \href{https://dx.doi.org/10.1103/PhysRev.131.2766}{{\em Phys. Rev.}
  {\bfseries 131} (1963) 2766}.

\bibitem{Sudarshan:1963ts}
E.~C.~G. Sudarshan, ``{Equivalence of semiclassical and quantum mechanical
  descriptions of statistical light beams},''
  \href{https://dx.doi.org/10.1103/PhysRevLett.10.277}{{\em Phys. Rev. Lett.}
  {\bfseries 10} (1963) 277}.

\bibitem{Bravyi:2004isx}
S.~Bravyi and A.~Kitaev, ``{Universal quantum computation with ideal Clifford
  gates and noisy ancillas},''
  \href{https://dx.doi.org/10.1103/PhysRevA.71.022316}{{\em Phys. Rev. A}
  {\bfseries 71} no.~2, (2005) 022316},
  \href{https://arxiv.org/abs/quant-ph/0403025}{{\ttfamily
  arXiv:quant-ph/0403025}}.

\bibitem{Bao:2025nsd}
Y.~Bao, D.~Y. Cheong, N.~L. Rodd, J.~Takach, L.-T. Wang, and K.~Zhou,
  ``{Intrinsically Quantum Effects of Axion Dark Matter Are Undetectable},''
  \href{https://dx.doi.org/10.1103/9mff-p5k6}{{\em Phys. Rev. Lett.} {\bfseries
  136} no.~17, (2026) 171601},
  \href{https://arxiv.org/abs/2510.05198}{{\ttfamily arXiv:2510.05198
  [hep-ph]}}.

\bibitem{Chakram:2021bxb}
S.~Chakram, A.~E. Oriani, R.~K. Naik, A.~V. Dixit, K.~He, A.~Agrawal, H.~Kwon,
  and D.~I. Schuster, ``{Seamless High-Q Microwave Cavities for Multimode
  Circuit Quantum Electrodynamics},''
  \href{https://dx.doi.org/10.1103/PhysRevLett.127.107701}{{\em Phys. Rev.
  Lett.} {\bfseries 127} no.~10, (2021) 107701}.

\bibitem{Dixit:2020ymh}
A.~V. Dixit, S.~Chakram, K.~He, A.~Agrawal, R.~K. Naik, D.~I. Schuster, and
  A.~Chou, ``{Searching for Dark Matter with a Superconducting Qubit},''
  \href{https://dx.doi.org/10.1103/PhysRevLett.126.141302}{{\em Phys. Rev.
  Lett.} {\bfseries 126} no.~14, (2021) 141302},
  \href{https://arxiv.org/abs/2008.12231}{{\ttfamily arXiv:2008.12231
  [hep-ex]}}.

\bibitem{Gu:2025pms}
{\bfseries RADES} Collaboration, Y.~Gu, ``{Dark matter detection with
  superconducting qubit in RADES experiment},''
  \href{https://dx.doi.org/10.22323/1.474.0067}{{\em PoS} {\bfseries
  COSMICWISPers2024} (2025) 067}.

\bibitem{Agrawal:2023umy}
A.~Agrawal, A.~V. Dixit, T.~Roy, S.~Chakram, K.~He, R.~K. Naik, D.~I. Schuster,
  and A.~Chou, ``{Stimulated Emission of Signal Photons from Dark Matter
  Waves},'' \href{https://dx.doi.org/10.1103/PhysRevLett.132.140801}{{\em Phys.
  Rev. Lett.} {\bfseries 132} no.~14, (2024) 140801},
  \href{https://arxiv.org/abs/2305.03700}{{\ttfamily arXiv:2305.03700
  [quant-ph]}}.

\bibitem{Zheng:2025qgv}
P.~Zheng {\em et~al.}, ``{Quantum-Enhanced Dark Matter Search Using Cat
  States},'' \href{https://dx.doi.org/10.1103/wbhn-v1sw}{{\em Phys. Rev. Lett.}
  {\bfseries 136} no.~17, (2026) 171002},
  \href{https://arxiv.org/abs/2507.23538}{{\ttfamily arXiv:2507.23538
  [quant-ph]}}.

\bibitem{continuous_paper}
Y.~Bao, D.~Y. Cheong, N.~L. Rodd, J.~Takach, L.-T. Wang, and K.~Zhou,
  ``Continuous measurement of quantum axion dark matter.'' 2026.
\newblock To appear.

\bibitem{Carney:2023nzz}
D.~Carney, V.~Domcke, and N.~L. Rodd, ``{Graviton detection and the
  quantization of gravity},''
  \href{https://dx.doi.org/10.1103/PhysRevD.109.044009}{{\em Phys. Rev. D}
  {\bfseries 109} no.~4, (2024) 044009},
  \href{https://arxiv.org/abs/2308.12988}{{\ttfamily arXiv:2308.12988
  [hep-th]}}.

\bibitem{Carney:2024dsj}
D.~Carney, \href{https://dx.doi.org/10.1007/978-3-031-91266-5_2}{``{Comments on
  Graviton Detection},''} in {\em {Proceedings of Gravity, Strings and Fields}:
  {A Conference in Honour of Gordon Semenoff}}, CRM Series in Mathematical
  Physics, pp.~11--28.
\newblock Springer Cham, 2025.
\newblock \href{https://arxiv.org/abs/2408.00094}{{\ttfamily arXiv:2408.00094
  [gr-qc]}}.

\bibitem{Brubaker:2017ohw}
B.~M. Brubaker, {\em {First results from the HAYSTAC axion search}}.
\newblock PhD thesis, Yale U., 2017.
\newblock \href{https://arxiv.org/abs/1801.00835}{{\ttfamily arXiv:1801.00835
  [astro-ph.CO]}}.

\bibitem{Kim:2021yyo}
H.~Kim and A.~Lenoci, ``{Gravitational focusing of wave dark matter},''
  \href{https://dx.doi.org/10.1103/PhysRevD.105.063032}{{\em Phys. Rev. D}
  {\bfseries 105} no.~6, (2022) 063032},
  \href{https://arxiv.org/abs/2112.05718}{{\ttfamily arXiv:2112.05718
  [hep-ph]}}.

\bibitem{Budker:2023sex}
D.~Budker, J.~Eby, M.~Gorghetto, M.~Jiang, and G.~Perez, ``{A generic formation
  mechanism of ultralight dark matter solar halos},''
  \href{https://dx.doi.org/10.1088/1475-7516/2023/12/021}{{\em JCAP} {\bfseries
  12} (2023) 021}, \href{https://arxiv.org/abs/2306.12477}{{\ttfamily
  arXiv:2306.12477 [hep-ph]}}.

\bibitem{Dror:2021nyr}
J.~A. Dror, H.~Murayama, and N.~L. Rodd, ``{Cosmic axion background},''
  \href{https://dx.doi.org/10.1103/PhysRevD.103.115004}{{\em Phys. Rev. D}
  {\bfseries 103} no.~11, (2021) 115004},
  \href{https://arxiv.org/abs/2101.09287}{{\ttfamily arXiv:2101.09287
  [hep-ph]}}. [Erratum: Phys.Rev.D 106, 119902 (2022)].

\bibitem{ADMX:2023rsk}
{\bfseries ADMX} Collaboration, T.~Nitta {\em et~al.}, ``{Search for a
  Dark-Matter-Induced Cosmic Axion Background with ADMX},''
  \href{https://dx.doi.org/10.1103/PhysRevLett.131.101002}{{\em Phys. Rev.
  Lett.} {\bfseries 131} no.~10, (2023) 101002},
  \href{https://arxiv.org/abs/2303.06282}{{\ttfamily arXiv:2303.06282
  [hep-ex]}}.

\bibitem{PhysRevA.87.033811}
E.~Agudelo, J.~Sperling, and W.~Vogel, ``Quasiprobabilities for multipartite
  quantum correlations of light,''
  \href{https://dx.doi.org/10.1103/PhysRevA.87.033811}{{\em Phys. Rev. A}
  {\bfseries 87} (Mar, 2013) 033811}.

\bibitem{Cushen:1971}
C.~D. Cushen and R.~L. Hudson, ``A quantum-mechanical central limit theorem,''
  \href{https://dx.doi.org/10.2307/3212170}{{\em J. Appl. Probab.} {\bfseries
  8} no.~3, (1971) 454--469}.

\bibitem{Becker:2020myv}
S.~Becker, N.~Datta, L.~Lami, and C.~Rouz{\'e}, ``{Convergence rates for the
  quantum central limit theorem},''
  \href{https://dx.doi.org/10.1007/s00220-021-03988-1}{{\em Commun. Math.
  Phys.} {\bfseries 383} (2021) 223--279},
  \href{https://arxiv.org/abs/1912.06129}{{\ttfamily arXiv:1912.06129
  [quant-ph]}}.

\bibitem{Bu:2023ssg}
K.~Bu, W.~Gu, and A.~Jaffe, ``{Quantum Entropy and Central Limit Theorem},''
  \href{https://dx.doi.org/10.1073/pnas.2304589120}{{\em Proc. Nat. Acad. Sci.}
  {\bfseries 120} (2023) e2304589120},
  \href{https://arxiv.org/abs/2302.07841}{{\ttfamily arXiv:2302.07841
  [quant-ph]}}.

\bibitem{PhysRev.136.A316}
P.~L. Kelley and W.~H. Kleiner, ``Theory of electromagnetic field measurement
  and photoelectron counting,''
  \href{https://dx.doi.org/10.1103/PhysRev.136.A316}{{\em Phys. Rev.}
  {\bfseries 136} (Oct, 1964) A316--A334}.

\bibitem{PhysRevA.6.2211}
F.~T. Arecchi, E.~Courtens, R.~Gilmore, and H.~Thomas, ``Atomic coherent states
  in quantum optics,'' \href{https://dx.doi.org/10.1103/PhysRevA.6.2211}{{\em
  Phys. Rev. A} {\bfseries 6} (Dec, 1972) 2211--2237}.

\bibitem{PhysRevA.46.485}
G.~S. Agarwal and K.~Tara, ``Nonclassical character of states exhibiting no
  squeezing or sub-{Poissonian} statistics,''
  \href{https://dx.doi.org/10.1103/PhysRevA.46.485}{{\em Phys. Rev. A}
  {\bfseries 46} (Jul, 1992) 485--488}.

\bibitem{Kullback:1951zyt}
S.~Kullback and R.~A. Leibler, ``{On Information and Sufficiency},''
  \href{https://dx.doi.org/10.1214/aoms/1177729694}{{\em Ann. Math. Statist.}
  {\bfseries 22} no.~1, (1951) 79--86}.

\bibitem{Cover:2005lom}
T.~M. Cover and J.~A. Thomas, \href{https://dx.doi.org/10.1002/047174882x}{{\em
  {Elements of Information Theory}}}.
\newblock Wiley, 2005.

\bibitem{PhysRev.177.1857}
K.~E. Cahill and R.~J. Glauber, ``Ordered expansions in boson amplitude
  operators,'' \href{https://dx.doi.org/10.1103/PhysRev.177.1857}{{\em Phys.
  Rev.} {\bfseries 177} (Jan, 1969) 1857--1881}.

\bibitem{PhysRevA.31.338}
M.~Hillery, ``Conservation laws and nonclassical states in nonlinear optical
  systems,'' \href{https://dx.doi.org/10.1103/PhysRevA.31.338}{{\em Phys. Rev.
  A} {\bfseries 31} (Jan, 1985) 338--342}.

\bibitem{AGARWAL1993109}
G.~Agarwal, ``Nonclassical characteristics of the marginals for the radiation
  field,'' \href{https://dx.doi.org/10.1016/0030-4018(93)90059-E}{{\em Opt.
  Commun.} {\bfseries 95} no.~1, (1993) 109--112}.

\bibitem{PhysRevA.71.011802}
E.~Shchukin, T.~Richter, and W.~Vogel, ``Nonclassicality criteria in terms of
  moments,'' \href{https://dx.doi.org/10.1103/PhysRevA.71.011802}{{\em Phys.
  Rev. A} {\bfseries 71} (Jan, 2005) 011802}.

\bibitem{PhysRevA.41.1721}
C.~T. Lee, ``Higher-order criteria for nonclassical effects in photon
  statistics,'' \href{https://dx.doi.org/10.1103/PhysRevA.41.1721}{{\em Phys.
  Rev. A} {\bfseries 41} (Feb, 1990) 1721--1723}.

\bibitem{PhysRevA.41.1569}
C.~T. Lee, ``Many-photon antibunching in generalized pair coherent states,''
  \href{https://dx.doi.org/10.1103/PhysRevA.41.1569}{{\em Phys. Rev. A}
  {\bfseries 41} (Feb, 1990) 1569--1575}.

\bibitem{KLYSHKO19967}
D.~Klyshko, ``Observable signs of nonclassical light,''
  \href{https://dx.doi.org/10.1016/0375-9601(96)00091-6}{{\em Phys. Lett. A}
  {\bfseries 213} no.~1, (1996) 7--15}.

\bibitem{PhysRevA.72.043808}
E.~V. Shchukin and W.~Vogel, ``Nonclassical moments and their measurement,''
  \href{https://dx.doi.org/10.1103/PhysRevA.72.043808}{{\em Phys. Rev. A}
  {\bfseries 72} (Oct, 2005) 043808}.

\bibitem{PhysRevLett.124.133601}
M.~Bohmann and E.~Agudelo, ``Phase-space inequalities beyond negativities,''
  \href{https://dx.doi.org/10.1103/PhysRevLett.124.133601}{{\em Phys. Rev.
  Lett.} {\bfseries 124} (Mar, 2020) 133601}.

\bibitem{PhysRevResearch.3.043116}
J.~Park, J.~Lee, and H.~Nha, ``Verifying single-mode nonclassicality beyond
  negativity in phase space,''
  \href{https://dx.doi.org/10.1103/PhysRevResearch.3.043116}{{\em Phys. Rev.
  Res.} {\bfseries 3} (Nov, 2021) 043116}.

\bibitem{PhysRevA.51.3340}
N.~L\"utkenhaus and S.~M. Barnett, ``Nonclassical effects in phase space,''
  \href{https://dx.doi.org/10.1103/PhysRevA.51.3340}{{\em Phys. Rev. A}
  {\bfseries 51} (Apr, 1995) 3340--3342}.

\bibitem{Ferrie:2011rce}
C.~Ferrie, ``{Quasi-probability representations of quantum theory with
  applications to quantum information science},''
  \href{https://dx.doi.org/10.1088/0034-4885/74/11/116001}{{\em Rept. Prog.
  Phys.} {\bfseries 74} no.~11, (2011) 116001},
  \href{https://arxiv.org/abs/1010.2701}{{\ttfamily arXiv:1010.2701
  [quant-ph]}}.

\bibitem{Veitch:2012ttw}
V.~Veitch, C.~Ferrie, D.~Gross, and J.~Emerson, ``{Negative quasi-probability
  as a resource for quantum computation},''
  \href{https://dx.doi.org/10.1088/1367-2630/14/11/113011}{{\em New J. Phys.}
  {\bfseries 14} no.~11, (2012) 113011},
  \href{https://arxiv.org/abs/1201.1256}{{\ttfamily arXiv:1201.1256
  [quant-ph]}}.

\bibitem{Mari:2012ypq}
A.~Mari and J.~Eisert, ``{Positive Wigner Functions Render Classical Simulation
  of Quantum Computation Efficient},''
  \href{https://dx.doi.org/10.1103/PhysRevLett.109.230503}{{\em Phys. Rev.
  Lett.} {\bfseries 109} no.~23, (2012) 230503},
  \href{https://arxiv.org/abs/1208.3660}{{\ttfamily arXiv:1208.3660
  [quant-ph]}}.

\bibitem{Wang:2019nve}
X.~Wang, M.~M. Wilde, and Y.~Su, ``{Quantifying the magic of quantum
  channels},'' \href{https://dx.doi.org/10.1088/1367-2630/ab451d}{{\em New J.
  Phys.} {\bfseries 21} no.~10, (2019) 103002},
  \href{https://arxiv.org/abs/1903.04483}{{\ttfamily arXiv:1903.04483
  [quant-ph]}}.

\bibitem{nielsen2010quantum}
M.~A. Nielsen and I.~L. Chuang, {\em Quantum Computation and Quantum
  Information}.
\newblock Cambridge University Press, 2010.

\bibitem{Chen:2022quj}
S.~Chen, H.~Fukuda, T.~Inada, T.~Moroi, T.~Nitta, and T.~Sichanugrist,
  ``{Detecting Hidden Photon Dark Matter Using the Direct Excitation of
  Transmon Qubits},''
  \href{https://dx.doi.org/10.1103/PhysRevLett.131.211001}{{\em Phys. Rev.
  Lett.} {\bfseries 131} no.~21, (2023) 211001},
  \href{https://arxiv.org/abs/2212.03884}{{\ttfamily arXiv:2212.03884
  [hep-ph]}}.

\bibitem{Chen:2023swh}
S.~Chen, H.~Fukuda, T.~Inada, T.~Moroi, T.~Nitta, and T.~Sichanugrist,
  ``{Quantum Enhancement in Dark Matter Detection with Quantum Computation},''
  \href{https://dx.doi.org/10.1103/PhysRevLett.133.021801}{{\em Phys. Rev.
  Lett.} {\bfseries 133} no.~2, (2024) 021801},
  \href{https://arxiv.org/abs/2311.10413}{{\ttfamily arXiv:2311.10413
  [hep-ph]}}.

\bibitem{Ito:2023zhp}
A.~Ito, R.~Kitano, W.~Nakano, and R.~Takai, ``{Quantum entanglement of ions for
  light dark matter detection},''
  \href{https://dx.doi.org/10.1007/JHEP02(2024)124}{{\em JHEP} {\bfseries 02}
  (2024) 124}, \href{https://arxiv.org/abs/2311.11632}{{\ttfamily
  arXiv:2311.11632 [hep-ph]}}.

\bibitem{Chen:2024aya}
S.~Chen, H.~Fukuda, T.~Inada, T.~Moroi, T.~Nitta, and T.~Sichanugrist,
  ``{Search for QCD axion dark matter with transmon qubits and quantum
  circuit},'' \href{https://dx.doi.org/10.1103/PhysRevD.110.115021}{{\em Phys.
  Rev. D} {\bfseries 110} no.~11, (2024) 115021},
  \href{https://arxiv.org/abs/2407.19755}{{\ttfamily arXiv:2407.19755
  [hep-ph]}}.

\bibitem{Fukuda:2025zcf}
H.~Fukuda, Y.~Matsuzaki, and T.~Sichanugrist, ``{Directional Searching for
  Light Dark Matter with Quantum Sensors},''
  \href{https://dx.doi.org/10.1103/cwx5-2n1y}{{\em Phys. Rev. Lett.} {\bfseries
  135} no.~24, (2025) 241802},
  \href{https://arxiv.org/abs/2506.19614}{{\ttfamily arXiv:2506.19614
  [hep-ph]}}.

\bibitem{Bodas:2025vff}
A.~Bodas, S.~Ghosh, and R.~Harnik, ``{On the Speed-up of Wave-like Dark Matter
  Searches with Entangled Qubits},''
  \href{https://arxiv.org/abs/2510.11795}{{\ttfamily arXiv:2510.11795
  [hep-ph]}}.

\bibitem{PhysRevA.60.2752}
S.~M. Tan, ``Confirming entanglement in continuous variable quantum
  teleportation,'' \href{https://dx.doi.org/10.1103/PhysRevA.60.2752}{{\em
  Phys. Rev. A} {\bfseries 60} (Oct, 1999) 2752--2758}.

\bibitem{PhysRevLett.84.2726}
R.~Simon, ``{Peres-Horodecki Separability Criterion for Continuous Variable
  Systems},'' \href{https://dx.doi.org/10.1103/PhysRevLett.84.2726}{{\em Phys.
  Rev. Lett.} {\bfseries 84} (Mar, 2000) 2726--2729}.

\bibitem{PhysRevLett.96.050503}
M.~Hillery and M.~S. Zubairy, ``Entanglement conditions for two-mode states,''
  \href{https://dx.doi.org/10.1103/PhysRevLett.96.050503}{{\em Phys. Rev.
  Lett.} {\bfseries 96} (Feb, 2006) 050503}.

\bibitem{PhysRevLett.84.2722}
L.-M. Duan, G.~Giedke, J.~I. Cirac, and P.~Zoller, ``Inseparability criterion
  for continuous variable systems,''
  \href{https://dx.doi.org/10.1103/PhysRevLett.84.2722}{{\em Phys. Rev. Lett.}
  {\bfseries 84} (Mar, 2000) 2722--2725}.

\bibitem{Foster:2020fln}
J.~W. Foster, Y.~Kahn, R.~Nguyen, N.~L. Rodd, and B.~R. Safdi, ``{Dark Matter
  Interferometry},'' \href{https://dx.doi.org/10.1103/PhysRevD.103.076018}{{\em
  Phys. Rev. D} {\bfseries 103} no.~7, (2021) 076018},
  \href{https://arxiv.org/abs/2009.14201}{{\ttfamily arXiv:2009.14201
  [hep-ph]}}.

\bibitem{Riedel:2012ur}
C.~J. Riedel, ``{Direct detection of classically undetectable dark matter
  through quantum decoherence},''
  \href{https://dx.doi.org/10.1103/PhysRevD.88.116005}{{\em Phys. Rev. D}
  {\bfseries 88} no.~11, (2013) 116005},
  \href{https://arxiv.org/abs/1212.3061}{{\ttfamily arXiv:1212.3061
  [quant-ph]}}.

\bibitem{Riedel:2016acj}
C.~J. Riedel and I.~Yavin, ``{Decoherence as a way to measure extremely soft
  collisions with dark matter},''
  \href{https://dx.doi.org/10.1103/PhysRevD.96.023007}{{\em Phys. Rev. D}
  {\bfseries 96} no.~2, (2017) 023007},
  \href{https://arxiv.org/abs/1609.04145}{{\ttfamily arXiv:1609.04145
  [quant-ph]}}.

\bibitem{Du:2022ceh}
Y.~Du, C.~Murgui, K.~Pardo, Y.~Wang, and K.~M. Zurek, ``{Atom interferometer
  tests of dark matter},''
  \href{https://dx.doi.org/10.1103/PhysRevD.106.095041}{{\em Phys. Rev. D}
  {\bfseries 106} no.~9, (2022) 095041},
  \href{https://arxiv.org/abs/2205.13546}{{\ttfamily arXiv:2205.13546
  [hep-ph]}}.

\bibitem{Badurina:2024nge}
L.~Badurina, C.~Murgui, and R.~Plestid, ``{Coherent collisional decoherence},''
  \href{https://dx.doi.org/10.1103/PhysRevA.110.033311}{{\em Phys. Rev. A}
  {\bfseries 110} no.~3, (2024) 033311},
  \href{https://arxiv.org/abs/2402.03421}{{\ttfamily arXiv:2402.03421
  [quant-ph]}}.

\bibitem{Badurina:2026owr}
L.~Badurina and K.~M. Zurek, ``{Matter-Wave Interferometers as Open-System Dark
  Matter Detectors},'' \href{https://arxiv.org/abs/2606.00237}{{\ttfamily
  arXiv:2606.00237 [hep-ph]}}.

\bibitem{Badurina:2025idj}
L.~Badurina, D.~Blas, J.~Ellis, and S.~A.~R. Ellis, ``{Ultralight dark matter
  detection with trapped-ion interferometry},''
  \href{https://dx.doi.org/10.1103/528c-xs6p}{{\em Phys. Rev. D} {\bfseries
  113} no.~9, (2026) 092004},
  \href{https://arxiv.org/abs/2507.17825}{{\ttfamily arXiv:2507.17825
  [hep-ph]}}.

\bibitem{Goldberg:2021syi}
A.~Z. Goldberg and K.~Heshami, ``{How squeezed states both maximize and
  minimize the same notion of quantumness},''
  \href{https://dx.doi.org/10.1103/PhysRevA.104.032425}{{\em Phys. Rev. A}
  {\bfseries 104} (2021) 032425},
  \href{https://arxiv.org/abs/2106.03862}{{\ttfamily arXiv:2106.03862
  [quant-ph]}}.

\bibitem{Gundhi:2025bwj}
A.~Gundhi and H.~Ulbricht, ``{Measuring Decoherence Due to Quantum Vacuum
  Fluctuations},'' \href{https://dx.doi.org/10.1103/s5c9-zjt9}{{\em Phys. Rev.
  Lett.} {\bfseries 135} no.~2, (2025) 020402},
  \href{https://arxiv.org/abs/2501.17928}{{\ttfamily arXiv:2501.17928
  [quant-ph]}}.

\bibitem{Allali:2021puy}
I.~J. Allali and M.~P. Hertzberg, ``{General Relativistic Decoherence with
  Applications to Dark Matter Detection},''
  \href{https://dx.doi.org/10.1103/PhysRevLett.127.031301}{{\em Phys. Rev.
  Lett.} {\bfseries 127} no.~3, (2021) 031301},
  \href{https://arxiv.org/abs/2103.15892}{{\ttfamily arXiv:2103.15892
  [gr-qc]}}.

\bibitem{Arvanitaki:2024taq}
A.~Arvanitaki, S.~Dimopoulos, and M.~Galanis, ``{Superradiant interactions of
  the cosmic neutrino background, axions, dark matter, and reactor
  neutrinos},'' \href{https://dx.doi.org/10.1103/PhysRevD.111.055015}{{\em
  Phys. Rev. D} {\bfseries 111} no.~5, (2025) 055015},
  \href{https://arxiv.org/abs/2408.04021}{{\ttfamily arXiv:2408.04021
  [hep-ph]}}.

\bibitem{Foster:2017hbq}
J.~W. Foster, N.~L. Rodd, and B.~R. Safdi, ``{Revealing the Dark Matter Halo
  with Axion Direct Detection},''
  \href{https://dx.doi.org/10.1103/PhysRevD.97.123006}{{\em Phys. Rev. D}
  {\bfseries 97} no.~12, (2018) 123006},
  \href{https://arxiv.org/abs/1711.10489}{{\ttfamily arXiv:1711.10489
  [astro-ph.CO]}}.

\bibitem{Holdom:1985ag}
B.~Holdom, ``{Two $U(1)$'s and $\epsilon$ Charge Shifts},''
  \href{https://dx.doi.org/10.1016/0370-2693(86)91377-8}{{\em Phys. Lett. B}
  {\bfseries 166} (1986) 196--198}.

\bibitem{ADMX:2010ubl}
{\bfseries ADMX} Collaboration, A.~Wagner {\em et~al.}, ``{A Search for Hidden
  Sector Photons with ADMX},''
  \href{https://dx.doi.org/10.1103/PhysRevLett.105.171801}{{\em Phys. Rev.
  Lett.} {\bfseries 105} (2010) 171801},
  \href{https://arxiv.org/abs/1007.3766}{{\ttfamily arXiv:1007.3766 [hep-ex]}}.

\bibitem{Chaudhuri:2014dla}
S.~Chaudhuri, P.~W. Graham, K.~Irwin, J.~Mardon, S.~Rajendran, and Y.~Zhao,
  ``{Radio for hidden-photon dark matter detection},''
  \href{https://dx.doi.org/10.1103/PhysRevD.92.075012}{{\em Phys. Rev. D}
  {\bfseries 92} no.~7, (2015) 075012},
  \href{https://arxiv.org/abs/1411.7382}{{\ttfamily arXiv:1411.7382 [hep-ph]}}.

\bibitem{Caputo:2021eaa}
A.~Caputo, A.~J. Millar, C.~A.~J. O'Hare, and E.~Vitagliano, ``{Dark photon
  limits: A handbook},''
  \href{https://dx.doi.org/10.1103/PhysRevD.104.095029}{{\em Phys. Rev. D}
  {\bfseries 104} no.~9, (2021) 095029},
  \href{https://arxiv.org/abs/2105.04565}{{\ttfamily arXiv:2105.04565
  [hep-ph]}}.

\bibitem{Cervantes:2022epl}
R.~Cervantes {\em et~al.}, ``{ADMX-Orpheus first search for
  70{\,}{\,}{\ensuremath{\mu}}eV dark photon dark matter: Detailed design,
  operations, and analysis},''
  \href{https://dx.doi.org/10.1103/PhysRevD.106.102002}{{\em Phys. Rev. D}
  {\bfseries 106} no.~10, (2022) 102002},
  \href{https://arxiv.org/abs/2204.09475}{{\ttfamily arXiv:2204.09475
  [hep-ex]}}.

\bibitem{BREAD:2023xhc}
{\bfseries BREAD} Collaboration, S.~Knirck {\em et~al.}, ``{First Results from
  a Broadband Search for Dark Photon Dark Matter in the 44 to
  52{\,}{\,}{\ensuremath{\mu}}eV Range with a Coaxial Dish Antenna},''
  \href{https://dx.doi.org/10.1103/PhysRevLett.132.131004}{{\em Phys. Rev.
  Lett.} {\bfseries 132} no.~13, (2024) 131004},
  \href{https://arxiv.org/abs/2310.13891}{{\ttfamily arXiv:2310.13891
  [hep-ex]}}.

\bibitem{Beadle:2025dgy}
C.~Beadle, S.~A.~R. Ellis, J.~M. Leedom, and N.~L. Rodd, ``{Nuclear magnetic
  resonance dark-matter searches are sensitive to dark photons and the
  axion-photon coupling},'' \href{https://dx.doi.org/10.1103/f9fj-6kkj}{{\em
  Phys. Rev. D} {\bfseries 113} no.~3, (2026) L031702},
  \href{https://arxiv.org/abs/2505.15897}{{\ttfamily arXiv:2505.15897
  [hep-ph]}}.

\bibitem{Kahn:2016aff}
Y.~Kahn, B.~R. Safdi, and J.~Thaler, ``{Broadband and Resonant Approaches to
  Axion Dark Matter Detection},''
  \href{https://dx.doi.org/10.1103/PhysRevLett.117.141801}{{\em Phys. Rev.
  Lett.} {\bfseries 117} no.~14, (2016) 141801},
  \href{https://arxiv.org/abs/1602.01086}{{\ttfamily arXiv:1602.01086
  [hep-ph]}}.

\bibitem{Berlin:2019ahk}
A.~Berlin, R.~T. D'Agnolo, S.~A.~R. Ellis, C.~Nantista, J.~Neilson,
  P.~Schuster, S.~Tantawi, N.~Toro, and K.~Zhou, ``{Axion Dark Matter Detection
  by Superconducting Resonant Frequency Conversion},''
  \href{https://dx.doi.org/10.1007/JHEP07(2020)088}{{\em JHEP} {\bfseries 07}
  (2020) 088}, \href{https://arxiv.org/abs/1912.11048}{{\ttfamily
  arXiv:1912.11048 [hep-ph]}}.

\bibitem{Berlin:2020vrk}
A.~Berlin, R.~T. D'Agnolo, S.~A.~R. Ellis, and K.~Zhou, ``{Heterodyne broadband
  detection of axion dark matter},''
  \href{https://dx.doi.org/10.1103/PhysRevD.104.L111701}{{\em Phys. Rev. D}
  {\bfseries 104} no.~11, (2021) L111701},
  \href{https://arxiv.org/abs/2007.15656}{{\ttfamily arXiv:2007.15656
  [hep-ph]}}.

\bibitem{Knapen:2017ekk}
S.~Knapen, T.~Lin, M.~Pyle, and K.~M. Zurek, ``{Detection of Light Dark Matter
  With Optical Phonons in Polar Materials},''
  \href{https://dx.doi.org/10.1016/j.physletb.2018.08.064}{{\em Phys. Lett. B}
  {\bfseries 785} (2018) 386--390},
  \href{https://arxiv.org/abs/1712.06598}{{\ttfamily arXiv:1712.06598
  [hep-ph]}}.

\bibitem{Mitridate:2020kly}
A.~Mitridate, T.~Trickle, Z.~Zhang, and K.~M. Zurek, ``{Detectability of Axion
  Dark Matter with Phonon Polaritons and Magnons},''
  \href{https://dx.doi.org/10.1103/PhysRevD.102.095005}{{\em Phys. Rev. D}
  {\bfseries 102} no.~9, (2020) 095005},
  \href{https://arxiv.org/abs/2005.10256}{{\ttfamily arXiv:2005.10256
  [hep-ph]}}.

\bibitem{Bloch:2024qqo}
I.~M. Bloch, S.~Knapen, A.~Madden, and G.~Marocco, ``{Broadband phonon
  production from axion absorption},''
  \href{https://dx.doi.org/10.1007/JHEP03(2025)080}{{\em JHEP} {\bfseries 03}
  (2025) 080}, \href{https://arxiv.org/abs/2411.10542}{{\ttfamily
  arXiv:2411.10542 [hep-ph]}}.

\bibitem{Hochberg:2016sqx}
Y.~Hochberg, T.~Lin, and K.~M. Zurek, ``{Absorption of light dark matter in
  semiconductors},'' \href{https://dx.doi.org/10.1103/PhysRevD.95.023013}{{\em
  Phys. Rev. D} {\bfseries 95} no.~2, (2017) 023013},
  \href{https://arxiv.org/abs/1608.01994}{{\ttfamily arXiv:1608.01994
  [hep-ph]}}.

\bibitem{Mitridate:2021ctr}
A.~Mitridate, T.~Trickle, Z.~Zhang, and K.~M. Zurek, ``{Dark matter absorption
  via electronic excitations},''
  \href{https://dx.doi.org/10.1007/JHEP09(2021)123}{{\em JHEP} {\bfseries 09}
  (2021) 123}, \href{https://arxiv.org/abs/2106.12586}{{\ttfamily
  arXiv:2106.12586 [hep-ph]}}.

\bibitem{Graham:2017ivz}
P.~W. Graham, D.~E. Kaplan, J.~Mardon, S.~Rajendran, W.~A. Terrano, L.~Trahms,
  and T.~Wilkason, ``{Spin Precession Experiments for Light Axionic Dark
  Matter},'' \href{https://dx.doi.org/10.1103/PhysRevD.97.055006}{{\em Phys.
  Rev. D} {\bfseries 97} no.~5, (2018) 055006},
  \href{https://arxiv.org/abs/1709.07852}{{\ttfamily arXiv:1709.07852
  [hep-ph]}}.

\bibitem{Berlin:2023ubt}
A.~Berlin, A.~J. Millar, T.~Trickle, and K.~Zhou, ``{Physical signatures of
  fermion-coupled axion dark matter},''
  \href{https://dx.doi.org/10.1007/JHEP05(2024)314}{{\em JHEP} {\bfseries 05}
  (2024) 314}, \href{https://arxiv.org/abs/2312.11601}{{\ttfamily
  arXiv:2312.11601 [hep-ph]}}.

\bibitem{Pang:2018eec}
B.~Pang and Y.~Chen, ``{Quantum interactions between a laser interferometer and
  gravitational waves},''
  \href{https://dx.doi.org/10.1103/PhysRevD.98.124006}{{\em Phys. Rev. D}
  {\bfseries 98} no.~12, (2018) 124006},
  \href{https://arxiv.org/abs/1808.09122}{{\ttfamily arXiv:1808.09122
  [quant-ph]}}.

\bibitem{Domcke:2024mfu}
V.~Domcke, S.~A.~R. Ellis, and N.~L. Rodd, ``{Magnets are Weber Bar
  Gravitational Wave Detectors},''
  \href{https://dx.doi.org/10.1103/966v-r5fm}{{\em Phys. Rev. Lett.} {\bfseries
  134} no.~23, (2025) 231401},
  \href{https://arxiv.org/abs/2408.01483}{{\ttfamily arXiv:2408.01483
  [hep-ph]}}.

\bibitem{Damour:1994zq}
T.~Damour and A.~M. Polyakov, ``{The string dilaton and a least coupling
  principle},'' \href{https://dx.doi.org/10.1016/0550-3213(94)90143-0}{{\em
  Nucl. Phys. B} {\bfseries 423} (1994) 532--558},
  \href{https://arxiv.org/abs/hep-th/9401069}{{\ttfamily
  arXiv:hep-th/9401069}}.

\bibitem{Damour:2010rp}
T.~Damour and J.~F. Donoghue, ``{Equivalence Principle Violations and Couplings
  of a Light Dilaton},''
  \href{https://dx.doi.org/10.1103/PhysRevD.82.084033}{{\em Phys. Rev. D}
  {\bfseries 82} (2010) 084033},
  \href{https://arxiv.org/abs/1007.2792}{{\ttfamily arXiv:1007.2792 [gr-qc]}}.

\bibitem{Arvanitaki:2014faa}
A.~Arvanitaki, J.~Huang, and K.~Van~Tilburg, ``{Searching for dilaton dark
  matter with atomic clocks},''
  \href{https://dx.doi.org/10.1103/PhysRevD.91.015015}{{\em Phys. Rev. D}
  {\bfseries 91} no.~1, (2015) 015015},
  \href{https://arxiv.org/abs/1405.2925}{{\ttfamily arXiv:1405.2925 [hep-ph]}}.

\bibitem{Stadnik:2015kia}
Y.~V. Stadnik and V.~V. Flambaum, ``{Can dark matter induce cosmological
  evolution of the fundamental constants of Nature?},''
  \href{https://dx.doi.org/10.1103/PhysRevLett.115.201301}{{\em Phys. Rev.
  Lett.} {\bfseries 115} no.~20, (2015) 201301},
  \href{https://arxiv.org/abs/1503.08540}{{\ttfamily arXiv:1503.08540
  [astro-ph.CO]}}.

\bibitem{Guerreiro:2019vbq}
T.~Guerreiro, ``{Quantum Effects in Gravity Waves},''
  \href{https://dx.doi.org/10.1088/1361-6382/ab9d5d}{{\em Class. Quant. Grav.}
  {\bfseries 37} no.~15, (2020) 155001},
  \href{https://arxiv.org/abs/1911.11593}{{\ttfamily arXiv:1911.11593
  [quant-ph]}}.

\bibitem{Hees:2018fpg}
A.~Hees, O.~Minazzoli, E.~Savalle, Y.~V. Stadnik, and P.~Wolf, ``{Violation of
  the equivalence principle from light scalar dark matter},''
  \href{https://dx.doi.org/10.1103/PhysRevD.98.064051}{{\em Phys. Rev. D}
  {\bfseries 98} no.~6, (2018) 064051},
  \href{https://arxiv.org/abs/1807.04512}{{\ttfamily arXiv:1807.04512
  [gr-qc]}}.

\bibitem{Banerjee:2022sqg}
A.~Banerjee, G.~Perez, M.~Safronova, I.~Savoray, and A.~Shalit, ``{The
  phenomenology of quadratically coupled ultra light dark matter},''
  \href{https://dx.doi.org/10.1007/JHEP10(2023)042}{{\em JHEP} {\bfseries 10}
  (2023) 042}, \href{https://arxiv.org/abs/2211.05174}{{\ttfamily
  arXiv:2211.05174 [hep-ph]}}.

\bibitem{Kim:2022ype}
H.~Kim and G.~Perez, ``{Oscillations of atomic energy levels induced by QCD
  axion dark matter},''
  \href{https://dx.doi.org/10.1103/PhysRevD.109.015005}{{\em Phys. Rev. D}
  {\bfseries 109} no.~1, (2024) 015005},
  \href{https://arxiv.org/abs/2205.12988}{{\ttfamily arXiv:2205.12988
  [hep-ph]}}.

\bibitem{Kim:2023pvt}
H.~Kim, A.~Lenoci, G.~Perez, and W.~Ratzinger, ``{Probing an ultralight QCD
  axion with electromagnetic quadratic interaction},''
  \href{https://dx.doi.org/10.1103/PhysRevD.109.015030}{{\em Phys. Rev. D}
  {\bfseries 109} no.~1, (2024) 015030},
  \href{https://arxiv.org/abs/2307.14962}{{\ttfamily arXiv:2307.14962
  [hep-ph]}}.

\bibitem{Beadle:2023flm}
C.~Beadle, S.~A.~R. Ellis, J.~Quevillon, and P.~N. Hoa~Vuong, ``{Quadratic
  coupling of the axion to photons},''
  \href{https://dx.doi.org/10.1103/PhysRevD.110.035019}{{\em Phys. Rev. D}
  {\bfseries 110} no.~3, (2024) 035019},
  \href{https://arxiv.org/abs/2307.10362}{{\ttfamily arXiv:2307.10362
  [hep-ph]}}.

\bibitem{Ma:2018bxa}
Y.~Ma, F.~Armata, K.~E. Khosla, and M.~S. Kim, ``{Optical squeezing for an
  optomechanical system without quantizing the mechanical motion},''
  \href{https://dx.doi.org/10.1103/PhysRevResearch.2.023208}{{\em Phys. Rev.
  Res.} {\bfseries 2} no.~2, (2020) 023208},
  \href{https://arxiv.org/abs/1811.04025}{{\ttfamily arXiv:1811.04025
  [quant-ph]}}.

\bibitem{Howl:2020isj}
R.~Howl, V.~Vedral, D.~Naik, M.~Christodoulou, C.~Rovelli, and A.~Iyer,
  ``{Non-Gaussianity as a Signature of a Quantum Theory of Gravity},''
  \href{https://dx.doi.org/10.1103/PRXQuantum.2.010325}{{\em PRX Quantum}
  {\bfseries 2} no.~1, (2021) 010325},
  \href{https://arxiv.org/abs/2004.01189}{{\ttfamily arXiv:2004.01189
  [quant-ph]}}.

\bibitem{Eberhardt:2022rcp}
A.~Eberhardt, M.~Kopp, and T.~Abel, ``{When quantum corrections alter the
  predictions of classical field theory for scalar field dark matter},''
  \href{https://dx.doi.org/10.1103/PhysRevD.106.103002}{{\em Phys. Rev. D}
  {\bfseries 106} no.~10, (2022) 103002},
  \href{https://arxiv.org/abs/2206.06519}{{\ttfamily arXiv:2206.06519
  [hep-ph]}}.

\bibitem{Eberhardt:2023axk}
A.~Eberhardt, A.~Zamora, M.~Kopp, and T.~Abel, ``{Classical field approximation
  of ultralight dark matter: Quantum break times, corrections, and
  decoherence},'' \href{https://dx.doi.org/10.1103/PhysRevD.109.083527}{{\em
  Phys. Rev. D} {\bfseries 109} no.~8, (2024) 083527},
  \href{https://arxiv.org/abs/2310.07119}{{\ttfamily arXiv:2310.07119
  [astro-ph.CO]}}.

\bibitem{Allali:2020shm}
I.~J. Allali and M.~P. Hertzberg, ``{Decoherence from General Relativity},''
  \href{https://dx.doi.org/10.1103/PhysRevD.103.104053}{{\em Phys. Rev. D}
  {\bfseries 103} no.~10, (2021) 104053},
  \href{https://arxiv.org/abs/2012.12903}{{\ttfamily arXiv:2012.12903
  [gr-qc]}}.

\bibitem{Anselm:1985obz}
A.~A. Anselm, ``{Arion $\leftrightarrow$ Photon Oscillations in a Steady
  Magnetic Field. (In Russian)},'' {\em Yad. Fiz.} {\bfseries 42} (1985)
  1480--1483.

\bibitem{VanBibber:1987rq}
K.~Van~Bibber, N.~R. Dagdeviren, S.~E. Koonin, A.~Kerman, and H.~N. Nelson,
  ``{Proposed experiment to produce and detect light pseudoscalars},''
  \href{https://dx.doi.org/10.1103/PhysRevLett.59.759}{{\em Phys. Rev. Lett.}
  {\bfseries 59} (1987) 759--762}.

\bibitem{Redondo:2010dp}
J.~Redondo and A.~Ringwald, ``{Light shining through walls},''
  \href{https://dx.doi.org/10.1080/00107514.2011.563516}{{\em Contemp. Phys.}
  {\bfseries 52} (2011) 211--236},
  \href{https://arxiv.org/abs/1011.3741}{{\ttfamily arXiv:1011.3741 [hep-ph]}}.

\bibitem{ALPS:2009des}
{\bfseries ALPS} Collaboration, K.~Ehret {\em et~al.}, ``{Resonant laser power
  build-up in ALPS: A 'Light-shining-through-walls' experiment},''
  \href{https://dx.doi.org/10.1016/j.nima.2009.10.102}{{\em Nucl. Instrum.
  Meth. A} {\bfseries 612} (2009) 83--96},
  \href{https://arxiv.org/abs/0905.4159}{{\ttfamily arXiv:0905.4159
  [physics.ins-det]}}.

\bibitem{Ehret:2010mh}
K.~Ehret {\em et~al.}, ``{New ALPS Results on Hidden-Sector Lightweights},''
  \href{https://dx.doi.org/10.1016/j.physletb.2010.04.066}{{\em Phys. Lett. B}
  {\bfseries 689} (2010) 149--155},
  \href{https://arxiv.org/abs/1004.1313}{{\ttfamily arXiv:1004.1313 [hep-ex]}}.

\bibitem{ALPSII:2025eri}
{\bfseries ALPS II} Collaboration, D.~C. Brotherton {\em et~al.}, ``{Any Light
  Particle Searches with ALPS II: first science results},''
  \href{https://arxiv.org/abs/2512.14110}{{\ttfamily arXiv:2512.14110
  [hep-ex]}}.

\bibitem{Grishchuk:1990bj}
L.~P. Grishchuk and Y.~V. Sidorov, ``{Squeezed quantum states of relic
  gravitons and primordial density fluctuations},''
  \href{https://dx.doi.org/10.1103/PhysRevD.42.3413}{{\em Phys. Rev. D}
  {\bfseries 42} (1990) 3413--3421}.

\bibitem{Albrecht:1992kf}
A.~Albrecht, P.~Ferreira, M.~Joyce, and T.~Prokopec, ``{Inflation and squeezed
  quantum states},'' \href{https://dx.doi.org/10.1103/PhysRevD.50.4807}{{\em
  Phys. Rev. D} {\bfseries 50} (1994) 4807--4820},
  \href{https://arxiv.org/abs/astro-ph/9303001}{{\ttfamily
  arXiv:astro-ph/9303001}}.

\bibitem{Polarski:1995jg}
D.~Polarski and A.~A. Starobinsky, ``{Semiclassicality and decoherence of
  cosmological perturbations},''
  \href{https://dx.doi.org/10.1088/0264-9381/13/3/006}{{\em Class. Quant.
  Grav.} {\bfseries 13} (1996) 377--392},
  \href{https://arxiv.org/abs/gr-qc/9504030}{{\ttfamily arXiv:gr-qc/9504030}}.

\bibitem{Kiefer:2008ku}
C.~Kiefer and D.~Polarski, ``{Why do cosmological perturbations look classical
  to us?},'' \href{https://dx.doi.org/10.1166/asl.2009.1023}{{\em Adv. Sci.
  Lett.} {\bfseries 2} (2009) 164--173},
  \href{https://arxiv.org/abs/0810.0087}{{\ttfamily arXiv:0810.0087
  [astro-ph]}}.

\bibitem{Laga:2026vwm}
F.~Laga and T.~Suyama, ``{Quantum description of gravitational waves generated
  by a classical source},'' \href{https://arxiv.org/abs/2604.20228}{{\ttfamily
  arXiv:2604.20228 [gr-qc]}}.

\bibitem{Kanno:2025how}
S.~Kanno, J.~Soda, and A.~Taniguchi, ``{Coherent State Description of
  Gravitational Waves from Binary Black Holes},''
  \href{https://dx.doi.org/10.1103/kv1t-j27m}{{\em Phys. Rev. Lett.} {\bfseries
  136} no.~6, (2026) 061404},
  \href{https://arxiv.org/abs/2508.17947}{{\ttfamily arXiv:2508.17947
  [gr-qc]}}.

\bibitem{Dorlis:2025zzz}
P.~Dorlis, N.~E. Mavromatos, S.~Sarkar, and S.-N. Vlachos, ``{Superradiant
  Axionic Black-Hole Clouds as Seeds for Graviton Squeezing},''
  \href{https://dx.doi.org/10.1103/9crd-zj6l}{{\em Phys. Rev. Lett.} {\bfseries
  135} no.~15, (2025) 151501},
  \href{https://arxiv.org/abs/2507.01689}{{\ttfamily arXiv:2507.01689
  [gr-qc]}}.

\bibitem{Dorlis:2025amf}
P.~Dorlis, N.~E. Mavromatos, S.~Sarkar, and S.-N. Vlachos, ``{Squeezed
  gravitons from superradiant axion fields around rotating black holes},''
  \href{https://dx.doi.org/10.1103/t822-86qp}{{\em Phys. Rev. D} {\bfseries
  113} no.~2, (2026) 026023},
  \href{https://arxiv.org/abs/2507.23475}{{\ttfamily arXiv:2507.23475
  [gr-qc]}}.

\bibitem{Loughlin:2025rih}
H.~A. Loughlin, G.~Tobar, E.~D. Hall, and V.~Sudhir, ``{Wave-particle duality
  in the measurement of gravitational radiation},''
  \href{https://dx.doi.org/10.1103/6dzy-4c45}{{\em Phys. Rev. Res.} {\bfseries
  7} no.~4, (2025) 043286}, \href{https://arxiv.org/abs/2504.03527}{{\ttfamily
  arXiv:2504.03527 [quant-ph]}}.

\bibitem{Gouin:2026mdz}
Y.~Gouin, S.~Kanno, and J.~Soda, ``{Intermittency in Quantum Graviton-Phonon
  Conversion},'' \href{https://arxiv.org/abs/2607.20107}{{\ttfamily
  arXiv:2607.20107 [gr-qc]}}.

\bibitem{Manikandan:2024fmf}
S.~K. Manikandan and F.~Wilczek, ``{Testing the Coherent State Description of
  Radiation Fields},''
  \href{https://dx.doi.org/10.1103/PhysRevA.111.033705}{{\em Phys. Rev. A}
  {\bfseries 111} no.~3, (2025) 033705},
  \href{https://arxiv.org/abs/2409.20378}{{\ttfamily arXiv:2409.20378
  [quant-ph]}}.

\bibitem{Manikandan:2025hlz}
S.~K. Manikandan and F.~Wilczek, ``{Probing quantum structure in gravitational
  radiation},'' \href{https://dx.doi.org/10.1142/S0218271825430011}{{\em Int.
  J. Mod. Phys. D} {\bfseries 34} no.~16, (2025) 2543001},
  \href{https://arxiv.org/abs/2505.11407}{{\ttfamily arXiv:2505.11407
  [gr-qc]}}.

\bibitem{Arani:2026jyz}
F.~S. Arani, B.~Lamine, and J.~Soda, ``{Cavity-QED Transducer of Gravitons},''
  \href{https://arxiv.org/abs/2603.27687}{{\ttfamily arXiv:2603.27687
  [quant-ph]}}.

\bibitem{Kanno:2018cuk}
S.~Kanno and J.~Soda, ``{Detecting nonclassical primordial gravitational waves
  with Hanbury-Brown{\textendash}Twiss interferometry},''
  \href{https://dx.doi.org/10.1103/PhysRevD.99.084010}{{\em Phys. Rev. D}
  {\bfseries 99} no.~8, (2019) 084010},
  \href{https://arxiv.org/abs/1810.07604}{{\ttfamily arXiv:1810.07604
  [hep-th]}}.

\bibitem{Kanno:2019gqw}
S.~Kanno, ``{Nonclassical primordial gravitational waves from the initial
  entangled state},''
  \href{https://dx.doi.org/10.1103/PhysRevD.100.123536}{{\em Phys. Rev. D}
  {\bfseries 100} no.~12, (2019) 123536},
  \href{https://arxiv.org/abs/1905.06800}{{\ttfamily arXiv:1905.06800
  [hep-th]}}.

\bibitem{Toccacelo:2026hcz}
K.~Toccacelo, T.~Beitel, U.~L. Andersen, and I.~Pikovski, ``{Quantum State
  Characterization of Gravitational Waves via Graviton Counting Statistics},''
  \href{https://arxiv.org/abs/2602.09125}{{\ttfamily arXiv:2602.09125
  [quant-ph]}}.

\bibitem{Manikandan:2025lfx}
S.~K. Manikandan and F.~Wilczek, ``{Detector correlations and null tests of the
  coherent state hypothesis},''
  \href{https://dx.doi.org/10.1142/s0217751x26420017}{{\em Int. J. Mod. Phys.
  A} {\bfseries 41} no.~11, (2026) 2642001},
  \href{https://arxiv.org/abs/2508.03367}{{\ttfamily arXiv:2508.03367
  [quant-ph]}}.

\bibitem{Athulya:2026wpl}
K.~P. Athulya and S.~K. Manikandan, ``{Correlated Quantum Sensing at the
  Seemingly Classical Limit},''
  \href{https://arxiv.org/abs/2606.01673}{{\ttfamily arXiv:2606.01673
  [quant-ph]}}.

\bibitem{Parikh:2020nrd}
M.~Parikh, F.~Wilczek, and G.~Zahariade, ``{The Noise of Gravitons},''
  \href{https://dx.doi.org/10.1142/S0218271820420018}{{\em Int. J. Mod. Phys.
  D} {\bfseries 29} no.~14, (2020) 2042001},
  \href{https://arxiv.org/abs/2005.07211}{{\ttfamily arXiv:2005.07211
  [hep-th]}}.

\bibitem{Parikh:2020fhy}
M.~Parikh, F.~Wilczek, and G.~Zahariade, ``{Signatures of the quantization of
  gravity at gravitational wave detectors},''
  \href{https://dx.doi.org/10.1103/PhysRevD.104.046021}{{\em Phys. Rev. D}
  {\bfseries 104} no.~4, (2021) 046021},
  \href{https://arxiv.org/abs/2010.08208}{{\ttfamily arXiv:2010.08208
  [hep-th]}}.

\bibitem{Kanno:2020usf}
S.~Kanno, J.~Soda, and J.~Tokuda, ``{Noise and decoherence induced by
  gravitons},'' \href{https://dx.doi.org/10.1103/PhysRevD.103.044017}{{\em
  Phys. Rev. D} {\bfseries 103} no.~4, (2021) 044017},
  \href{https://arxiv.org/abs/2007.09838}{{\ttfamily arXiv:2007.09838
  [hep-th]}}.

\bibitem{Guerreiro:2021qgk}
T.~Guerreiro, F.~Coradeschi, A.~M. Frassino, J.~R. West, and E.~J. Schioppa,
  ``{Quantum signatures in nonlinear gravitational waves},''
  \href{https://dx.doi.org/10.22331/q-2022-12-19-879}{{\em Quantum} {\bfseries
  6} (2022) 879}, \href{https://arxiv.org/abs/2111.01779}{{\ttfamily
  arXiv:2111.01779 [gr-qc]}}.

\bibitem{Manikandan:2025qgv}
S.~K. Manikandan and F.~Wilczek, ``{Complementary probes of gravitational
  radiation states},'' \href{https://dx.doi.org/10.1103/83tt-tt57}{{\em Phys.
  Rev. A} {\bfseries 112} no.~4, (2025) 043716},
  \href{https://arxiv.org/abs/2505.11422}{{\ttfamily arXiv:2505.11422
  [gr-qc]}}.

\bibitem{Dorlis:2026gth}
P.~Dorlis, N.~E. Mavromatos, S.~Sarkar, and S.-N. Vlachos, ``{How Much Can
  Gravitons Be Squeezed?},'' \href{https://arxiv.org/abs/2605.14797}{{\ttfamily
  arXiv:2605.14797 [gr-qc]}}.

\bibitem{Cho:2021gvg}
H.-T. Cho and B.-L. Hu, ``{Quantum noise of gravitons and stochastic force on
  geodesic separation},''
  \href{https://dx.doi.org/10.1103/PhysRevD.105.086004}{{\em Phys. Rev. D}
  {\bfseries 105} no.~8, (2022) 086004},
  \href{https://arxiv.org/abs/2112.08174}{{\ttfamily arXiv:2112.08174
  [gr-qc]}}.

\bibitem{Parikh:2023zat}
M.~Parikh and F.~Setti, ``{Quantum-gravitational noise correlation in nearby
  detectors},'' \href{https://dx.doi.org/10.1103/PhysRevD.111.046004}{{\em
  Phys. Rev. D} {\bfseries 111} no.~4, (2025) 046004},
  \href{https://arxiv.org/abs/2312.17335}{{\ttfamily arXiv:2312.17335
  [gr-qc]}}.

\bibitem{Hertzberg:2021rbl}
M.~P. Hertzberg and J.~A. Litterer, ``{Bound on quantum fluctuations in
  gravitational waves from LIGO-Virgo},''
  \href{https://dx.doi.org/10.1088/1475-7516/2023/03/009}{{\em JCAP} {\bfseries
  03} (2023) 009}, \href{https://arxiv.org/abs/2112.12159}{{\ttfamily
  arXiv:2112.12159 [gr-qc]}}.

\bibitem{Sen:2024nhb}
S.~Sen and S.~Gangopadhyay, ``{Probing the quantum nature of gravity using a
  Bose-Einstein condensate},''
  \href{https://dx.doi.org/10.1103/PhysRevD.110.026014}{{\em Phys. Rev. D}
  {\bfseries 110} no.~2, (2024) 026014},
  \href{https://arxiv.org/abs/2403.18460}{{\ttfamily arXiv:2403.18460
  [hep-th]}}.

\bibitem{Carney:2024wnp}
D.~Carney, M.~Karydas, and A.~Sivaramakrishnan, ``{Response of interferometers
  to the vacuum of quantum gravity},''
  \href{https://dx.doi.org/10.1103/j5kj-zdky}{{\em Phys. Rev. D} {\bfseries
  113} no.~10, (2026) 106002},
  \href{https://arxiv.org/abs/2409.03894}{{\ttfamily arXiv:2409.03894
  [hep-th]}}.

\bibitem{Sen:2024rot}
S.~Sen and S.~Gangopadhyay, ``{Quantum nature of gravity in a Bose-Einstein
  condensate},'' \href{https://dx.doi.org/10.1103/PhysRevD.111.066002}{{\em
  Phys. Rev. D} {\bfseries 111} no.~6, (2025) 066002},
  \href{https://arxiv.org/abs/2410.05184}{{\ttfamily arXiv:2410.05184
  [hep-th]}}.

\bibitem{Nandi:2026sww}
P.~Nandi, S.~Sahu, B.~R. Majhi, and F.~Petruccione, ``{State-Selective
  Signatures of Quantum and Classical Gravitational Environments},''
  \href{https://arxiv.org/abs/2603.05731}{{\ttfamily arXiv:2603.05731
  [gr-qc]}}.

\bibitem{Marshman:2019sne}
R.~J. Marshman, A.~Mazumdar, and S.~Bose, ``{Locality and entanglement in
  table-top testing of the quantum nature of linearized gravity},''
  \href{https://dx.doi.org/10.1103/PhysRevA.101.052110}{{\em Phys. Rev. A}
  {\bfseries 101} no.~5, (2020) 052110},
  \href{https://arxiv.org/abs/1907.01568}{{\ttfamily arXiv:1907.01568
  [quant-ph]}}.

\bibitem{Maity:2021zng}
D.~Maity and S.~Pal, ``{Probing non-classicality of primordial gravitational
  waves and magnetic field through quantum Poincare sphere},''
  \href{https://dx.doi.org/10.1016/j.physletb.2022.137503}{{\em Phys. Lett. B}
  {\bfseries 835} (2022) 137503},
  \href{https://arxiv.org/abs/2107.12793}{{\ttfamily arXiv:2107.12793
  [gr-qc]}}.

\bibitem{Kanno:2021vwu}
S.~Kanno and J.~Soda, ``{Squeezed quantum states of graviton and axion in the
  universe},'' \href{https://dx.doi.org/10.1142/S0218271822500985}{{\em Int. J.
  Mod. Phys. D} {\bfseries 31} no.~13, (2022) 2250098},
  \href{https://arxiv.org/abs/2112.14496}{{\ttfamily arXiv:2112.14496
  [gr-qc]}}.

\bibitem{Ikeda:2025uae}
T.~Ikeda, Y.~Kaku, S.~Kanno, and J.~Soda, ``{Toward graviton detection via
  photon-graviton quantum state conversion},''
  \href{https://dx.doi.org/10.1103/kxs5-yp85}{{\em Phys. Rev. D} {\bfseries
  112} no.~10, (2025) 103504},
  \href{https://arxiv.org/abs/2507.01609}{{\ttfamily arXiv:2507.01609
  [quant-ph]}}.

\bibitem{Lentz:2025lkg}
E.~W. Lentz, ``{Signatures from Non-classical Features of Axion Dark Matter in
  Cavity Haloscopes},'' \href{https://arxiv.org/abs/2509.03877}{{\ttfamily
  arXiv:2509.03877 [hep-ph]}}.

\bibitem{Mavromatos:2026nin}
N.~E. Mavromatos and S.~Sarkar, ``{The $\omega$-Effect from a Multimode
  Squeezed Graviton State},''
  \href{https://arxiv.org/abs/2606.24613}{{\ttfamily arXiv:2606.24613
  [gr-qc]}}.

\bibitem{milonni1976semiclassical}
P.~W. Milonni, ``Semiclassical and quantum-electrodynamical approaches in
  nonrelativistic radiation theory,''
  \href{https://dx.doi.org/10.1016/0370-1573(76)90037-5}{{\em Phys. Rept.}
  {\bfseries 25} no.~1, (1976) 1--81}.

\bibitem{Britto:2021pud}
R.~Britto, R.~Gonzo, and G.~R. Jehu, ``{Graviton particle statistics and
  coherent states from classical scattering amplitudes},''
  \href{https://dx.doi.org/10.1007/JHEP03(2022)214}{{\em JHEP} {\bfseries 03}
  (2022) 214}, \href{https://arxiv.org/abs/2112.07036}{{\ttfamily
  arXiv:2112.07036 [hep-th]}}.

\bibitem{Cristofoli:2021jas}
A.~Cristofoli, R.~Gonzo, N.~Moynihan, D.~O'Connell, A.~Ross, M.~Sergola, and
  C.~D. White, ``{The uncertainty principle and classical amplitudes},''
  \href{https://dx.doi.org/10.1007/JHEP06(2024)181}{{\em JHEP} {\bfseries 06}
  (2024) 181}, \href{https://arxiv.org/abs/2112.07556}{{\ttfamily
  arXiv:2112.07556 [hep-th]}}.

\bibitem{Dyson:2013hbl}
F.~Dyson, ``{Is a graviton detectable?},''
  \href{https://dx.doi.org/10.1142/S0217751X1330041X}{{\em Int. J. Mod. Phys.
  A} {\bfseries 28} (2013) 1330041}.

\bibitem{Rothman:2006fp}
T.~Rothman and S.~Boughn, ``{Can gravitons be detected?},''
  \href{https://dx.doi.org/10.1007/s10701-006-9081-9}{{\em Found. Phys.}
  {\bfseries 36} (2006) 1801--1825},
  \href{https://arxiv.org/abs/gr-qc/0601043}{{\ttfamily arXiv:gr-qc/0601043}}.

\bibitem{Marletto:2017kzi}
C.~Marletto and V.~Vedral, ``{Gravitationally-induced entanglement between two
  massive particles is sufficient evidence of quantum effects in gravity},''
  \href{https://dx.doi.org/10.1103/PhysRevLett.119.240402}{{\em Phys. Rev.
  Lett.} {\bfseries 119} no.~24, (2017) 240402},
  \href{https://arxiv.org/abs/1707.06036}{{\ttfamily arXiv:1707.06036
  [quant-ph]}}.

\bibitem{Bose:2017nin}
S.~Bose, A.~Mazumdar, G.~W. Morley, H.~Ulbricht, M.~Toros, M.~Paternostro,
  A.~Geraci, P.~Barker, M.~S. Kim, and G.~Milburn, ``{Spin Entanglement Witness
  for Quantum Gravity},''
  \href{https://dx.doi.org/10.1103/PhysRevLett.119.240401}{{\em Phys. Rev.
  Lett.} {\bfseries 119} no.~24, (2017) 240401},
  \href{https://arxiv.org/abs/1707.06050}{{\ttfamily arXiv:1707.06050
  [quant-ph]}}.

\bibitem{Carney:2018ofe}
D.~Carney, P.~C.~E. Stamp, and J.~M. Taylor, ``{Tabletop experiments for
  quantum gravity: a user's manual},''
  \href{https://dx.doi.org/10.1088/1361-6382/aaf9ca}{{\em Class. Quant. Grav.}
  {\bfseries 36} no.~3, (2019) 034001},
  \href{https://arxiv.org/abs/1807.11494}{{\ttfamily arXiv:1807.11494
  [quant-ph]}}.

\bibitem{COLPA1978327}
J.~Colpa, ``Diagonalization of the quadratic boson hamiltonian,''
  \href{https://dx.doi.org/10.1016/0378-4371(78)90160-7}{{\em Physica A}
  {\bfseries 93} no.~3, (1978) 327--353}.

\bibitem{PhysRevA.94.013814}
J.~Sperling, ``Characterizing maximally singular phase-space distributions,''
  \href{https://dx.doi.org/10.1103/PhysRevA.94.013814}{{\em Phys. Rev. A}
  {\bfseries 94} (Jul, 2016) 013814}.

\bibitem{mehta1967diagonal}
C.~L. Mehta, ``Diagonal coherent-state representation of quantum operators,''
  \href{https://dx.doi.org/10.1103/PhysRevLett.18.752}{{\em Phys. Rev. Lett.}
  {\bfseries 18} (1967) 752}.

\bibitem{Cramer_1970}
H.~Cram{\'e}r, \href{https://dx.doi.org/10.1017/CBO9780511470936}{{\em Random
  Variables and Probability Distributions}}.
\newblock No.~36 in Cambridge Tracts in Mathematics. Cambridge University
  Press, Cambridge, 3~ed., 1970.

\bibitem{HILLERY1985409}
M.~Hillery, ``Classical pure states are coherent states,''
  \href{https://dx.doi.org/10.1016/0375-9601(85)90483-9}{{\em Phys. Lett. A}
  {\bfseries 111} no.~8, (1985) 409--411}.

\end{thebibliography}\endgroup

\end{document}